\documentclass[mnsc,nonblindrev]{informs3noheader}

\usepackage{mathtools}
\usepackage{endnotes}
\usepackage{mathrsfs}
\usepackage{multirow}
\usepackage{subfigure}
\usepackage{xcolor}
\usepackage{bm}
\usepackage{dsfont}
\usepackage{amsmath}
\usepackage{amssymb}
\usepackage{graphicx}
\usepackage[colorlinks=TRUE,citecolor=MyDarkBlue,urlcolor=MyDarkBlue,linkcolor=MyDarkBlue]{hyperref}
\definecolor{MyDarkBlue}{RGB}{158,0,0}
\newcolumntype{L}[1]{>{\raggedright\let\newline\\\arraybackslash\hspace{0pt}}m{#1}}
\newcolumntype{C}[1]{>{\centering\let\newline\\\arraybackslash\hspace{0pt}}m{#1}}
\newcolumntype{R}[1]{>{\raggedleft\let\newline\\\arraybackslash\hspace{0pt}}m{#1}}

\usepackage{xurl}
\usepackage{ulem}

\usepackage{algorithm}
\usepackage{algpseudocode}
\usepackage{tikz}

\usepackage{natbib}
 \bibpunct[, ]{(}{)}{,}{a}{}{,}%
 \def\bibfont{\footnotesize}%
\EquationsNumberedThrough    

\TheoremsNumberedThrough     
\ECRepeatTheorems  %

\MANUSCRIPTNO{}

\newcommand{\tradition}{\textsf{ITR}}
\newcommand{\our}{\textsf{DPTR}}
\newcommand{\ourP}{\textsf{DPTR-P}}

\newcommand*{\QED}{\hfill\ensuremath{\square}}

\begin{document}


\RUNAUTHOR{Peng et al.}

\RUNTITLE{Data-Pooling for Treatment Selection}
\TITLE{Synthesizing Evidence: Data-Pooling as a Tool for Treatment Selection in Online Experiments}

\ARTICLEAUTHORS{%
\AUTHOR{Zhenkang Peng}
\AFF{The Chinese University of Hong Kong, \EMAIL{zhenkang.peng@cuhk.edu.hk}}

\AUTHOR{Chengzhang Li}
\AFF{Shanghai Jiao Tong University, \EMAIL{cz.li@sjtu.edu.cn}}

\AUTHOR{Ying Rong}
\AFF{Shanghai Jiao Tong University, \EMAIL{yrong@sjtu.edu.cn}}

\AUTHOR{Renyu (Philip) Zhang}
\AFF{The Chinese University of Hong Kong, \EMAIL{philipzhang@cuhk.edu.hk}}
} 

\ABSTRACT{%
Randomized experiments are the gold standard for causal inference but face significant challenges in business applications, including limited traffic allocation, the need for heterogeneous treatment effect estimation, and the complexity of managing overlapping experiments. These factors lead to high variability in treatment effect estimates, making data-driven policy roll-out difficult. To address these issues, we introduce the data-pooling treatment roll-out (\our) framework, which enhances policy roll-out by pooling data across experiments rather than focusing narrowly on individual ones. We establish formal theoretical guarantees for \our\ in non-overlapping experiments under linear specifications, and evaluate its performance under overlapping traffic, rich covariates, and nonlinear specifications through synthetic simulations and real-world applications. We demonstrate the framework’s robustness through a three-pronged validation: (a) theoretical analysis shows that \our\ surpasses the traditional difference-in-means and ordinary least squares methods  under non-overlapping experiments, particularly when the number of experiments is large; (b) synthetic simulations confirm its adaptability in complex scenarios with overlapping traffic, rich covariates and nonlinear specifications; and (c) empirical applications to two experimental datasets from real-world platforms, demonstrate its effectiveness in guiding customized policy roll-outs for subgroups within a single experiment and coordinating policy deployments across multiple experiments with overlapping scenarios. By reducing estimation variability to improve decision-making effectiveness, \our\ provides a scalable, practical solution for online platforms to better leverage their experimental data in today’s increasingly complex business environments.
}%





\maketitle

\section{Introduction}

Randomized experiments have long been regarded as the gold standard for estimating causal effects across a wide range of scientific disciplines. The adoption of randomized experiments in business, especially in the tech sector, has gained significant momentum in recent years. Online platforms routinely use randomized control trials (RCTs) to shape a wide array of decisions, including the product design, UI, recommendation algorithms, ad placement, and pricing strategies \citep{LucaBazerman2021}. The need for rapid validation and deployment has led companies to run hundreds or even thousands of RCTs concurrently. At Bing, for example, the number of completed experiments increased from fewer than 50 per week in 2008 to more than 300 per week by 2014 \citep{KohaviThomke2017}. Booking runs in excess of 1,000 concurrent experiments at any given moment across different products and target groups \citep{Booking2019}. The widespread implementation of RCTs is not incidental: \citet{KohaviTangXu2020} report that companies such as Microsoft, Google, and LinkedIn now conduct over 20,000 experiments annually. 

A central issue faced by randomized experiments is data scarcity. Despite the large overall user base of many platforms, the total amount of traffic that can be allocated to experiments is often constrained, either due to operational limitations or risk concerns, which further reduces the effective sample per experiment as the number of concurrent experiments increases.  For instance, \citet{LewisRao2015} examine 25 digital advertising RCTs conducted by large retailers and financial service firms. Their findings show that the median confidence interval for return on investment (ROI) is over 100 percentage points wide, making it nearly impossible for advertisers to distinguish between campaigns with a 50\% ROI difference. This problem becomes more acute in the context of targeted experimentation and personalization, which are now standard practices in digital marketing. When experiments are stratified by user attributes to tailor interventions to specific segments, the resulting sample sizes for each subgroup can become vanishingly small, undermining the statistical power of the experiment \citep[e.g.,][]{SusanGuido2016,LadaPeysakhovichAparicioBailey2019}. Furthermore, orthogonal experimental designs, often used to enable efficient estimation across high-dimensional treatment spaces, can exacerbate the issue by creating treatment combinations that are either underrepresented or completely unobserved in practice \citep[see,][]{ye2023deep}. 
These factors together pose a major barrier to identifying effective policy interventions and learning from past experiments, especially in fast-paced environments where decisions must be made with limited data and high uncertainty.

Given these challenges, our central research question is: {\it when running a large number of experiments and observing limited data for each, how should we improve the experiment roll-out decisions under data scarcity? }
We propose a novel data-pooling treatment roll-out framework (\our). Instead of analyzing each experiment in isolation, \our\ integrates data across multiple experiments to enhance experiment roll-out decisions, where the ``roll-out decision" refers to the policy selection in this paper, aiming at reducing variance with a tolerable increase in bias. In other words, this approach determines whether a policy should be implemented by leveraging both the data collected from its own experiment and the pooled data from other experiments. Specifically, the treatment effect estimation is conducted by combining an individual estimator and an anchor estimator, defined as the average of all individual estimators, via a data-driven scale parameter. More importantly, the proposed \our\ framework is flexible and can be adapted to handle various experimentation scenarios, which differ along two dimensions: whether the experiments involve overlapping traffic and whether the underlying model is nonlinear. Beyond estimating average treatment effects (ATEs) by pooling data across experiments, the \our\ framework, by design, is also capable of accounting for heterogeneous treatment effects (HTEs) within subgroups in a single experiment. This enables the \our\ framework to be broadly applicable to real-world business contexts, where experimentation scenarios are increasingly complex and data availability is limited.
\vspace{-0.4in}
\begin{figure}[htbp!]
\centering
\subfigure[Averaged total reward over 1000 instances.]{
\begin{minipage}[t]{0.5\linewidth}
\centering
\includegraphics[width=2.5in,height=1.8in]{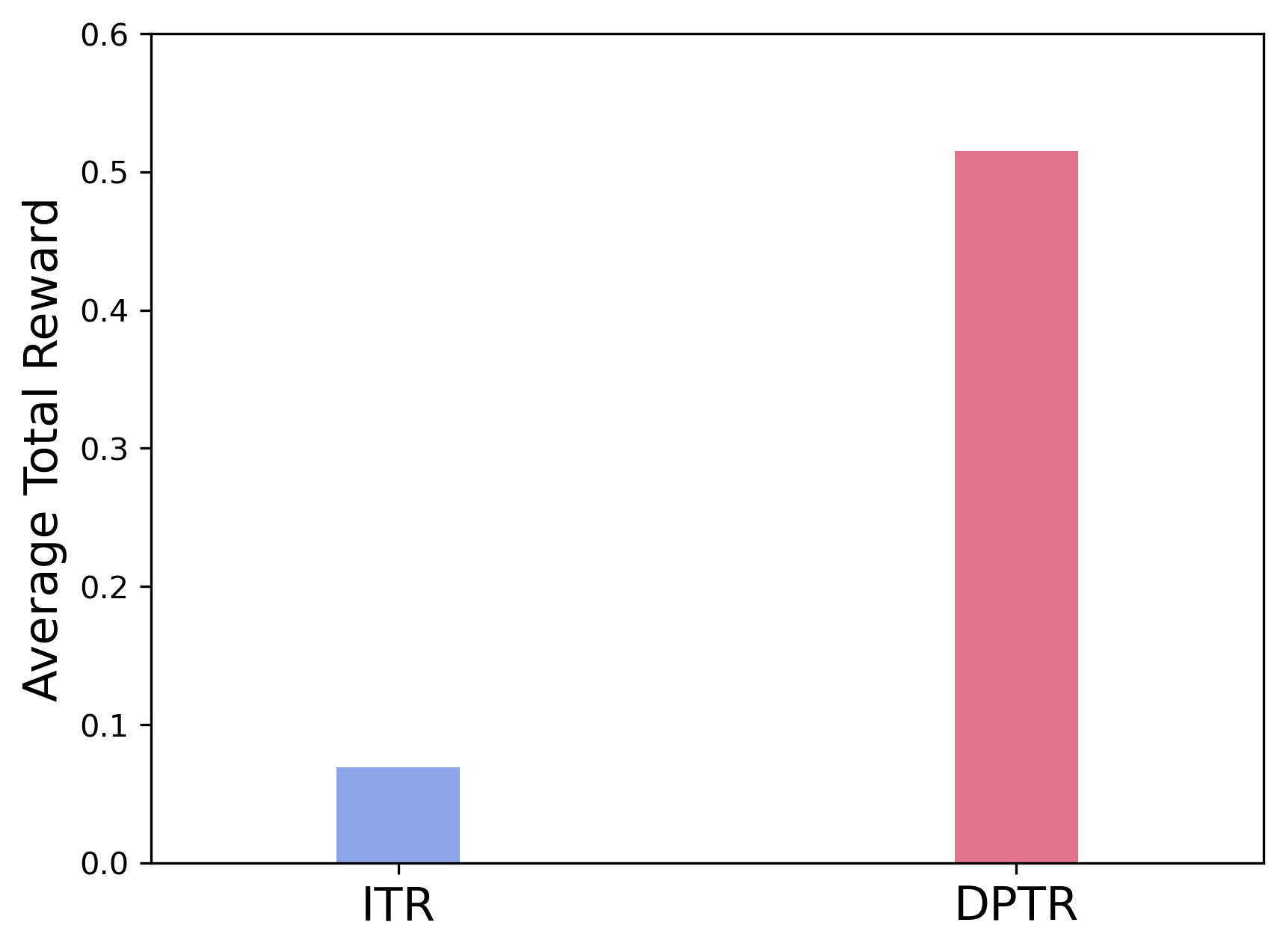}
\end{minipage}
}%
\subfigure[Decision-bound analysis in one instance.]{
\begin{minipage}[t]{0.5\linewidth}
\centering
\includegraphics[width=2.5in,height=1.8in]{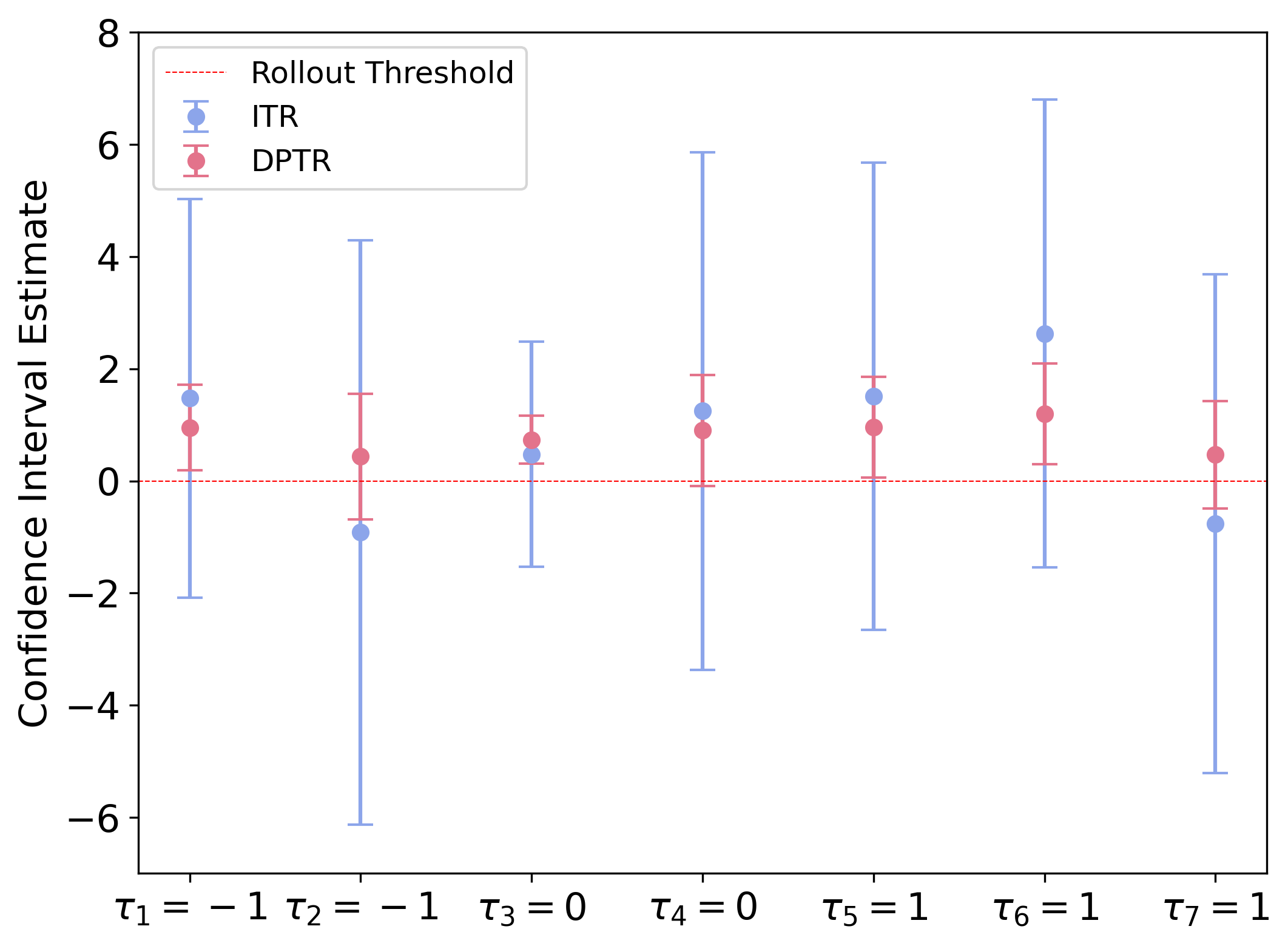}
\end{minipage}%
}%
\caption{A simple numerical case study with seven experiments.}
\label{fig:case_study}
\parbox[t]{0.95\linewidth}{\footnotesize Notes. Panel (a) reports the average total reward of \tradition\ and \our\ over 1000 simulated instances, each consisting of seven experiments. For each instance, each experiment contains 10 observations, with 5 assigned to the control group and 5 assigned to the treatment group. Panel (b) illustrates the analysis for one representative instance. The $x$-axis labels in Panel (b) report the true average treatment effect $\tau_k$ of each individual experiment. The noise term in the realized outcome has a standard deviation of 3. We use the difference-in-means method to obtain the initial point estimates and confidence.} 
\vspace{-0.3in}
\end{figure}

{\bf A case study.} We illustrate the advantage of \our\ framework over the conventional individual treatment roll-out (\tradition) through a running example where seven experiments are conducted simultaneously. The \tradition\ method rolls out each treatment based on the treatment effect estimation individually. Each experiment collects 10 observations, and we consider a heterogeneous setting where ATEs differ across experiments,
including both positive and negative values. We report the average total reward over 1{,}000 instances in Figure \ref{fig:case_study}(a) and find that \our\ greatly improves the roll-out decision reward. Next, we compare the classical confidence intervals used by \tradition\ with the shrunken decision bounds used by \our\ in one instance to illustrate why \our\ can obtain a higher reward. As shown in Figure \ref{fig:case_study}(b), the limited sample size results in excessively wide confidence intervals for the \tradition\ method, giving rise to low statistical power and unreliable roll-out decisions. In contrast, our \our\ method substantially tightens the shrunken decision bounds, thereby improving the ability to identify promising treatments. Although our method may introduce some bias and may occasionally misclassify non-positive ATEs as positive, the resulting roll-out decisions still yield higher overall reward compared to those based on the \tradition\ method. In particular, compared to \tradition, \our\ successfully rolls out the 5th and 6th experiments with positive ATEs, although it also mistakenly rolls out the 1st experiment, whose ATE is in fact negative.


The Bayesian framework offers a natural and principled method for aggregating evidence, and thus, similar to the proposed \our\ framework, can be employed to support data pooling across experiments to improve estimation efficiency at the level of individual experiments or targeted subgroups. Despite this alignment, the two frameworks differ fundamentally in their conceptual basis. The proposed \our\ framework is implemented using frequentist point estimators and roll-out rules, in contrast to the Bayesian approach that models ATEs as random variables and makes decisions from posterior probabilities. More importantly, the \our\ framework is decision-aware. That is, estimation and inference are directly aligned with downstream decisions, such as treatment roll-out decisions, by incorporating thresholds or significance levels into the analysis. This decision-centric design can lead to substantial performance gains in real-world deployment scenarios, where the goal is not merely to estimate effects accurately but to identify and implement effective interventions.

We demonstrate the value of the proposed \our\ framework from three perspectives. First, we theoretically examine \our's performance in the case with non-overlapping experiments and linear model specifications. Under the maintained regularity conditions, we show that \our\ weakly improves on the conventional \tradition\ approach and strictly improves it in the positive-anchor regime emphasized in the paper. 
Second, we evaluate \our's performance using synthetic data across a variety of complex scenarios, including those involving nonlinear model specifications and overlapping experiments.
Finally, we apply \our\ to two experimental datasets from real-world platforms, demonstrating its effectiveness in guiding customized policy roll-outs for subgroups within a single experiment, as well as in coordinating policy deployments across multiple experiments under overlapping scenarios.

Our contributions are threefold:
\begin{enumerate}
    \item {\bf A general data-pooling treatment roll-out framework}. We propose a novel framework that enhances treatment selection by integrating data across multiple experiments. The treatment roll-out decision leverages both the data collected from its own experiment and the pooled data from other experiments. More importantly, the \our\ framework is decision-aware. That is, estimation and inference are directly aligned with downstream decisions, such as treatment roll-out decisions, by incorporating significance levels into the analysis. This framework is implemented in a broad set of settings: our formal guarantees cover non-overlapping experiments under linear specifications, while overlapping traffic, nonlinear specifications, and personalized roll-out decisions are evaluated through simulations and empirical applications. This general and flexible design makes \our\ well-suited for modern online platforms that face data scarcity and conduct high-dimensional experiments.
    \item {\bf Analysis and performance guarantee}. We provide theoretical guarantees that \our\ outperforms \tradition, particularly as the number of experiments increases. Specifically, in the positive-anchor regime emphasized in our theoretical analysis, the average rewards associated with roll-out decisions derived by \our\ with the optimal scale parameter are strictly higher than those of the \tradition\ method. This implies that \our\ strikes a delicate balance between bias and variance by deriving the optimal scale parameter for shrinkage.
    We further construct a consistent estimator of the optimal scale parameter, which guarantees the superior performance of the proposed approach as the number of experiments increases. This ensures the robust performance of the proposed approach under data-driven settings. These analyses establish \our\ as a theoretically grounded method for improving the roll-out of experiments.
    \item {\bf Empirical validation and case studies with real-world data}. Through extensive numerical studies using synthetic data, we demonstrate the practical value of \our. Across diverse scenarios, \our\ consistently yields higher rewards and better decision quality than \tradition\ and Bayesian benchmarks. In particular, the benefit of \our\ becomes more pronounced when the expected ATE is small, the number of experiments is large, and the sample size is small. In scenarios with covariate information and non-overlapping experiments, the personalized estimators prescribed by \our\ can yield additional improvements. Furthermore, we demonstrate the robustness of the \our\ method under model misspecification. Finally, the application of \our\ method to two datasets from real-world platforms shows that \our\ enables effective subgroup targeting and coordination across concurrent experiments, confirming its scalability and impact in real-world experimentation.
\end{enumerate}

The rest of the paper is organized as follows. In Section \ref{sec:literature}, we review the related literature. In Section \ref{sec:framework}, we present our data-pooling treatment roll-out framework and prescribe implementation details of \our\ in four experimentation scenarios. In Section \ref{sec:theoretical_valid}, we theoretically validate the proposed \our\ in the case with non-overlapping experiments under linear model specifications. In Section \ref{sec:Synthetic_valid}, we conduct comprehensive synthetic experiments to demonstrate the superior and robust performance of our proposed framework. In Section \ref{sec:realdata}, we implement the framework to analyze real-world experiments. Section \ref{sec:Conclusion} concludes the paper. 

\section{Literature Review}\label{sec:literature}

Our work is related to two streams of literature: causal inference and its applications on online platforms, and small-data prediction and decision-making.
\subsection{Causal inference and its application on online platforms}
Randomized experiments have long been considered as the gold standard for estimating causal effects in social science research \citep[e.g.,][]{AngristPischke2009}. However, their implementation can be costly, and firms may sometimes face challenges in obtaining large sample sizes. To improve the efficiency of treatment effect estimation and reduce mean squared error (MSE), \citet{RosenmanBasseOwen2023Baiocchi2023} and \citet{Gui2024} propose a weighted sum that combines estimates from randomized experiments with secondary data. 
In addition, integrating experimental and secondary data offers additional advantages, such as aiding identification for estimating long-term effects \citep{AtheyChettyImbens2020, ImbensKallusMaoWang2024}. While this line of research focuses on leveraging diverse data sources tied to a single policy, our approach aggregates data across multiple experimental policies.

Another approach to addressing the challenge of small sample sizes is Bayesian inference. The Bayesian approach offers several advantages, including its ability to handle uncertainty, incorporate prior knowledge, and model complex data structures. Recently, Bayesian approaches for causal inference have attracted growing attention \citep[e.g.,][]{ImbensRubin2015,HahnMurrayCarvalho2020}. A number of studies have applied Bayesian methods to guide roll-out decisions \citep[e.g.,][]{AbadieAgarwalImbens2023, SimesterTimoshenkoZoumpoulis2025, Tetenov2016}. To ensure comparability with hypothesis testing from the frequentist perspective, \citet{Raftery1995} proposed that a policy should be implemented in the Bayesian setting only when the posterior evidence is sufficiently strong, analogous to meeting a predefined significance level in the frequentist approach. The key distinction between our \our\ method and the Bayesian approach lies in the implementation of shrinkage:  our method adopts a frequentist perspective, and explicitly incorporates the significance level into the shrinkage procedure applied to an unbiased estimator.

Our work is also related to the stream of studies on inference with multiple experiments. Conventional approaches for analyzing multiple experiments rely on factorial designs \citep[e.g.,][]{Box1978, WuHamada2011}. Recent works propose the potential outcome framework to facilitate causal inference across multiple experiments \citep{DasguptaPillaiRubin2015, PashleyBind2023}. However, as pointed out by \cite{ye2023deep}, factorial designs become impractical in modern large-scale A/B testing environments, where the number of experiments is hundreds or thousands. Even if one adopts the fractional factorial design, only a limited number of treatment combinations are testable. To address this challenge, \cite{ye2023deep} proposes a double machine learning framework that can infer all $2^m$ treatment combinations using $m+2$ observed combinations. However, even when the ATEs are linear, the growing number of experiments on online platforms often leads to insufficient traffic per experiment. To address this, our paper focuses on how to make roll-out decisions with a large number of experiments and limited observations per experiment.

Our paper also contributes to the applications of causal inference to online platforms. The recent decade has witnessed a growing body of research on this topic.  
From an empirical perspective, field experiments on large-scale online platforms enable causal inference to empower decision making in a wide variety of business settings \citep[e.g.,][]{cheung2017dynamic,cui2020reducing,zeng2023impact, zhan2024estimating}.  
On the theoretical side, scholars develop novel methods to overcome challenges arising from experimentation and causal inference on online platforms, such as two-sided randomization \citep[e.g.,][]{nandy2021b, johari2022experimental,ye2022cold}, sequential experiments \citep[e.g.,][]{song2021ensembling,BojinovSimchi-LeviZhao2023,xiong2022bias, NiBojinovZhao2023, Ni2025}, block randomization \citep[][]{candogan2021near}, multiple experiments \citep[][]{ye2023deep}, and personalized policy learning \citep[][]{Zhang2025}. We contribute to this literature by proposing a new method to effectively pool data from multiple experiments and improve experiment roll-out decisions for online platforms.
\subsection{Small-data prediction and decision-making}
The pioneering work by \cite{Stein1956} introduces the idea of data pooling for the simultaneous estimation of multiple Gaussian means and demonstrates its benefit over the decoupled approach, a result known as Stein's phenomenon. This finding has spurred extensive follow-up research aimed at explaining and contextualizing Stein's result \citep[e.g.,][]{Brown1971, EfronMorris1977}. Building on this foundation, \cite{GuptaKallus2022} extend Stein’s method to data-driven optimization problems, proposing a shrinkage-based approach that improves upon the decoupled approach by shrinking individual-level data to an anchor distribution. \cite{LeiQiLiuGengZhangHuSheng2024} proposes to treat the aggregated top-level sales information as a regularization for fitting the individual-level prediction model, which improves forecasting performance. \cite{chen2024role} empirically investigate how data aggregation and sharing via a digital platform can enhance the analytics based on individual-level data for small retailers.
A critical distinction lies in the nature of the data used: while problems like the newsvendor problem rely on observable labels such as random demand and cost parameters as direct inputs for optimization, our framework addresses situations where individual treatment effects are inherently unobservable. The lack of labeled outcomes calls for a tailored methodological approach to address the unique challenges of causal estimation and policy decision-making.

Our work is also related to multitask learning, which aims to learn both shared and task-specific representations across different tasks \citep{Caruana1997}. In a similar vein, our work leverages observations across multiple experiments to improve estimation accuracy. While multi-task learning is primarily designed for predictive tasks, our method focuses on causal inference and is tailored towards decision-rule optimization. A related paradigm is transfer learning, a special case of multi-task learning, that improves a learner from one target domain by transferring information from a related source domain. Recently, transfer learning has been adopted to enhance the efficiency of operational decision-making \citep{Bastani2021, NabiNassifHongMamanimbens2022, FengLiShanthikumar2023}. The transfer learning approach relies on sufficient information from the related source domain to enhance predictions or decisions in the target domain. In contrast, our work involves multiple experiments, each with limited observations, and aims to improve the roll-out decisions for all experiments.

\section{General Framework of Data Pooling in Experiment Roll-out}\label{sec:framework}
\enlargethispage{\baselineskip}

We now develop a new framework leveraging data pooling to improve the effectiveness of experiment roll-out decisions for online platforms. Suppose that a platform runs $K$ independent A/B tests concurrently, each designed to evaluate the ATE of a distinct policy\footnote{Alternatively, one could consider a single experiment designed to estimate HTEs across $K$ different subgroups, aiming to determine whether the policy should be rolled out for each subgroup. Throughout this paper, except in Section~\ref{subset:real_non_overlapping}, we adopt the notation of $K$ experiments to illustrate \our\ framework. In Section~\ref{subset:real_non_overlapping}, we demonstrate how this framework can be applied to roll out decisions across different subgroups using data from a single experiment.}. Without loss of generality, we assume that each experiment is fully randomized, with users assigned to treatment and control groups with equal probability. We further assume that the standard Stable Unit Treatment Value Assumption (SUTVA) holds.

The goal of the platform is to identify and roll out \textit{all} policies with a positive ATE. Let $Y\in \mathbb R$ denote a key outcome variable the platform cares about (e.g., whether the user clicks the recommended advertisement), and $D_k\in\{0,1\}$ denote the treatment assignment of experiment $k$, capturing whether \textit{all} the users are under the treatment or control condition. Define $\bm{X}\in\mathbb R^{d_x}$ as the covariate vector of a user, where $d_x$ denotes the number of covariates. We impose the following Assumption \ref{assum:linear_additive} throughout this paper.

\begin{assumption} \label{assum:linear_additive}
For each experiment $k \in [K]$, there exists a scalar $\tau_k \in \mathbb{R}$ such that
\begin{equation*}
    \mathbb{E}[Y|D_k = 1,\boldsymbol{D}_{-k}^{(a)}] - \mathbb{E}[Y|D_k = 0,\boldsymbol{D}_{-k}^{(a)}] = \mathbb{E}[Y|D_k = 1,\boldsymbol{D}_{-k}^{(b)}] - \mathbb{E}[Y|D_k = 0,\boldsymbol{D}_{-k}^{(b)}] = \tau_k, \quad \forall\ \boldsymbol{D}_{-k}^{(a)} \neq \boldsymbol{D}_{-k}^{(b)},
\end{equation*}
where the expectation is taken with respect to $(Y,\bm{X})$, and $\boldsymbol{D}_{-k}^{(a)},\boldsymbol{D}_{-k}^{(b)}$ denote any two distinct treatment assignment vectors for the other $K-1$ experiments excluding experiment $k$.
\end{assumption}

An immediate implication of Assumption~\ref{assum:linear_additive} is that the marginal effect of experiment $k$ on the outcome is invariant to the treatment assignment of the other $K-1$ experiments, so the average marginal effects across experiments are additive at the population level. Accordingly, $\tau_k$ is well defined as the average treatment effect (ATE) of policy $k$ and captures its full causal effect on the outcome.

Equivalently, in potential-outcome notation, if $Y_i(\bm d)$ denotes the outcome that user $i$ would realize under treatment vector $\bm d\in\{0,1\}^K$, Assumption~\ref{assum:linear_additive} requires that $\mathbb{E}[Y_i(1,\bm d_{-k})-Y_i(0,\bm d_{-k})]=\tau_k$ for every assignment vector $\bm d_{-k}$ of the other experiments. This potential-outcome formulation makes explicit that the causal effect of experiment $k$ is invariant to the other treatment assignments.

Assumption~\ref{assum:linear_additive} is a commonly used simplifying approximation in large-scale online experimentation, especially when interactions across treatments are believed to be rare or small. Prior evidence from online experimentation platforms suggests that interactions across overlapping treatments are rare \citep{Kohavi2013, Chan2021, Microsoft2023} and typically small in magnitude, and thus do not materially affect roll-out decisions \citep{Chan2021}. When Assumption~\ref{assum:linear_additive} is mildly violated, the \our\ framework remains effective. Specifically, following common practice in online platforms to balance speed and accuracy, we stress-test the method and demonstrate its robustness under a logistic transformation of treatment effects that introduces smooth and moderate nonlinearity (see Section~\ref{subset:real_ovelapping} and Appendix~\ref{subset: robust_model_misspecification}). More generally, in Appendix~\ref{app:limit_nonlinear}, we delineate regimes under which \our\ performs well versus poorly, providing guidance on when the method is expected to be reliable.

If the platform knows the ground-truth ATEs, $\tau_1,\tau_2,...\tau_K$, it will roll out policy $k$ if and only if $\tau_k>0$. Hence, under Assumption~\ref{assum:linear_additive}, the optimal per-experiment reward of the platform is:
\begin{equation*}
    r^* = \frac{1}{K}\sum_{k:\tau_k>0} \tau_k.
\end{equation*}

In practice, the ground-truth ATEs, $\tau_1,\cdots,\tau_K$, are unobservable to the platform, so it runs A/B tests to estimate them and make roll-out decisions accordingly. Let $\mathcal{S}$ denote the resulting experimental dataset, whose generation depends on the problem setting and experimentation method. For instance, when the \(K\) experiments are conducted independently, each user is assigned to either the treatment or the control condition of one experiment. In this case, if each experiment has \(N\) observations, the dataset can be represented as \(\mathcal{S} = \{(Y_{k,i}, D_{k,i},X_{k,i}): 1 \leq k \leq K, 1 \leq i \leq N\}\) where $Y_{k,i}$, $D_{k,i}$ and $X_{k,i}$ represent the individual outcome, treatment assignment status and covariate vector for subject $i$ in experiment $k$, respectively. As another example, the platform may adopt an orthogonal experiment design in which a user may be simultaneously targeted by multiple experiments \citep[e.g.,][]{tang2010overlapping,xiong2020orthogonal}. With a total of \(N\) users in the experiments, the dataset can be represented as \(\mathcal{S} = \{(Y_i, \bm{D}_i,X_i): 1 \leq i \leq  N\}\), where \(\bm{D}_i = (D_{1,i}, D_{2,i}, \dots, D_{K,i})\) is the vector of treatment assignments across all $K$ experiments for user $i$. Let $Y_i$ and $X_i$ denote the observed outcome and covariate vector for subject $i$ under the joint realization of treatment assignments from all experiments. 

Given the dataset $\mathcal{S}$, the platform typically relies on classical hypothesis testing to decide whether or not to roll out each experiment \(k\). Specifically, the commonly adopted approaches to estimate and infer the ATEs of the policies include, e.g., difference-in-means (DM) \citep[e.g., Section 1.1 in][]{wager2024causal}, ordinary least squares (OLS) \citep[e.g., Section 1.2 in][]{wager2024causal}, double machine learning (DML) \citep[e.g.,][]{chernozhukov2018double,farrell2020deep,ye2023deep,ShiMaoYangLi2025}, etc. We provide the general procedure of such a standard decision-making framework for online platforms in Algorithm~\ref{alg:general_dm}. Without loss of generality, we denote $\mathcal M(\cdot)$ as a general estimation method  for the null hypothesis $H_0$: $\tau_k = 0$, which maps the experimental dataset $\mathcal S$ and significance level $\alpha$ to the point estimate $\hat \tau_k$ and the (1-$\alpha$)-confidence interval $[\hat\tau_k^{\text{lb}}, \hat\tau_k^{\text{ub}}]$. For example, in the special case where $\hat\tau_k \sim \mathcal{N}(\tau_k,\sigma_k^2)$ and $\sigma_k^2$ is known, we have $\hat\tau_k^{\text{lb}} = \hat\tau_k -z_{1-\alpha/2}\sigma_k$ and $\hat\tau_k^{\text{ub}} = \hat\tau_k + z_{1-\alpha/2}\sigma_k$, where $z_{1-\alpha/2}$ is the (1-$\frac{\alpha}{2}$)-fractile quantile $z$-score of a standard normal distribution.



\begin{algorithm}[!ht]
    \begin{algorithmic}[1]
	\Require Set of policies $[K] = \{1,2,\cdots,K\}$; experimental dataset $\mathcal{S}$; significance level $\alpha$; the classical estimation method $\mathcal{M}(\cdot)$.
	
	\State {$\hat{\mathcal A}_{\tradition} \leftarrow \{\}$;} //{\footnotesize Initialize the roll-out decision as an empty set.}
    \For{$k \in [K]$}
    \State{
Run $\mathcal{M}(\mathcal{S},k,\alpha)$ to obtain the point estimate $\hat\tau_k$ and the (1-$\alpha$)-confidence interval $[\hat\tau_k^{\text{lb}}, \hat\tau_k^{\text{ub}}]$}. 

     \If{$ \hat\tau_k^{\text{lb}}>0$}

     \State{$\hat{\mathcal A}_{\tradition} \leftarrow \hat{\mathcal A}_{\tradition} \bigcup \{k\} $}

     \EndIf
   
    \EndFor
    \Ensure $\hat{\mathcal A}_{\tradition}$: Roll out policy $k$ if and only if $k\in \hat{\mathcal A}_{\tradition}$.
    \end{algorithmic}
	\caption{Individual Treatment Roll-Outs (\tradition)}
	\label{alg:general_dm}
\end{algorithm}


The experiment roll-out strategy based on the \tradition\ method (Algorithm~\ref{alg:general_dm}) generates a realized per-experiment reward, which is given by:
 \begin{equation*}
    \hat{r}_{\tradition} = \frac{1}{K}\sum_{k \in \hat{\mathcal A}_{\tradition}} \tau_k.
 \end{equation*}

The \tradition\ will have great performance if the sample size of each experiment, $N$, is large. However, if $N$ is small, the variance of ATE estimator $\hat{\tau}_k$ of  experiment $k$, will be too large, resulting in a poor performance of per-experiment reward for the \tradition\ method. To address this challenge, we design a new estimator that combines data from different experiments to lower its variance, at the cost of a higher bias. Based on the idea of shrinkage \citep[e.g.,][]{GuptaKallus2022}, the new estimator of $\tau_k$ is parametrized by an anchor $\tau$ and a scale parameter $\beta \ge 0$:
\begin{equation}\label{eq:DPTR-estimator}
    \bar{\tau}_k = \frac{N}{N + \beta}\hat{\tau}_k + \frac{\beta}{N + \beta}\tau. 
\end{equation}
Based on the new estimator \eqref{eq:DPTR-estimator}, we devise the platform roll-out decision according to a new estimator $\bar{\tau}_k$ at the significance level $\alpha$. The corresponding shrunken decision bounds for $\bar{\tau}_k$ are $[\bar\tau_k^{\text{lb}}, \bar\tau_k^{\text{ub}}]$, where $\bar\tau_k^{\text{lb}} = \frac{N}{N+\beta}\hat\tau_k^{\text{lb}} + \frac{\beta}{N+\beta}\tau$ and $\bar\tau_k^{\text{ub}} = \frac{N}{N+\beta}\hat\tau_k^{\text{ub}} + \frac{\beta}{N+\beta}\tau$. We use these as decision bounds for the roll-out rule rather than as frequentist confidence intervals for the original $\tau_k$, since the shrinkage step introduces a controlled bias relative to $\tau_k$ unless $\tau_k=\tau_0$, while substantially reducing the variance. Here, we set $\tau = \hat\tau_0 := \frac{1}{K}\sum_k \hat\tau_k$, as this choice is a least squares estimator of central tendency among all individual estimators $\hat\tau_k$. Next, we are ready to propose a general framework to identify proper values for \(\beta\), so as to optimize the experiment roll-out decisions, as detailed in Algorithm~\ref{alg:general_dp}.

\begin{algorithm}[!ht]

    \begin{algorithmic}[1]
	\Require Set of policies $[K] = \{1,2,\cdots,K\}$; experimental dataset $\mathcal{S}$; significance level $\alpha$; the classical estimation method $\mathcal{M}(\cdot)$.
	
	\State {$\hat{\mathcal{A}}_{ \our} \leftarrow \{\}$;} //{\footnotesize Initialize the roll-out decision as an empty set.}

    \State{$\{(\hat\tau_k,\hat\tau_k^{\text{lb}},\hat\tau_k^{\text{ub}}): k\in [K]\}  \leftarrow \{\mathcal{M}(\mathcal{S},k,\alpha):k \in [K] \} $}. //{\footnotesize Run the baseline estimation method}

    \State{Obtain data-driven parameters $\hat\tau_0 \leftarrow \frac{1}{K}\sum_k \hat\tau_k$ and $\hat{\beta}(\mathcal{S}, \mathcal{M})$}
    
    \For{$k \in [K]$}
     \State{$ \bar\tau_k^{\text{lb}} \leftarrow \frac{N}{N + \max(\hat{\beta}(\mathcal{S}, \mathcal{M}),0)}\hat\tau_k^{\text{lb}} + \frac{\max(\hat{\beta}(\mathcal{S}, \mathcal{M}),0)}{N + \max(\hat{\beta}(\mathcal{S}, \mathcal{M}),0)}\hat\tau_0, \bar\tau_k^{\text{ub}} \leftarrow \frac{N}{N + \max(\hat{\beta}(\mathcal{S}, \mathcal{M}),0)}\hat\tau_k^{\text{ub}} + \frac{\max(\hat{\beta}(\mathcal{S}, \mathcal{M}),0)}{N + \max(\hat{\beta}(\mathcal{S}, \mathcal{M}),0)}\hat\tau_0$. //{\footnotesize Construct the shrunken decision lower and upper bounds} }

     \If{$ \bar\tau_k^{\text{lb}} > 0$}

     \State{$\hat{\mathcal{A}}_{\our} \leftarrow \hat{\mathcal{A}}_{\our} \bigcup \{k\} $}

     \EndIf
   
    \EndFor
    \Ensure $\hat{\mathcal A}_{\our}$: Roll out policy $k$ if and only if $k\in \hat{\mathcal A}_{\our}$. 
    \end{algorithmic}
	\caption{Data-Pooling Treatment Roll-Outs (\our)}
	\label{alg:general_dp}
\end{algorithm}

Algorithm \ref{alg:general_dp} provides a general  procedure with data pooling to roll out experiments. In particular, \(\hat{\beta}(\mathcal{S}, \mathcal{M})\) denotes the scale parameter which depends on the aggregated historical dataset across all experiments, \(\mathcal{S}\), and the estimation method \(\mathcal{M}\). In Algorithm \ref{alg:general_dp}, when constructing the shrunken decision bounds for $\bar{\tau}_k$, we ignore the randomness of \(\hat{\tau}_0\) and \(\hat{\beta}(\mathcal{S},\mathcal{M})\), which are obtained from the data of all \(K\) experiments. When \(K\) is large, the variances of \(\hat{\tau}_0\) and \(\hat{\beta}(\mathcal{S},\mathcal{M})\) are orders of magnitude smaller than that of $\hat\tau_k$. Moreover, Algorithm~\ref{alg:general_dp} does not specify the formula of \(\hat{\beta}(\mathcal{S},\mathcal{M})\), which depends on the specific context, the dataset $\mathcal S$, and the estimation method $\mathcal M$. Intuitively, $\hat{\beta}(\mathcal{S},\mathcal{M})$ is larger when (1) The variation in individual treatment effects within each experiment is large, i.e., the individual ATE estimates ($\hat\tau_k$'s) are more volatile and less credible; (2) The treatment effects of different experiments ($\tau_k$'s) are concentrated, so that the data-driven anchor $
\hat\tau_0$ effectively aggregates information across different experiments, thus significantly enhancing the reliability of $\bar\tau_k$. This intuition is formally derived in Theorem~\ref{theorem:AP} and Theorem~\ref{theorem:data_driven_para}, where we consider the simplest case with no overlapping traffic, no covariate information and linear model specifications. Furthermore, $\hat{\beta}(\mathcal{S}, \mathcal{M})$ may vary across experiments due to differences in experiment-specific covariate information, assuming there is no overlapping traffic. This intuition is formally established in Theorem~\ref{theorem:AP_feature} and Theorem~\ref{theorem:data_driven_para_feature}. In addition, throughout Sections~\ref{subsec:scenario-1} to \ref{subsec:scenario-4}, we present the corresponding formulas of \(\hat{\beta}(\mathcal{S},\mathcal{M})\) for experiments with overlapping and nonoverlapping subjects as well as linear and nonlinear model specifications. Given \(\hat{\beta}(\mathcal{S},\mathcal{M})\) and $\hat\tau_0$, the experiment roll-out strategy based on the \our\ method (Algorithm~\ref{alg:general_dp}) generates a realized per-experiment reward:
 \begin{equation*}
     \hat{r}_\our = \frac{1}{K}\sum_{k \in \hat{\mathcal{A}}_{\our}} \tau_k.
 \end{equation*}


We would like to clarify that our \our\ method is specifically designed for settings involving multiple parallel experiments without capacity constraints. In other settings, such as multi-arm experiments or parallel experiments with capacity constraints that limit the number of roll-out decisions, our method may not offer a clear advantage over alternative approaches. We have clarified the scope and limitations of our framework in Appendix \ref{app:limit_capacity}.



\subsection{Scenario 1: Non-Overlapping Experiments With Linear Specifications} \label{subsec:scenario-1}
We begin by examining a scenario where $K$ experiments are conducted in $K$ separate subject pools, with each pool exclusively assigned to a single experiment. In this scenario, we consider linear model specifications. Specifically, we assume the following data-generating process (DGP):
\begin{equation}\label{eq:simple ols}
    Y_{k,i} = a_k + \tau_kD_{k,i} + \epsilon_{k,i}, \ k = 1,2,\dots,K,\ i = 1,\dots,N,
\end{equation}
where $\tau_k$ represents the ATE of policy $k$, $\epsilon_{k,i}$ is the i.i.d. random noise with zero mean and variance $\sigma_k^2$, and $a_k$ denotes the expected outcome under control condition for experiment $k$.

For each experiment, suppose the platform allocates exactly $N$ users exclusively to it, with \( N/2 \) randomly assigned to the treatment condition and \( N/2 \) to the control condition. The total dataset in this scenario can be represented as \( \mathcal{S} = \{\mathcal{S}_1, \mathcal{S}_2, \dots, \mathcal{S}_K\} \), where \( \mathcal{S}_k = \{(Y_{k,i}, D_{k,i}): 1 \leq i \leq N\} \). In this case, the classic estimation method \( \mathcal{M}(\cdot) \) can be the unbiased DM estimator:
\begin{equation}\label{eq:DM}
    \hat{\tau}_k := \frac{2}{N}\sum_{D_{k,i} = 1}Y_{k,i}- \frac{2}{N}\sum_{D_{k,i} = 0}Y_{k,i}.
\end{equation}
It is straightforward to derive that $\mathrm{Var}(\hat\tau_k) = \frac{4\sigma_k^2}{N}$. Furthermore, the unbiased estimator for variance $\sigma_k^2$ in experiment $k$ can be expressed as:
\begin{equation*}
    s_k^2 = \frac{1}{N-2}\sum_{j \in \{0,1\}}\sum_{D_{k,i} = j}(Y_{k,i} - \frac{2}{N}\sum_{D_{k,i} = j}Y_{k,i})^2.
\end{equation*}
Thus, for any experiment $k$, we have the central limit theorem (CLT):
\begin{equation*}
    \sqrt{N}(4s_k^2)^{-1/2}(\hat\tau_k - \tau_k) \to_d \mathcal{N}(0,1),
\end{equation*}
where $\to_d$ refers to convergence in distribution. Based on the estimators $\{\hat\tau_1,\cdots,\hat\tau_K\}$ and $\{4s_1^2,\cdots,4s_K^2\}$, the clean plug-in scale is:
\begin{equation} \label{eq:beta_form}
    \hat\beta = \frac{\frac{1}{K}\sum_{k}4s_k^2}{\frac{1}{K}\sum_{k}(\hat\tau_k - \hat\tau_0)^2 - \frac{1}{KN}\sum_{k}4s_k^2} + \frac{z_{1 - \alpha/2}\sqrt{N\frac{1}{K}\sum_{k}4s_k^2}}{\hat\tau_0}.
\end{equation}
In the first term of the plug-in \(\hat{\beta}\), Eqn \eqref{eq:beta_form}, the numerator \(\frac{1}{K}\sum_{k}4s_k^2\) is an unbiased estimator of \(\frac{1}{K}\sum_{k}4\sigma_k^2\), capturing the average variation of all individual estimators \(\hat{\tau}_k\). Hence, more variable estimations of the treatment effects for individual experiments lead to a larger \(\hat{\beta}\), which in turn shrinks the new estimator $\bar\tau_k$ (recall Eqn.~\eqref{eq:DPTR-estimator}) more towards \(\hat{\tau}_0\). The denominator \(\frac{1}{K}\sum_{k}(\hat{\tau}_k - \hat{\tau}_0)^2 - \frac{1}{KN}\sum_{k}4s_k^2\) is an unbiased estimator for \(\frac{1}{K}\sum_{k}(\tau_k - \frac{1}{K}\sum_k \tau_k)^2\) (see Theorem~\ref{theorem:data_driven_para}), capturing the variability of different experiments' ATEs. When the ATEs across different experiments are more concentrated, the aggregated information provided by \(\hat{\tau}_0\) becomes more valuable, so the scale parameter \(\hat{\beta}\) is larger and $\bar\tau_k$ is shrunk to $\hat\tau_0$ further. The second term of \(\hat{\beta}\) is a decision-aware adjustment for the roll-out threshold. Without it, $\hat{\beta}$ would minimize the estimation loss rather than maximize the reward (see Theorem~\ref{theorem:mse_optimal}). We relegate the derivation for \(\hat{\beta}\)'s formula as Eqn.~\eqref{eq:beta_form} to Theorem~\ref{theorem:AP} (see Section~\ref{subset:theoretical_without_covariates}). The exact theoretical guarantees in Section~\ref{sec:theoretical_valid} are derived under the homoskedastic Gaussian model in Assumption~\ref{assum:1}; when within-experiment variances $\sigma_k^2$ differ across experiments, the formula above should be interpreted as a moment-based heuristic that replaces the common variance by an average. A fully optimal heteroskedastic rule would generally use experiment-specific shrinkage scales.

Next, we incorporate covariate information into the OLS model specification and DGP as follows: 
\begin{equation} \label{eq:simple ols_feature}
    Y_{k,i} = a_k + \tau_kD_{k,i}+ \bm\theta_k^{\top}X_{k,i} + \epsilon_{k,i}, 1\le k \le K, 1 \le i \le N,
\end{equation}
where $\bm\theta_k$ denotes the parameter vector associated with the covariates in experiment $k$. We denote $\bm I = [0,1,\cdots,0] \in \mathbb{R}^{2+d_x}$, $\bm Y_k = [Y_{k,1},\cdots,Y_{k,N}]^{\top}$, $\bm \epsilon_k = [\epsilon_{k,1},\cdots,\epsilon_{k,N}]^{\top}$ and $\bm t_k = [(1,D_{k,1},\bm X_{k,1}),\cdots,(1,D_{k,N}, \bm X_{k,N})]^{\top}$. Thus, we can obtain the estimator $\hat\tau_k$ by OLS as follows:
\begin{equation}\label{eq: ols_tau_k_hat}
    \hat{\tau}_k = \bm I^{\top}(\bm t_k^{\top}\bm t_k)^{-1}\bm t_k^{\top}\bm Y_k = \tau_k + \bm I^{\top}(\bm t_k^{\top}\bm t_k)^{-1}\bm t_k^{\top}\bm \epsilon_k.
\end{equation}
It is straightforward to derive that $\mathrm{Var}(\hat\tau_k) = \sigma_k^2\bm I^{\top}(\bm t_k^{\top}\bm t_k)^{-1}\bm I$. Furthermore, the unbiased estimator for variance $\sigma_k^2$ in experiment $k$ can be expressed as:
\begin{equation*}
    s_k^2  = \frac{1}{N-2-d_x}(\bm Y_k - \bm t_k(\bm t_k^{\top}\bm t_k)^{-1}\bm t_k^{\top}\bm Y_k)^{\top}(\bm Y_k - \bm t_k(\bm t_k^{\top}\bm t_k)^{-1}\bm t_k^{\top}\bm Y_k).
\end{equation*}
For any experiment $k$, we have the following CLT:
\begin{equation} \label{eq:on_overlapping_feature}
    \sqrt{N}(b_k^2s_k^2)^{-1/2}(\hat\tau_k - \tau_k) \to_d \mathcal{N}(0,1),
\end{equation}
where $b_k = \sqrt{N \bm{I}^{\top} (\bm{t}_k^{\top} \bm{t}_k)^{-1} \bm{I}}$. Similarly, according to the Eqn.~\eqref{eq:beta_form}, one can construct the same $\hat\beta(\mathcal{S},\mathcal{M})$ for all experiments as follows:
\begin{equation} \label{eq:beta_form_nonpersonlized}
    \hat\beta = \frac{\frac{1}{K}\sum_{k}b_k^2s_k^2}{\frac{1}{K}\sum_{k}(\hat\tau_k - \hat\tau_0)^2 - \frac{1}{KN}\sum_{k}b_k^2s_k^2} + \frac{z_{1 - \alpha/2}\sqrt{N\frac{1}{K}\sum_{k}b_k^2s_k^2}}{\hat\tau_0}.
\end{equation}


Furthermore, we can observe that $\{b_k^2: k\in[K]\}$ are different across experiments and they can be  derived from the training covariate vectors, which are known prior to making the roll-out decision. Thus, we can derive a personalized  $\hat\beta(\mathcal{S},\mathcal{M})$ for experiment $k$ as follows:
\begin{equation} \label{eq:beta_form_personlized}
    \hat\beta_k = \frac{b_k^2(\frac{1}{K}\sum_{k^{'}}s_{k^{'}}^2)}{\frac{1}{K}\sum_{k^{'}}(\hat\tau_{k^{'}} - \hat\tau_0)^2 - \frac{1}{N}(\frac{1}{K}\sum_{k^{'}}b_{k^{'}}^2\cdot \frac{1}{K}\sum_{k^{'}}s_{k^{'}}^2)} + \frac{z_{1 - \alpha/2}b_k\sqrt{N\frac{1}{K}\sum_{k^{'}}s_{k^{'}}^2}}{\hat\tau_0}.
\end{equation}
We defer the derivation of the formula for $\hat{\beta}_k$, presented in Eqn.~\eqref{eq:beta_form_personlized}, to Theorem~\ref{theorem:AP_feature} (see Section~\ref{subset:theoretical_with_covariates}), and demonstrate that the personalized $\hat{\beta}_k$ performs even better than the shared $\hat\beta$ in Eqn.~\eqref{eq:beta_form_nonpersonlized} using simulations with synthetic data (see Section~\ref{subsec:no-overlap-no-feature}).

\subsection{Scenario 2: Non-Overlapping Experiments With Nonlinear Specifications}  \label{subsec:scenario-2}
Building upon the scenario outlined in Section \ref{subsec:scenario-1}, we now introduce a second scenario that considers nonlinear model specifications, extending the framework to a partial linear model. Specifically, we assume the following DGP:
\begin{equation*}
    Y_{k,i} =  g_k(X_{k,i})^{\top}\boldsymbol{t}_{k,i} + \epsilon_{k,i},\ k = 1,2,\dots,K,\ i = 1,\dots,N,
\end{equation*}
where \( g_k(\cdot): \mathbb{R}^{d_{x}} \to \mathbb{R}^2 \) represents the true response function for experiment \( k \), and \( \boldsymbol{t}_{k,i} = [1, D_{k,i}]^{\top} \) is the treatment vector, which includes the constant term. The term \( \epsilon_{k,i} \) denotes i.i.d. random noise with zero mean and variance \( \sigma_k^2 \). All functions \( \{g_1, g_2, \dots, g_K\} \) belong to the same function class \( \mathcal{F} \). Consequently, the total dataset in this scenario is given by \( \mathcal{S} = \{\mathcal{S}_1, \mathcal{S}_2, \dots, \mathcal{S}_K\} \), where \( \mathcal{S}_k = \{(Y_{k,i}, D_{k,i}, X_{k,i}): i = 1, \dots, N\} \). The ATE for experiment \( k \) is denoted as \( \tau_k = \mathbb{E}[g_k(X_k)^{\top} \boldsymbol{t}^*] \), where \( \boldsymbol{t}^* = [0, 1]^{\top} \).  

Building on the partial linear framework, we employ the double machine learning method \citep{farrell2020deep} as \( \mathcal{M} \) in this scenario. Here, for each experiment \( k \), we apply the cross-fitting techniques \citep[][]{chernozhukov2018double,farrell2020deep} to obtain the estimator \( \hat{\tau}_k \) and $\hat{\Psi}_k$.  The detailed estimation procedures for \( \hat{\tau}_k \) and \( \hat{\Psi}_k \) are provided in Appendix \ref{app:dml_scenario2}.

Based on Theorem 3 of \citet{farrell2020deep}, as long as the nuisance parameter estimator $\hat g_k(\cdot)$ converges to $g_k(\cdot)$ sufficiently fast, we have:
\begin{equation} \label{eq: DML}
    \sqrt{N}\hat{\Psi}_k^{-1/2}(\hat\tau_k - \tau_k) \to_d \mathcal{N}(0,1).
\end{equation}

Based on the estimators $\{\hat\tau_1,\cdots,\hat\tau_K\}$ and $\{\hat{\Psi}_1,\cdots,\hat{\Psi}_K\}$, we can construct the scale parameter $\hat\beta(\mathcal{S},\mathcal{M})$ following the same intuition as Eqn. \eqref{eq:beta_form_nonpersonlized}:
\begin{equation}\label{eq:scale-s2}
    \hat\beta = \frac{\frac{1}{K}\sum_{k}\hat{\Psi}_k}{\frac{1}{K}\sum_k(\hat\tau_k - \hat\tau_0)^2 - \frac{1}{KN}\sum_{k}\hat{\Psi}_k} + \frac{z_{1 - \alpha/2}\sqrt{N\frac{1}{K}\sum_{k}\hat{\Psi}_k}}{\hat\tau_0}.
\end{equation}
Similar to Eqn.~\eqref{eq:beta_form_personlized}, we can also construct the heuristic personalized scale parameter $\hat\beta_k$'s in this setting. First, we can compute $b_k = \sqrt{N \bm{I}^{\top} (\bm{t}_k^{\top} \bm{t}_k)^{-1} \bm{I}}$ using the covariate information. Second, by comparing Eqn.~\eqref{eq:on_overlapping_feature} and~\eqref{eq: DML}, we can find that $b_k^2s_k^2$ and $\hat{\Psi}_k$ play the same role and we can construct $s_k^2 = \frac{\hat{\Psi}_k}{b_k^2}$. Thus, we can derive a personalized  $\hat\beta(\mathcal{S},\mathcal{M})$ for experiment $k$ as follows:
\begin{equation} \label{eq:beta_form_personlized_DML}
    \hat\beta_k = \frac{b_k^2(\frac{1}{K}\sum_{k^{'}}\frac{\hat{\Psi}_{k^{'}}}{b_{k^{'}}^2})}{\frac{1}{K}\sum_{k^{'}}(\hat\tau_{k^{'}} - \hat\tau_0)^2 - \frac{1}{N}(\frac{1}{K}\sum_{k^{'}}b_{k^{'}}^2\cdot \frac{1}{K}\sum_{k^{'}}\frac{\hat{\Psi}_{k^{'}}}{b_{k^{'}}^2})} + \frac{z_{1 - \alpha/2}b_k\sqrt{N\frac{1}{K}\sum_{k^{'}}\frac{\hat{\Psi}_{k^{'}}}{b_{k^{'}}^2}}}{\hat\tau_0}.
\end{equation}

\subsection{Scenario 3: Overlapping Experiments With Linear Specifications}  \label{subsec:scenario-3}
In this subsection, we examine the scenario that a user may be simultaneously targeted by multiple experiments. Similarly, we first examine the scenario with linear model  specifications, followed by an extension that considers nonlinear model  specifications in the next subsection. Specifically, we assume the following DGP:
\begin{equation}\label{eq:DGP}
    Y_{i} = a + \sum_{k \in K}\tau_k D_{k,i} + \epsilon_{i}, \mbox{ }i=1,2,...,N,
\end{equation}
where $\tau_k$ represents the ATE of policy $k$, $\epsilon_{i}$ is the i.i.d. random noise with zero mean and variance $\sigma^2$, and $a$ represents the expected outcome if a subject receives the control status in all experiments. Furthermore, $D_{k,i}$'s are i.i.d. Bernoulli random variables with $\mathbb P[D_{k,i}=1]=\mathbb P[D_{k,i}=0]=0.5$.

In this scenario, the dataset can be represented as $\mathcal{S} = \{(Y_i,\bm D_i): i = 1, \cdots, N\}$ where $\bm D_i = [D_{1,i},D_{2,i},\dots,D_{K,i}]^\top$ denotes the treatment status vector across all experiments. We define $\bm \tau = [a,\tau_1,\tau_2,\cdots,\tau_K]^{\top}$ as the average treatment effect vector. The classical estimation method $\mathcal{M}(\cdot)$ in this scenario adopts OLS to estimate $\bm{\tau}$.

To proceed, we define $\bm{t}_i = [1, \bm{D}_i]^{\top}$. Let $\bm{Y} = [Y_1, Y_2, \dots, Y_N]^{\top}$ and $\mathcal{T} = [\bm{t}_1, \bm{t}_2, \dots, \bm{t}_N]^{\top}$. Thus, the OLS estimator of $\bm{\tau}$ is given by:
\begin{equation*}
    \hat {\bm \tau} = (\mathcal{T}^{\top}\mathcal{T})^{-1}\mathcal{T}^{\top}\bm Y.
\end{equation*}
In addition, we have, $\mathbb{E}(\hat {\bm \tau}) = \bm \tau$ and $\mathbb{V}(\hat {\bm \tau} |\mathcal{T}) = \sigma^2(\mathcal{T}^{\top}\mathcal{T})^{-1}$. To estimate the variance $\sigma^2$, we use $\hat{\sigma}^2 = \frac{\sum_{i} (Y_i - \hat{\bm{\tau}}^{\top} \bm{t}_i)^2}{N - K - 1}$. Next, let $I_k$ be the $(K+1)$-dimensional vector where the $(k+1)$th component is equal to 1, and all other components are equal to 0. The estimate of the average treatment effect for experiment $k$ is given by, $\hat{\tau}_k = I_k^{\top} \hat{\bm{\tau}}.$ The standard error (SE) of $\hat{\tau}_k$ is, $
\text{SE}_k = \sqrt{I_k^{\top} \hat{\sigma}^2 (\mathcal{T}^{\top} \mathcal{T})^{-1}  I_k}$, and we denote $s_k^2 = \frac{N}{4}\text{SE}_k^2$. Similarly, by \cite{Greene2003}, we can conclude that:
\begin{equation*}
    \sqrt{N}(4s_k^2)^{-1/2}(\hat\tau_k - \tau_k) \to_d \mathcal{N}(0,1).
\end{equation*}
Similar to Section \ref{subsec:scenario-1}, we can derive the scale parameter $\hat\beta(\mathcal{S},\mathcal{M})$ as:
\begin{equation}\label{eq:scale-s3}
    \hat\beta = \frac{\frac{1}{K}\sum_{k}4s_k^2}{\frac{1}{K}\sum_{k}(\hat\tau_k - \hat\tau_0)^2 - \frac{1}{KN}\sum_{k}4s_k^2} + \frac{z_{1 - \alpha/2}\sqrt{N\frac{1}{K}\sum_{k}4s_k^2}}{\hat\tau_0}.
\end{equation}
When covariate vectors are incorporated in the OLS model with the following DGP: 
\begin{equation} \label{eq:DGP_OLS_feature}
    Y_i = a + \sum_k \tau_k D_{k,i} + \bm\theta^\top X_i + \epsilon_i,  1 \le i \le N,
\end{equation}
where $\bm\theta$ denotes the parameter vector associated with the covariates. The OLS estimator for each experiment $k$, $\hat\tau_k$, is asymptotically normal estimator, so Algorithm \ref{alg:general_dp} can be applied with the scale parameter defined by Eqn.~\eqref{eq:scale-s3}.


\subsection{Scenario 4: Overlapping Experiments With Nonlinear Specifications}  \label{subsec:scenario-4}

Finally, we consider the setting with overlapping experiments and nonlinear model specifications. Similar to Section~\ref{subsec:scenario-2}, we adopt the partial linear model framework, and the DGP is given by:
\begin{equation} \label{eq:DML_DGP_overlapping}
    Y_{i} =  \bm g(X_{i})^{\top}\bm t_i + \epsilon_{i},\ i = 1,\dots,N,\end{equation}
where $\bm g(\cdot): \mathbb{R}^{d_x}\to \mathbb{R}^{K+1}$ is the true response function and $\bm t_i = [1,\bm D_i]^\top$ is the treatment vector including the intercept. $\epsilon_{i}$ denotes the i.i.d. random noise with zero mean and variance $\sigma^2$. The dataset in this scenario can be represented as $\mathcal{S} = \{(Y_{i},X_{i},\bm t_i):i = 1,\cdots,N \}$. The ATE for experiment $k$ can be denoted as $\tau_k = \mathbb{E}[\bm g(X)^{\top} \bm t^*_k]$ where $\bm t^*_k$ is a $K + 1$ dimension vector of which the $(k + 1)$th component is equal to 1 and other components are equal to zero.

For fixed $K$ and under the standard cross-fitting, overlap, moment, and nuisance-rate regularity conditions in \citet{farrell2020deep}, the overlapping assignment structure does not by itself invalidate the asymptotic normality result. Similar to Section~\ref{subsec:scenario-2}, we obtain the ATE estimator $\hat{\tau}_k$ and the corresponding variance estimator $\hat{\Psi}_k$, along with the asymptotic normality result:
\begin{equation*}
    \sqrt{N}\hat{\Psi}_k^{-1/2}(\hat\tau_k - \tau_k) \to_d \mathcal{N}(0,1). 
\end{equation*}

Then, we can construct the scale parameter similar to Eqn.~\eqref{eq:scale-s2}:
\begin{equation*}
\hat\beta(\mathcal{S}) = \frac{\frac{1}{K}\sum_{k}\hat{\Psi}_k}{\frac{1}{K}\sum_k(\hat\tau_k - \hat\tau_0)^2 - \frac{1}{KN}\sum_{k}{\hat{\Psi}_k}} + \frac{z_{1 - \alpha/2}\sqrt{N\frac{1}{K}\sum_{k}\hat{\Psi}_k}}{\hat\tau_0}.
\end{equation*}

For the rest of this paper, we demonstrate that the roll-out decisions derived from the shrunken decision bounds in Algorithm~\ref{alg:general_dp} can outperform those obtained using the traditional \tradition\ method in Algorithm~\ref{alg:general_dm}. Formal guarantees are provided in Section~\ref{sec:theoretical_valid} for the non-overlapping linear setting, while overlapping and nonlinear settings are evaluated through numerical experiments and empirical applications. 

\section{Theoretical Analysis} \label{sec:theoretical_valid}
In this section, we derive the optimal scale parameter $\beta$ and provide the theoretical justification for the \our\ framework. To this end, we focus on the simplest scenario, non-overlapping experiments under linear model specifications (see Section~\ref{subsec:scenario-1} for details). Specifically, we first prove that, in the case without covariate information, i.e., the DGP follows Eqn.~\eqref{eq:simple ols}, the \our\ experiment roll-out method weakly improves the expected reward relative to the \tradition\ method, and strictly improves it in the positive-anchor regime maintained for the data-driven results. We then extend this result to the setting with covariates (see Eqn.~\eqref{eq:simple ols_feature}). The following assumption is made throughout this section.



\begin{assumption} \label{assum:1}
    $\tau_k$ is drawn from a normal distribution, $\mathcal{N}(\tau_0, \sigma_0^2)$. Furthermore, the i.i.d. random noise $\epsilon_{k,i}$ follows a normal distribution, $\mathcal{N}(0, \sigma^2)$.
\end{assumption}

\subsection{Model Without Covariates} \label{subset:theoretical_without_covariates}
In this subsection, we theoretically justify our proposed method for the setting without covariates. In this setting, the DM estimator (see Eqn.~\eqref{eq:DM}) follows a normal distribution with mean $\tau_k$ and variance $4\sigma^2/N$ under Assumption \ref{assum:1}.


Suppose that the variance $\sigma^2$ is known. The platform uses $z$-statistics to construct the classical interval in $\mathcal{M}(\cdot)$. We define the per-experiment reward of \our\ with scale $\beta$ and anchor $\tau$ as follows:
\begin{equation}\label{eq:objective-known-sigma}
\tilde{r}(\beta,\tau) = \frac{1}{K}\sum_{k:\bar\tau_k>\frac{N}{N+\beta}\frac{2\sigma z_{1 - \alpha/2}}{\sqrt{N}}} \tau_k.
\end{equation}
It follows that $\tilde{r}(0, \cdot) = \tilde{r}(0, \tau_0) = \hat{r}_{\tradition}$. Define $\mathcal{R}(\beta,\tau) := \lim\limits_{K\uparrow+\infty} {\mathbb{E}[\tilde{r}(\beta,\tau)]}$ as the expected per experiment reward when the number of experiments $K \to \infty$. Our analysis begins with identifying the optimal value of $\beta$ that maximizes the expected per-experiment reward when $\tau = \tau_0$. 

\begin{theorem} \label{theorem:AP}
Suppose that Assumptions~\ref{assum:linear_additive} and~\ref{assum:1} hold and $\sigma,\sigma_0$ are known. If $\tau = \tau_0$, the optimal value of $\beta$ is:
    \begin{equation}\label{eq:beta*}
        \beta^* = \begin{cases}
 \dfrac{4\sigma^2}{\sigma_0^2} + \dfrac{2\sqrt{N}\,z_{1 - \alpha/2}\,\sigma}{\tau_0}, & \tau_0 > 0\ \text{or}\ \tau_0 < -\dfrac{\sqrt{N}\,z_{1-\alpha/2}\,\sigma_0^2}{2\sigma}, \\[6pt]
0, & -\dfrac{\sqrt{N}\,z_{1-\alpha/2}\,\sigma_0^2}{2\sigma} \le \tau_0 < 0.
\end{cases}
    \end{equation}
When $\tau_0=0$, $\mathcal{R}(\beta,0)$ is invariant to $\beta$, so $\beta=0$ is selected as a canonical optimum. In particular, $\mathcal{R}(\beta^*,\tau_0) \ge \mathcal{R}(0,\tau_0) = \lim\limits_{K\uparrow+\infty} \mathbb{E}[\hat{r}_{\tradition}]$, with the inequality holding strictly when $\tau_0 > 0$ or $\tau_0 < -\dfrac{\sqrt{N}\,z_{1-\alpha/2}\,\sigma_0^2}{2\sigma}$.
\end{theorem}


Theorem \ref{theorem:AP} characterizes the optimal scale parameter $\beta^*$ that maximizes the expected per-experiment reward of the \our\ roll-out method when the anchor is set at $\tau = \tau_0$. As a consequence, our proposed \our\ method with the optimal scale parameter weakly improves on the classical DM method, and strictly improves on it in the positive and sufficiently negative regimes identified in the theorem, highlighting the potential value of data pooling for multiple A/B tests. 

The optimal scale parameter takes a piecewise form. When $\tau_0$ is positive or sufficiently negative, $\beta^*$ admits the closed-form expression in Eqn.~\eqref{eq:beta*} and \our\ strictly improves upon the \tradition\ baseline; when $\tau_0$ falls in the intermediate negative range $\big[-\tfrac{\sqrt{N}\,z_{1-\alpha/2}\,\sigma_0^2}{2\sigma},\,0\big)$, no shrinkage is preferred and $\beta^*=0$, so \our\ coincides with \tradition\ and no pooling benefit arises. At $\tau_0=0$, all $\beta\ge0$ yield the same reward and we use $\beta=0$ as a canonical choice. This asymmetry in the favorable region of our method arises from the incorporation of the significance level \(\alpha\), which implies that our roll-out criterion for the estimator is not merely that the estimate itself exceeds zero, but rather that the lower bound of the corresponding confidence set exceeds zero.

Motivated by this structural result, we maintain the working assumption $\tau_0 > 0$ throughout the remainder of Section~\ref{sec:theoretical_valid}, justified on two grounds. Practically, when the platform considers a randomized field experiment, it typically conducts a pilot study to assess the potential value and proceeds only if the policy's ATE is likely to be positive; more broadly, firms routinely perform ex ante screening when evaluating a portfolio of experiments, so the expected ATE across the retained portfolio is generally positive, i.e., $\tau_0 > 0$. Without such screening, running experiments indiscriminately would be inefficient and wasteful. Theoretically, Theorem~\ref{theorem:AP} shows that $\tau_0 > 0$ is precisely the practically dominant regime in which \our\ yields a strict benefit over \tradition, making it the regime of primary interest for the subsequent analysis. 

We can treat $\hat{\tau}_k$ as the signal from experiment $k$, and $\tau_0$ as the aggregate information across all experiments. Based on Eqn.~\eqref{eq:DPTR-estimator}, we observe that, when $N$ is fixed, a larger value of $\beta$ results in less weight being assigned to the individual signal under the \our\ method. Eqn.~\eqref{eq:beta*} prescribes that $\beta^*$ consists of two positive terms $\frac{4\sigma^2}{\sigma_0^2}$ and $\frac{2\sqrt{N}z_{1 - \alpha/2}\sigma}{\tau_0}$. The first term is proportional to the ratio of the variance of noise within an experiment to the variance of treatment effects across experiments. When this ratio is large, it indicates that the aggregate information from all experiments is more reliable than the individual signal. In this case, less weight should be placed on the individual signal.
The second term may seem counterintuitive at the first glance: why does a smaller $\tau_0$ imply less weight being placed on the individual signal? In fact, the weight on the individual signal depends on the relative ratio of $\tau_0$ to $\sigma$. If the individual signals have a small variance, they may be more informative than the aggregate signal $\tau_0$. Therefore, when the individual signal is more precise, it remains more beneficial to assign greater weight to it, even when $\tau_0$ is small. Although $\beta^*$ is increasing in $N$, Eqn.~\eqref{eq:DPTR-estimator} also implies that the weight placed on the individual signal also increases with $N$ under \our.

To further demonstrate the role of the second term of $\beta^*$, we contrast our decision-aware criterion $\mathcal{R}(\beta,\tau_0)$ with a purely estimation-based criterion, namely the mean squared error $c(\beta,\tau) = \frac{1}{K}\sum_{k \in [K]}(\bar{\tau}_k - \tau_k)^2$, and, analogous to the construction above, we define $\text{MSE}(\beta,\tau_0) := \lim_{K\uparrow+\infty}\mathbb{E}[c(\beta,\tau_0)]$. The following theorem characterizes the MSE-optimal scale parameter.

\begin{theorem} \label{theorem:mse_optimal}
 Suppose that Assumptions~\ref{assum:linear_additive} and~\ref{assum:1} hold and $\sigma,\sigma_0$ are known. If $\tau = \tau_0$, the MSE-optimal scale parameter is
\begin{equation} \label{eq:best_beta_mse}
    \beta_{\mathrm{MSE}}^* := \argmin_{\beta \ge 0} \mathrm{MSE}(\beta,\tau_0) = \frac{4\sigma^2}{\sigma_0^2}.
\end{equation}
\end{theorem}
A head-to-head comparison of Eqns.~\eqref{eq:beta*} and~\eqref{eq:best_beta_mse} reveals that the MSE-optimal scale $\beta_{\mathrm{MSE}}^*$ coincides with the first term of the decision-aware scale $\beta^*$. This first term reflects the variance-versus-heterogeneity trade-off between the individual signal $\hat{\tau}_k$ and the aggregate anchor $\hat{\tau}_0$, and it is all that is needed when the objective is accurate estimation of $\tau_k$. The second term of $\beta^*$, namely $\frac{2\sqrt{N}\,z_{1-\alpha/2}\,\sigma}{\tau_0}$, is the decision-specific correction: it captures the distortion introduced by the roll-out threshold $\frac{N}{N+\beta}\frac{2\sigma z_{1 - \alpha/2}}{\sqrt{N}}$  in Eqn.\eqref{eq:objective-known-sigma}, and it aligns the shrinkage level with the platform's reward objective rather than with estimation accuracy. Omitting the second term would therefore yield a scale that is optimal for minimizing MSE but suboptimal for maximizing roll-out reward, a clean contrast that highlights the decision-aware nature of the \our\ framework.


While the parameters $\tau_0$, $\sigma^2$, and $\sigma_0^2$, which are used for deriving the optimal scale parameter, are unobservable in practice, one can leverage the pooled data from all experiments to estimate them. Thus, a natural estimator of the optimal scale parameter is one that replaces these parameters in Eqn.~\eqref{eq:beta*} by their estimates, as formally stated in Theorem \ref{theorem:data_driven_para}.

\begin{theorem} \label{theorem:data_driven_para}
Suppose Assumptions~\ref{assum:linear_additive} and~\ref{assum:1} hold and $\tau_0>0$.
   Define $s_k^2 := \frac{1}{N-2}\sum_{j \in \{0,1\}}\sum_{D_{k,i} = j}(Y_{k,i} - \frac{2}{N}\sum_{D_{k,i} = j}Y_{k,i})^2$ and
\begin{equation}\label{eq:data-driven-beta-S1}
       \hat{\beta}^* := \frac{\frac{1}{K}\sum_{k}4s_k^2}{\frac{1}{K}\sum_{k}(\hat{\tau}_k - \hat{\tau}_0)^2 - \frac{1}{KN}\sum_{k}4s_k^2} + \frac{z_{1 - \alpha/2}\sqrt{N\frac{1}{K}\sum_{k}4s_k^2}}{\hat{\tau}_0}.
   \end{equation}
We have, as $K \uparrow \infty$, $\hat{\tau}_0 \to_p \tau_0$ and $\hat{\beta}^* \to_p \beta^*$, where $\to_p$ refers to convergence in probability.
\end{theorem}

As shown in Theorem \ref{theorem:data_driven_para}, the proposed data-driven estimator for the optimal scale parameter is consistent even when $\sigma$ is unknown.  With finite samples, the plug-in scale parameter in \eqref{eq:data-driven-beta-S1} may be negative or undefined, since the denominator can be non-positive and $\hat{\tau}_0$ can be close to zero. Our implementation accordingly truncates $\hat{\beta}^*$ at zero. Under the maintained regime $\tau_0>0$ and $\sigma_0^2>0$, both safeguards are asymptotically inactive and therefore do not affect the consistency claim.
Next, we show that the \our\ method, as specified in Algorithm~\ref{alg:general_dp}, with $\hat{\tau}_0$ and $\hat\beta(\mathcal{S},\mathcal{M}) = \hat{\beta}^*$, will generate the same expected reward per experiment as the baseline case where $\sigma$, $\sigma_0$, and $\tau_0$ were known. We now introduce the expected per-experiment reward of the \our\ method with the estimated variance $\hat\sigma^2 = \frac{1}{K}\sum_{k}s_k^2$:
\begin{equation*} 
    \bar{r}(\beta,\tau) = \frac{1}{K}\sum_{k:\bar\tau_k>\frac{N}{N+\beta}\frac{2\hat\sigma z_{1 - \alpha/2}}{\sqrt{N}}} \tau_k.
\end{equation*}

\begin{theorem} \label{thm:cost} Suppose Assumptions~\ref{assum:linear_additive} and~\ref{assum:1} hold and $\tau_0>0$.
     As $K \uparrow \infty$, we have $\bar{r}(\hat{\beta}^*,\hat{\tau}_0) \to_p \mathcal{R}(\beta^*,\tau_0) > \mathcal{R}(0,\tau_0)$.
\end{theorem}

Theorem~\ref{thm:cost} implies that the \our\ method with data-driven parameters can achieve an even higher reward than the \tradition\ method with known $\sigma$, as long as the number of experiments $K$ is sufficiently large. The key driving force behind this result is the delicate balance between bias and variance achieved by the \our\ method. The anchor $\hat{\tau}_0$ leverages pooled data from a large number of experiments, significantly reducing variance compared to the individual signal $\hat{\tau}_k$. At the same time, our proposed \our\ method carefully controls bias through the optimally chosen scale parameter $\hat{\beta}^*$, ensuring an effective trade-off between bias and variance. Specifically, our method  optimizes the roll-out decision to maximize the expected reward.

We conclude the analysis of the non-overlapping linear setting by characterizing the roll-out probability induced by Algorithm~\ref{alg:general_dp}. Under the algorithm, experiment $k$ is selected into the roll-out set if and only if $\bar{\tau}_k^{\mathrm{lb}} > 0$. The following theorem characterizes the asymptotic selection probability.
\begin{theorem} \label{theorem:select_pro}
Suppose Assumptions~\ref{assum:linear_additive} and~\ref{assum:1} hold. Then,
\begin{equation*}
    \lim_{K\to\infty}\mathbb{P}(\bar{\tau}_k^{\mathrm{lb}} > 0) = \Phi\Bigg(\frac{N\tau_k+\tau_0\beta^*}{2\sqrt{N}\sigma} - z_{1-\alpha/2}\Bigg),
\end{equation*}
where $\Phi(\cdot)$ is the cumulative distribution function of the standard normal distribution.
\end{theorem}

Theorem~\ref{theorem:select_pro} shows that \our\ implements a probabilistic rather than deterministic roll-out rule: the asymptotic probability of rolling out experiment $k$ varies smoothly with its true effect $\tau_k$, approaching one as $\tau_k$ grows sufficiently positive and vanishing as $\tau_k$ grows sufficiently negative, while remaining non-negligible when $\tau_k$ is near zero. The smoothness is inherited from the incorporation of estimation uncertainty via the lower shrunken decision bound $\bar{\tau}_k^{\mathrm{lb}}$. Moreover, since $\beta^*$ is of order $\sqrt{N}$, $N\tau_k$ is the dominating term in the numerator, so the argument of $\Phi(\cdot)$ becomes increasingly sensitive to the sign of $\tau_k$ as $N$ grows. Consequently, larger sample sizes sharpen the selection rule toward a threshold limit in which experiments with $\tau_k > 0$ are rolled out with probability approaching one and those with $\tau_k < 0$ with probability approaching zero.

\subsection{Model with Covariates} \label{subset:theoretical_with_covariates}
In this subsection, we prove the \our\ method yields a higher expected reward than \tradition\ for the OLS model with covariates. In this setting, based on Eqn. \eqref{eq: ols_tau_k_hat}, the ATE estimator of treatment $k$ is given by
\begin{equation*}
    \hat{\tau}_k = \bm I^{\top}(\bm t_k^{\top}\bm t_k)^{-1}\bm t_k^{\top}\bm Y_k = \tau_k + \bm I^{\top}(\bm t_k^{\top}\bm t_k)^{-1}\bm t_k^{\top}\bm \epsilon_k.
\end{equation*}
Under Assumption \ref{assum:1}, $\hat\tau_k$ follows a normal distribution with mean $\tau_k$ and variance $\sigma^2b_k^2/N$, where \( b_k = \sqrt{N \bm{I}^{\top} (\bm{t}_k^{\top} \bm{t}_k)^{-1} \bm{I}} \). 
By Algorithm \ref{alg:general_dp}, we construct the new ATE estimator $\bar{\tau}_k = \frac{N}{N+\beta(b_k)}\hat\tau_k + \frac{\beta(b_k)}{N+\beta(b_k)}\tau$ parametrized by scale $\beta(b_k)$ and anchor $\tau$. The \our\ method then determines whether policy $k$ will be rolled out based on $\bar{\tau}_k$. Different from the setting without covariates, the scale parameter depends on the treatment and covariate vector ${\bm t}_k$ through the parameter $b_k$.

Similar to the model without covariates, we first assume that the variance \( \sigma^2 \) is known. The platform uses $z$-statistics to construct the classical bounds with the method $\mathcal{M}(\cdot)$. We define the per-experiment reward in this setting:
\begin{equation}\label{eq:objective-known-sigma-feature}
\tilde{r}(\beta(\cdot),\tau) = \frac{1}{K}\sum_{k:\bar\tau_k>\frac{N}{N+\beta(b_k)}\frac{\sigma b_k}{\sqrt{N}}z_{1 - \alpha/2 }}\tau_k.
\end{equation}
It follows that $\tilde{r}(0, \cdot) = \tilde{r}(0, \tau_0) = \hat{r}_{\tradition}$. Define $\mathcal{R}(\beta(\cdot),\tau) := \lim\limits_{K\uparrow+\infty} {\mathbb{E}[\tilde{r}(\beta(\cdot),\tau)]}$ as the expected per-experiment reward when $\tau=\tau_0$. Our analysis begins with identifying the optimal scale parameter $\beta(\cdot)$ that maximizes the expected per-experiment reward when $\tau = \tau_0$ and $K \to \infty$.

\begin{assumption}\label{assum:design_regular}
Conditional on the design matrices $\{\bm t_k\}_{k=1}^K$, the leverage factors $b_k=\sqrt{N\bm I^\top(\bm t_k^\top\bm t_k)^{-1}\bm I}$ are deterministic and satisfy $\sup_k b_k\le \bar b<\infty$, $\inf_k b_k\ge \underline b>0$, and $\frac{1}{K}\sum_{k=1}^K b_k^2=O(1)$ and $\frac{1}{K}\sum_{k=1}^K b_k^4=O(1)$ as $K\to\infty$.
\end{assumption}

\begin{theorem}\label{theorem:AP_feature}
Suppose Assumptions~\ref{assum:linear_additive},~\ref{assum:1}, and~\ref{assum:design_regular} hold , $\sigma,\sigma_0$ are known, and $\tau_0>0$. If $\tau = \tau_0$, the optimal scale function $\beta(\cdot)$ is:
    \begin{equation*}\label{eq:beta*_feature}
        \beta^{*}(b_k)  = \frac{\sigma^2 b_k^2}{\sigma_0^2} + \frac{\sqrt{N}\,z_{1 - \alpha/2}\,\sigma\, b_k}{\tau_0}.
    \end{equation*}
In particular, $\mathcal{R}(\beta^{*}(\cdot),\tau_0) >\mathcal{R}(0,\tau_0) = \lim\limits_{K\uparrow+\infty} \mathbb{E}[\hat{r}_{\tradition}]$ under the maintained regime $\tau_0>0$.
\end{theorem}

Theorem~\ref{theorem:AP_feature} characterizes the optimal scale parameter function $\beta^*(\cdot)$ that maximizes the expected per-experiment reward of the \our\ roll-out method when the anchor is set at $\tau = \tau_0$. Unlike in Theorem~\ref{theorem:AP}, the scale parameter in this setting varies across experiments and depends on \( b_k \). This dependency ensures that the scale parameter effectively leverages the diverse heterogeneous information ($b_k^2$) obtained from different experiments. The sensitivity analysis for the parameters \( \sigma^2 \), \( \sigma_0^2 \), and \( \tau_0 \) remains the same as in the discussion following Theorem~\ref{theorem:AP}. Readers may refer to the previous subsection for details. 

Similarly, in practice, the parameters $\tau_0$, $\sigma^2$, and $\sigma_0^2$ are unobservable by the platform. Hence, we estimate these parameters with the pooled data from all experiments and derive the estimator for function $\beta^*(\cdot)$. With estimated variance $\hat\sigma^2 = \frac{1}{K}\sum_{k}s_k^2$, we define the expected per-experiment reward under the \our\ method:
\begin{equation*} 
    \bar{r}(\beta(\cdot),\tau) = \frac{1}{K}\sum_{k:\bar\tau_k>\frac{N}{N+\beta(b_k)}\frac{\hat{\sigma} b_k}{\sqrt{N}}z_{1 - \alpha/2 }}\tau_k.
\end{equation*}

\begin{theorem} \label{theorem:data_driven_para_feature}
Suppose Assumptions~\ref{assum:linear_additive},~\ref{assum:1}, and~\ref{assum:design_regular} hold and $\tau_0>0$.
   Define $s_k^2 := \frac{1}{N-2-d_x}(\bm Y_k - \bm t_k(\bm t_k^{\top}\bm t_k)^{-1}\bm t_k^{\top}\bm Y_k)^{\top}(\bm Y_k - \bm t_k(\bm t_k^{\top}\bm t_k)^{-1}\bm t_k^{\top}\bm Y_k)$, $\hat{\tau}_0 = \frac{1}{K}\sum_{k}\hat{\tau}_k$  and,
   \begin{footnotesize}
        \begin{equation*}
       \hat{\beta}^*(b_k) := \frac{b_k^2(\frac{1}{K}\sum_{k^{'}}s_{k^{'}}^2)}{ \frac{1}{K}\sum_{k^{'}}(\hat{\tau}_{k^{'}} - \hat{\tau}_0)^2 -  \frac{1}{N}\Big(\frac{1}{K}\sum_{k^{'}}s_{k^{'}}^2 \cdot \frac{1}{K}\sum_{k^{'}}b_{k^{'}}^2 \Big)} + \frac{z_{1 - \alpha/2}b_k\sqrt{N\frac{1}{K}\sum_{k^{'}}s_{k^{'}}^2}}{\hat{\tau}_0}.
   \end{equation*}
   \end{footnotesize}
We have, as $K \uparrow \infty$, $\hat{\tau}_0 \to_p \tau_0$ and for any $b_k$, $\hat{\beta}^*(b_k) \to_p \beta^*(b_k)$. Furthermore, we have, $\bar{r}(\hat{\beta}^*(\cdot),\hat{\tau}_0) \to_p \mathcal{R}(\beta^*(\cdot),\tau_0)$. 
\end{theorem}


Theorem~\ref{theorem:AP_feature} and Theorem~\ref{theorem:data_driven_para_feature} together prove that, under the maintained positive-anchor and design-regularity assumptions, the data-driven \our\ method converges to the optimal covariate-adjusted shrinkage rule and strictly outperforms the \tradition\ method with covariates under the OLS specification when $K$ is sufficiently large. 

\section{Synthetic Experiments}
\label{sec:Synthetic_valid}
In this section, we demonstrate the advantage of our proposed \our\ method using synthetic experiments across various scenarios outlined in Section~\ref{subsec:scenario-1} to Section~\ref{subsec:scenario-4}. All relevant code can be found at GitHub.\footnote{See \url{https://github.com/shoucheng666/Data-Pooling-Treatment-Roll-Outs}.}

\textbf{Benchmarks:} In addition to the traditional \tradition\ method, we consider a widely used alternative, the Bayesian approach \citep{AbadieAgarwalImbens2023}, which also leverages information pooling across multiple experiments. Details on the data pooling procedure under the Bayesian framework are provided in Appendix \ref{app:bayesian_method}.


\textbf{Metrics:} We introduce two key metrics to evaluate the our methods: Optimality Ratio (OR) and Value of Data Pooling (VDP). OR measures the relative performance of a roll-out method (e.g, \our, Bayesian method or \tradition) compared to the oracle per-experiment reward $r^*$, while VDP quantifies the relative reward improvement of \our\ or Bayesian method over \tradition. Formally, we define:
\begin{equation*}
    \text{Optimality Ratio (OR) of Method $\mathsf{z}$} = \frac{\hat{r}_{\mathsf{z}}}{r^*},\ \text{Value of Data Pooling (VDP)} = \frac{\hat{r}_{\mathsf{z}}}{\hat{r}_{\tradition}} - 1,
\end{equation*}
where $\hat{r}_{\mathsf{z}}$ is the per-experiment reward generated by method $\mathsf{z}$.

To provide a more comprehensive evaluation of our proposed method, we also frame the treatment roll-out decision problem as a classification task. Specifically, for each experiment $k$, if $\tau_k > 0$, it is labeled as a positive case; otherwise, it is labeled as a negative case. This classification perspective allows us to analyze \our\ and \tradition\ as different classification algorithms. Thus, we further evaluate both methods using four standard classification metrics: Accuracy, Recall, Specificity, and Precision, each derived from the confusion matrix. These metrics are formally defined as:
\begin{align*}
    &\text{Accuracy} = \frac{TP+TN}{TP+TN+FP+FN},\ \text{Recall} = \frac{TP}{TP+FN},\\ &\text{Specificity} = \frac{TN}{TN+FP},\ \text{Precision} = \frac{TP}{TP+FP},
\end{align*}
where TP (True Positives) denotes the number of correctly identified positive cases; TN (True Negatives) denotes the number of correctly identified negative cases; FP (False Positives) denotes the number of negative cases incorrectly classified as positive; FN (False Negatives) denotes the number of positive cases incorrectly classified as negative. Unless otherwise specified, all experiments in this paper use a default significance level of $\alpha = 0.05$.

\subsection{Non-overlapping Experiments and Linear Specification} \label{subsec:no-overlap-no-feature}
When experiments are non-overlapping and model specifications are linear, results in Section~\ref{sec:theoretical_valid} have already theoretically demonstrated how the \our\ method outperforms the \tradition\ method by effectively balancing the bias-variance tradeoff in experiment roll-out decisions. 
In this subsection, we use synthetic experiments to illustrate the substantial edge of our proposed method even when the number of experiments is small or moderate. 



\subsubsection{Without Covariate Information.}\label{sec:s1 - without covariate} We set up the basic experimental setting as follows: the platform conducts $K = 100$ experiments. The ground-truth ATE, $\tau_k$, is randomly sampled from a normal distribution $\mathcal{N}(1, 3^2)$. Each experiment has $N = 10$ observations. The noise term $\epsilon_{k,i}$ follows a normal distribution, $\epsilon_{k,i} \sim \mathcal{N}(0, 3^2)$.\footnote{In reality, each experiment typically contains a much larger number of observations, at the magnitude of hundreds of thousands or even millions, for a large-scale online platform~\citep{KohaviTangXu2020}. In this case, due to significant heterogeneity among users, $\sigma$ is also orders of magnitude higher than $\sigma_0$. In our experiments, we proportionally scale down $N$ and $\sigma$, while still capturing the key characteristics of the real-world scenario with reduced computational burden.} Hereafter, we refer to this configuration of ATE and outcome both following normal distributions as the normal-normal setting. 
We first compare \our\ (Bayesian method) and \tradition\ with respect to different metrics: OR, Accuracy, Recall, Specificity, and Precision. We simulate 1,000 iterations and apply both methods in each iteration for the roll-out decision. 

\begin{figure}[htbp!]
\vspace{-0.2in}
\centering
\subfigure[Metric differences with 95\% middle results: \our\ (Bayesian method)  minus \tradition.]{
\begin{minipage}[t]{0.4\linewidth}
\centering
\includegraphics[width=2.5in,height=2.0in]{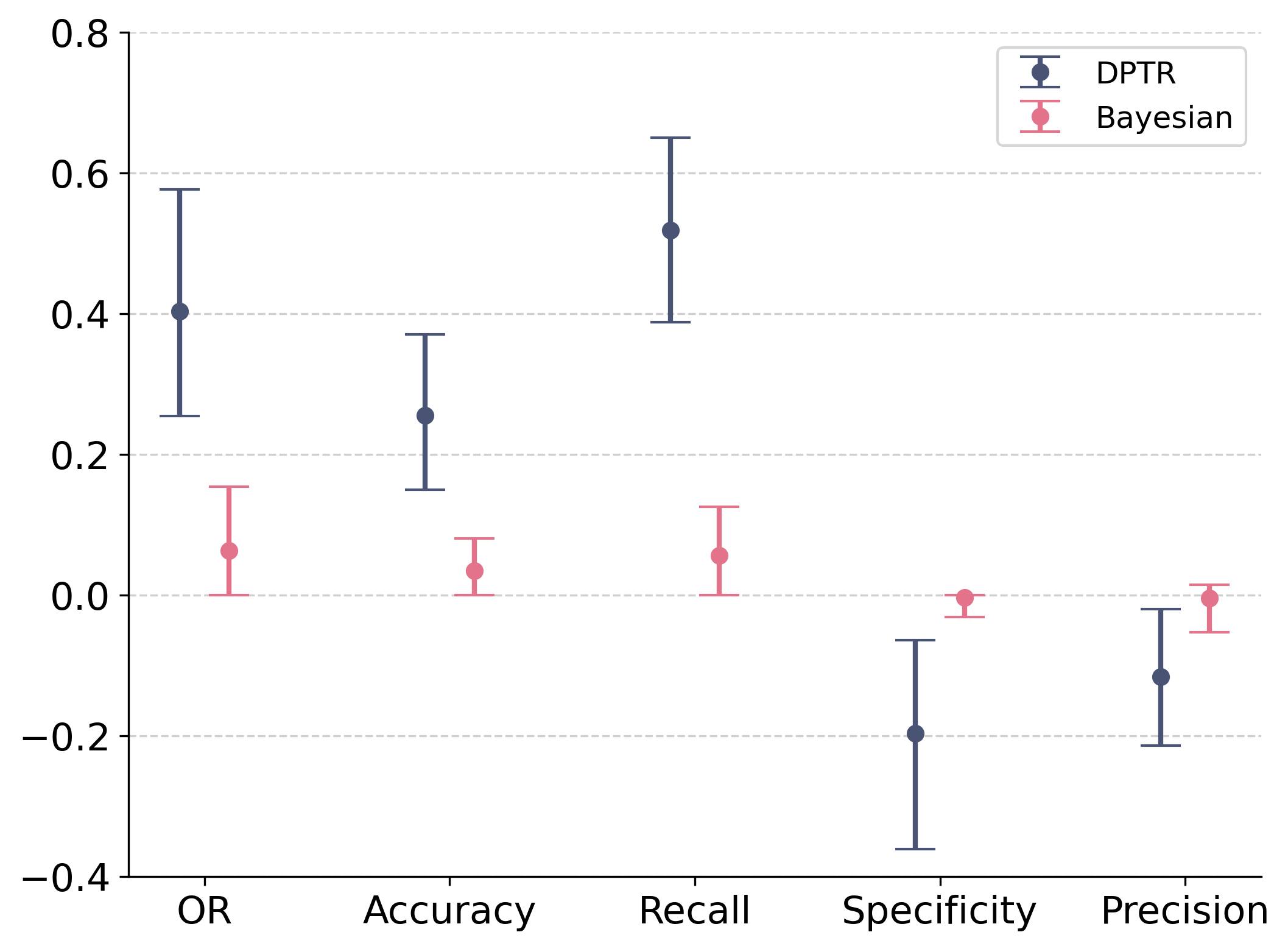}
\end{minipage}
}%
\subfigure[Performance comparisons for different significance level $\alpha$]{
\begin{minipage}[t]{0.55\linewidth}
\centering
\includegraphics[width=2.5in,height=2.0in]{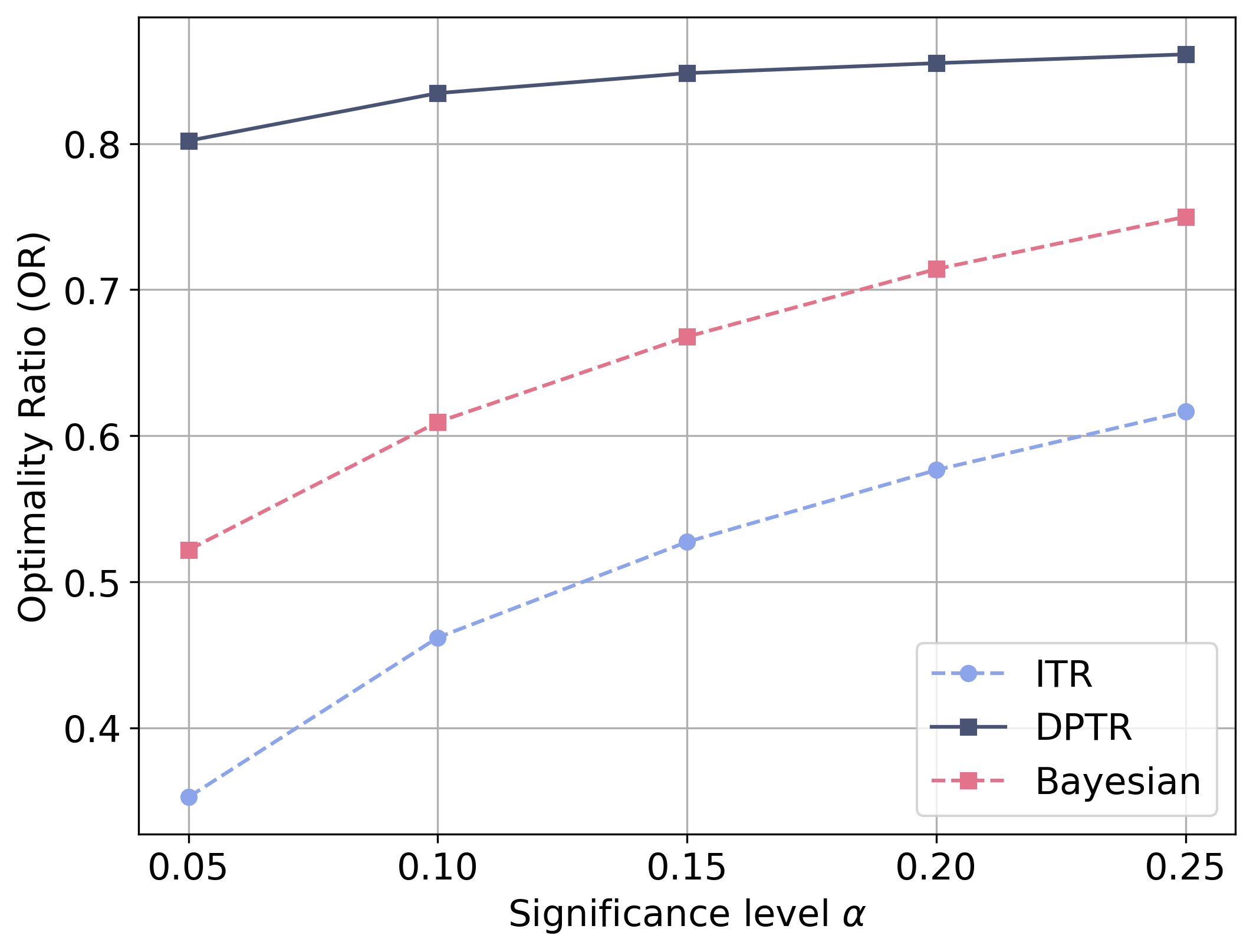}
\end{minipage}%
}%
\caption{Performance comparisons for non-overlapping experiments without covariates.}
\label{fig:s1-basic}
\vspace{-0.15in}
\end{figure}

Figure~\ref{fig:s1-basic}(a) depicts the intervals that cover the middle 95\% of the differences in the five metrics between the \our\ (Bayesian method) and \tradition\ methods across 1,000 iterations.
A positive difference indicates that \our\ (Bayesian method) outperforms \tradition, and vice versa.
We find that \our\ and Bayesian methods consistently generate higher rewards than \tradition, suggesting the superior performance of data pooling methods in roll-out decisions. A more careful look at the performance metrics reveals that the \our\ and Bayesian methods are more likely to make correct experiment roll-out decisions. Furthermore, \our\ and Bayesian methods significantly improve the recall. Compared to \tradition, they can better identify the experiments that should be rolled out. On the other hand, such improvement is also at the cost of lower specificity and precision. This is because \our\ and Bayesian methods may mistakenly roll out some experiments with a negative treatment effect. 

We compare the performance of the \our\ method with the Bayesian method. First, we observe that both data-pooling methods exhibit similar trends across all metrics, indicating that the core ideas and insights behind \our\ method are closely aligned with those of the Bayesian method. Second, the \our\ method achieves greater improvements in the optimality ratio, accuracy, and recall, albeit at the cost of slightly lower specificity and precision. This trade-off arises because our method is more decision-aware and places a stronger emphasis on maximizing reward, as discussed in Section~\ref{subset:theoretical_without_covariates}.






It is useful to investigate the robustness of our method with respect to different significance levels $\alpha$. Specifically, we vary $\alpha$ from 0.05 to 0.25 in increments of 0.05 and plot the OR metric for \our, \tradition\ and the Bayesian method. As shown in Figure~\ref{fig:s1-basic}(b), our \our\ method consistently achieves higher OR values across all tested $\alpha$'s and, notably, the performance gap remains significant regardless of the significance levels. Therefore, our proposed method strikes a delicate balance between statistical power and decision quality.

\begin{figure}[htbp!]
\vspace{-0.15 in}
\centering
\subfigure[Normal-normal distribution]{
\begin{minipage}[t]{0.5\linewidth}
\centering
\includegraphics[width=2.5in,height=1.8in]{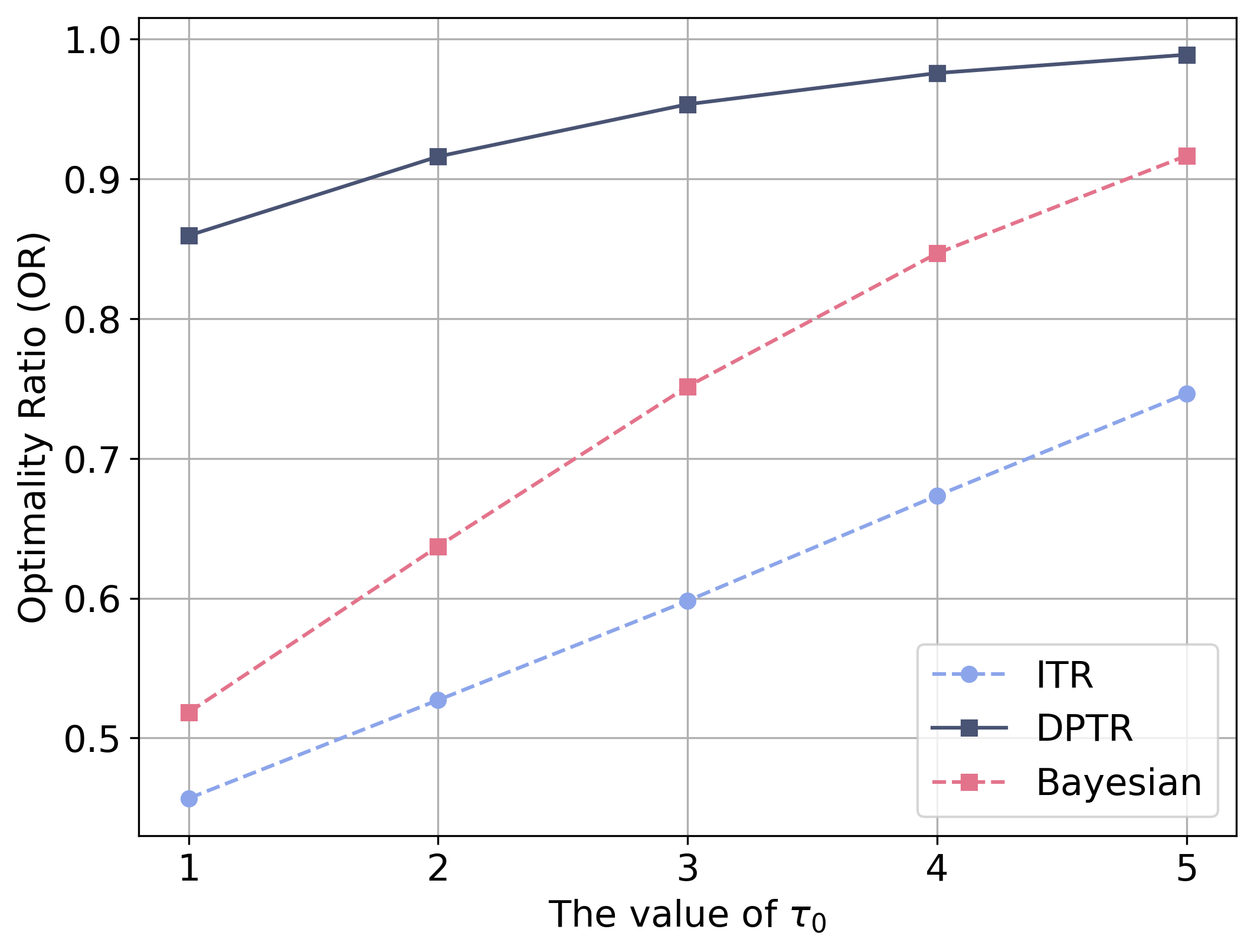}
\end{minipage}
}%
\subfigure[Uniform-uniform distribution]{
\begin{minipage}[t]{0.5\linewidth}
\centering
\includegraphics[width=2.5in,height=1.8in]{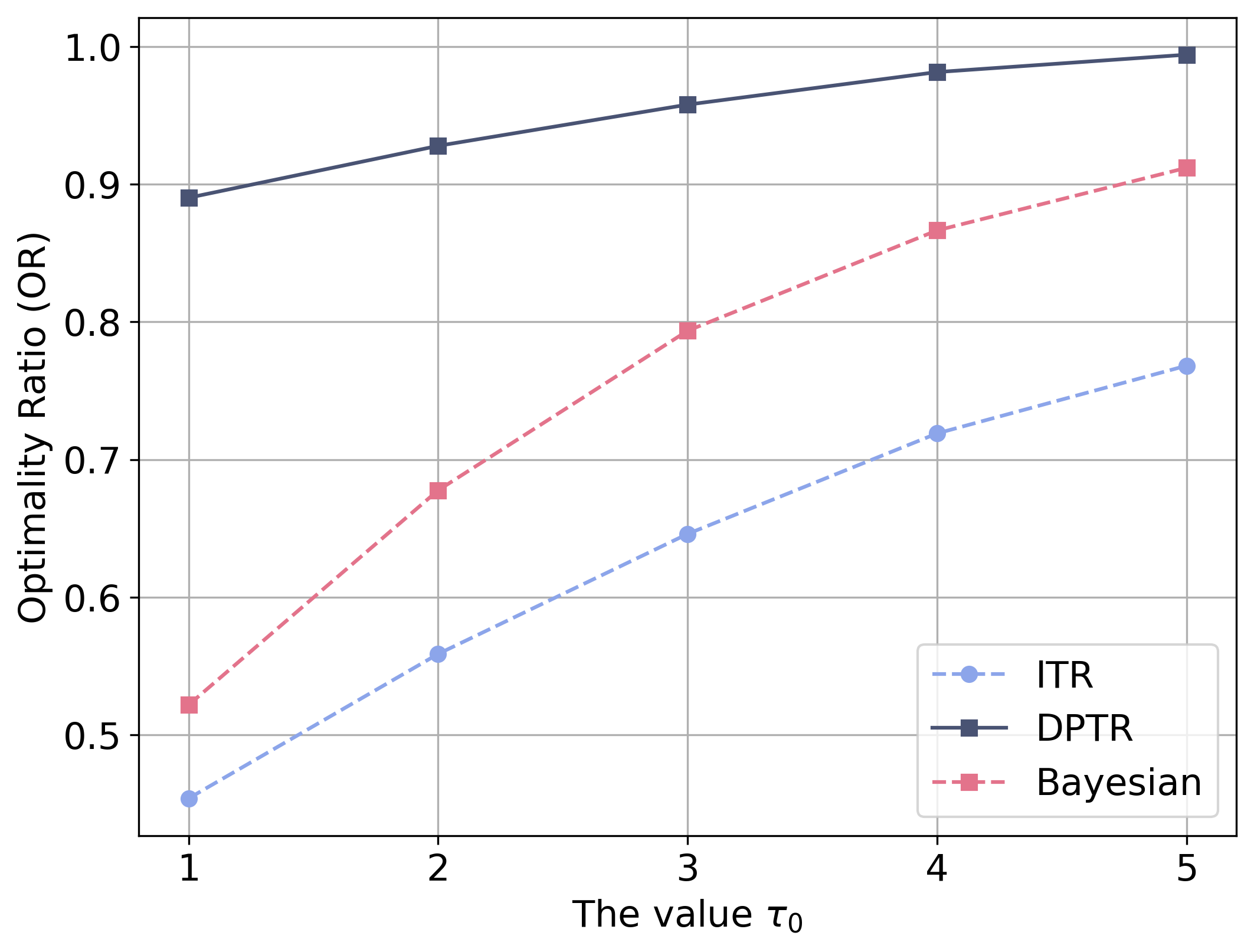}
\end{minipage}%
}%

\subfigure[Different number of experiments $K$]{
\begin{minipage}[t]{0.5\linewidth}
\centering
\includegraphics[width=2.5in,height=1.8in]{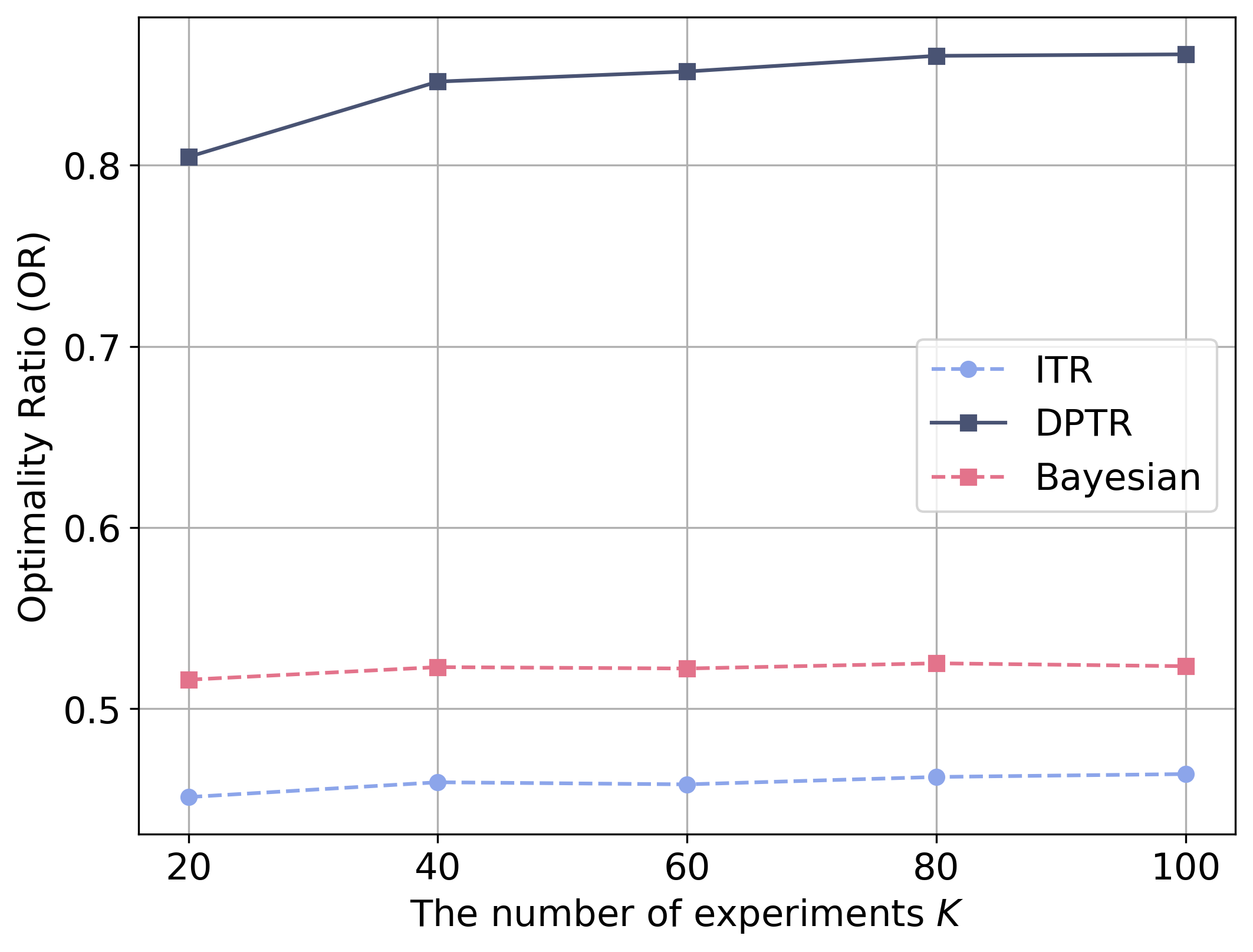}
\end{minipage}%
}%
\subfigure[Different sample size $N$]{
\begin{minipage}[t]{0.5\linewidth}
\centering
\includegraphics[width=2.5in,height=1.8in]{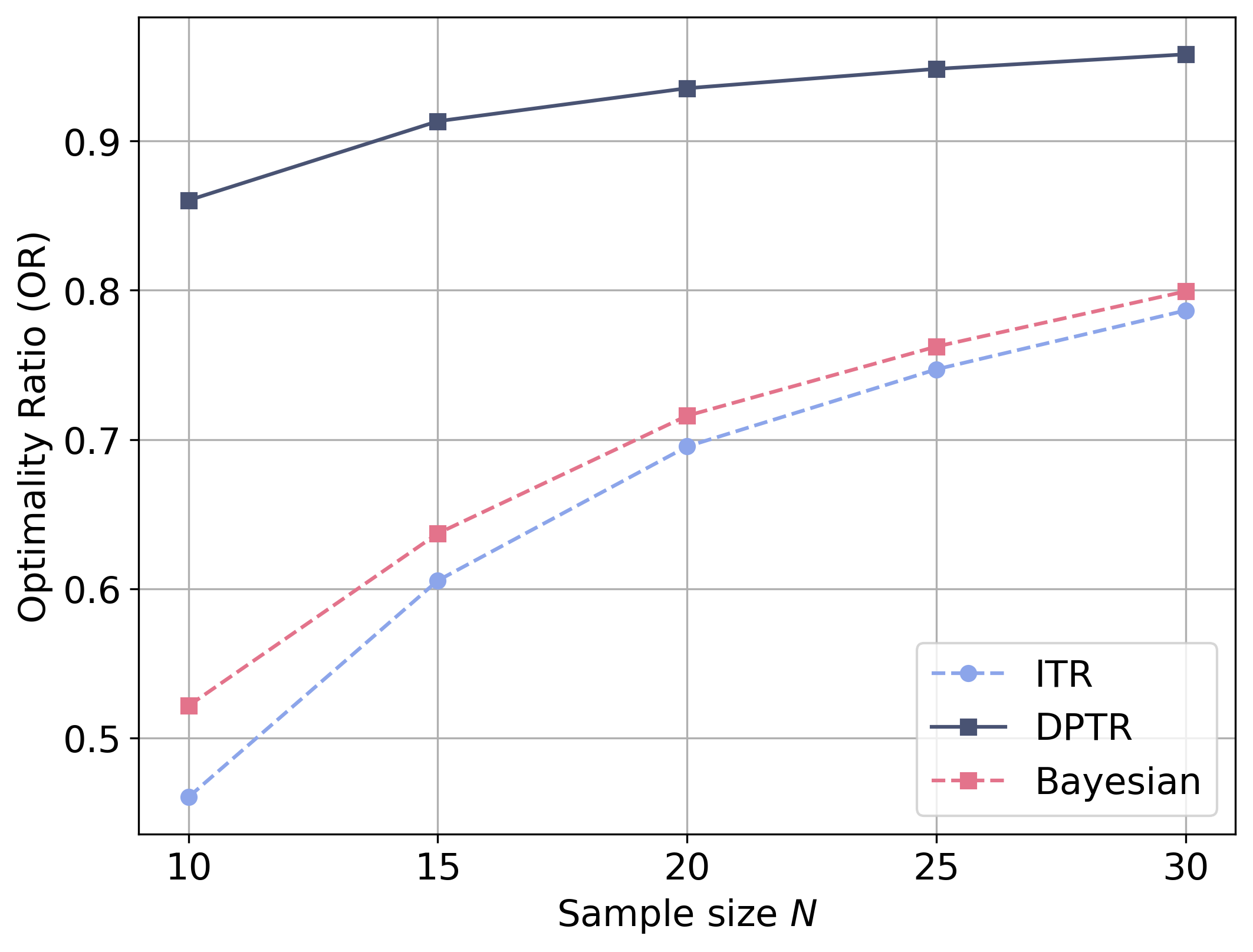}
\end{minipage}%
}%
\centering
\caption{Performance comparisons for non-overlapping experiments without covariates}
\label{fig:s1-sensitivity}
\vspace{-0.15 in}
\end{figure}

We proceed to further examine the robustness of \our\, under varying distributions, numbers of experiments and sample sizes. As shown in Figure~\ref{fig:s1-sensitivity}, \our\ outperforms the \tradition\ and Bayesian benchmarks regardless of prior ATE mean $\tau_0$, the distributions of prior ATE and noise terms, the number of experiments $K$, and the sample size $N$. This sensitivity analysis reveals our proposed method is particularly effective when the prior ATE mean $\tau_0$ is small, the sample size $N$ is small, or the number of experiments $K$ is large. In these cases, \our\ assigns a high weight on the anchor $\hat\tau_0$ estimated from data of multiple experiments, fully leveraging the benefit of data pooling. In particular, the finding related to sample size $N$ provides a theoretical explanation for the empirical evidence presented in \cite{chen2024role}, which shows that incorporating aggregate market information benefits small retailers more than large ones, likely due to the limited data available to smaller retailers.

\subsubsection{With Covariate Information.}\label{sec:covariate} We set up the experimental setting as follows: the number of covariates is set to \( d_x = 4 \), with covariate values sampled from \( x_{k,i}^j \sim U(0,1) \). The intercept \( a_k \) and the coefficients \( \boldsymbol{\theta}_k \) are randomly drawn from \( U(-0.3, 0.5) \). The number of experiments (\( K \)), the sample size (\( N \)), and the distributions of \( \tau_k \) and noise \( \epsilon_{k,i} \) are the same as in the previous subsection on non-overlapping experiments without covariates.

\begin{figure}[!ht]
\vspace{-0.2 in}
\centering
\subfigure[Normal-normal distribution]{
\begin{minipage}[t]{0.5\linewidth}
\centering
\includegraphics[width=2.5in,height=1.8in]{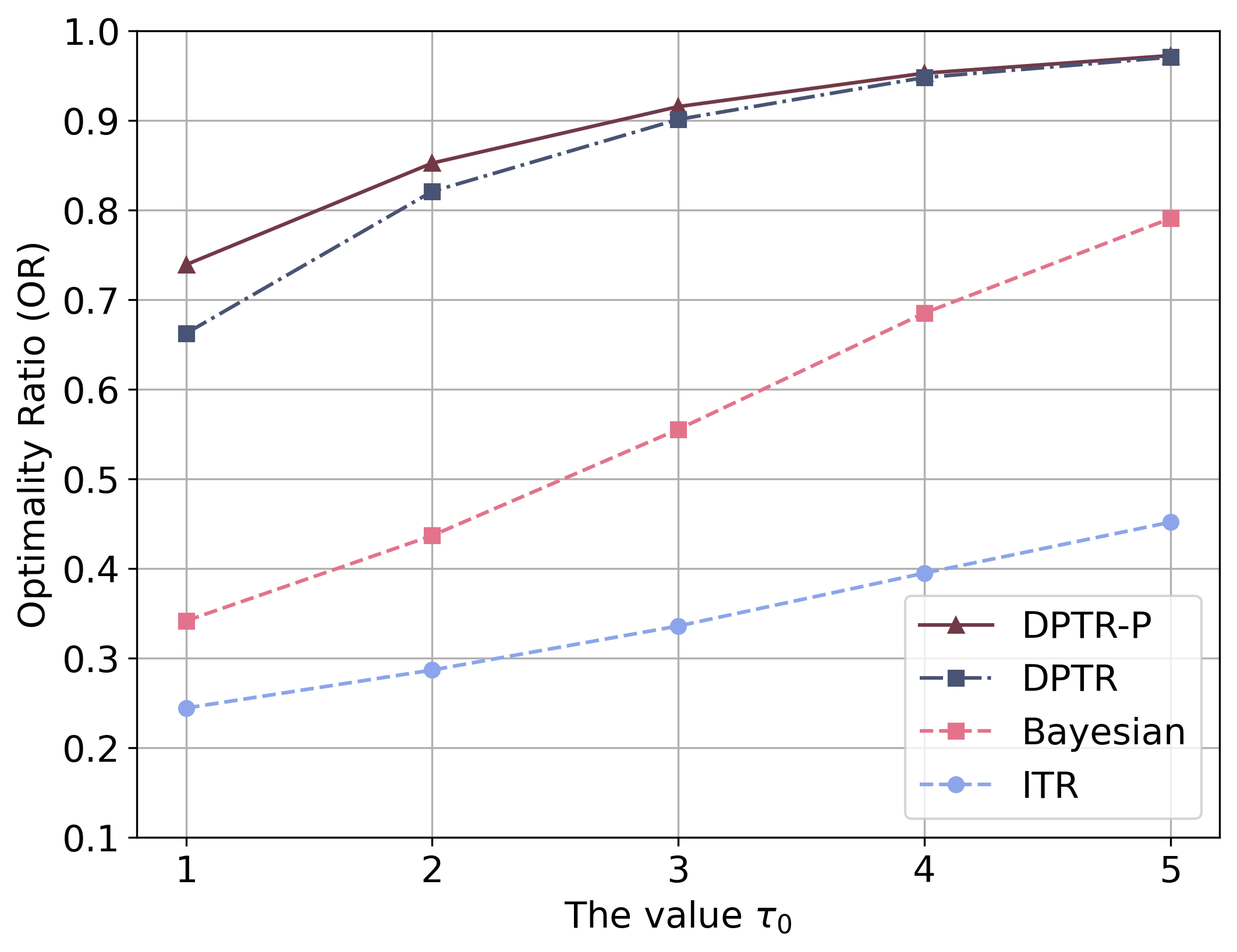}
\end{minipage}
}%
\subfigure[Uniform-uniform distribution]{
\begin{minipage}[t]{0.5\linewidth}
\centering
\includegraphics[width=2.5in,height=1.8in]{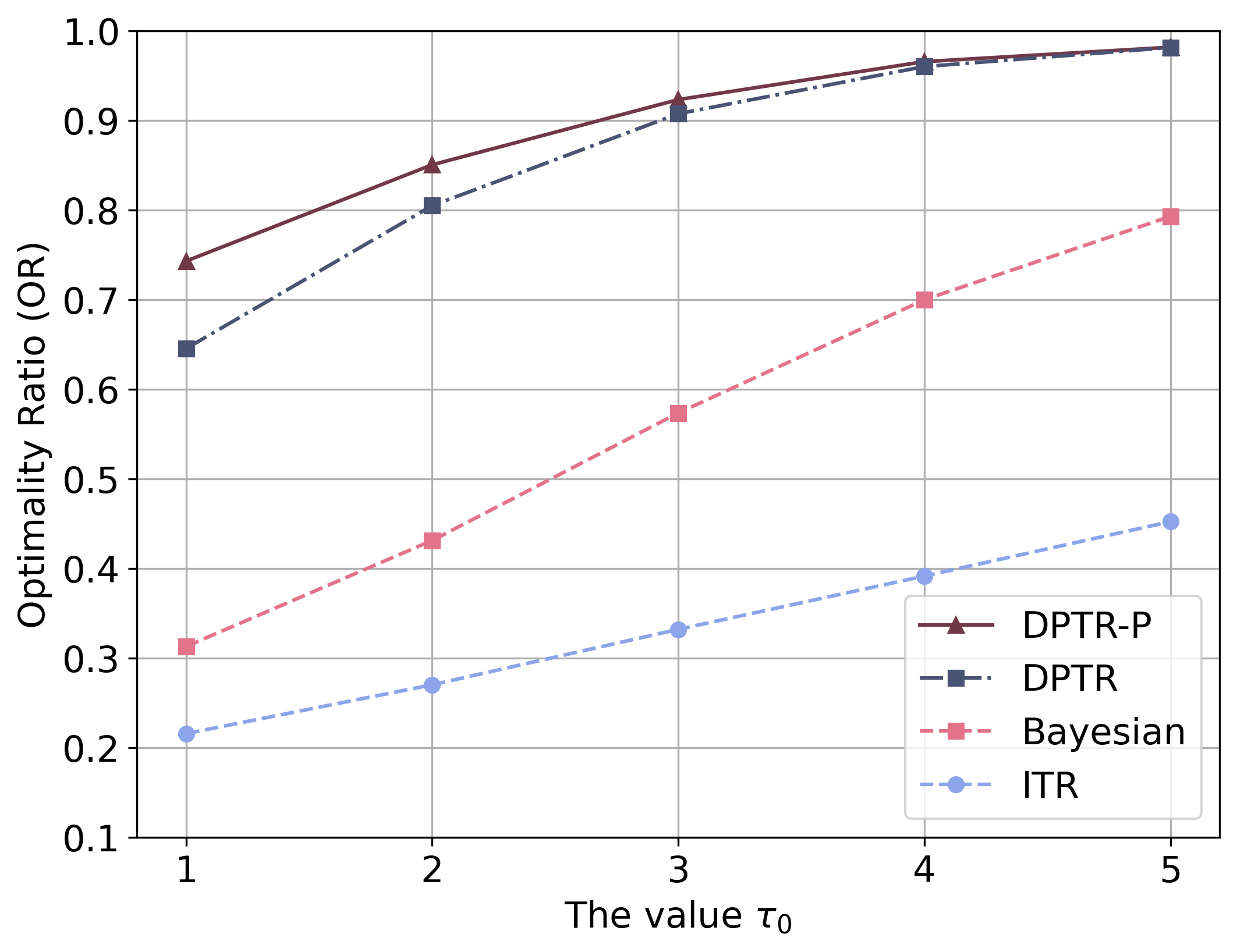}
\end{minipage}%
}%

\subfigure[Different number of experiments $K$]{
\begin{minipage}[t]{0.5\linewidth}
\centering
\includegraphics[width=2.5in,height=1.8in]{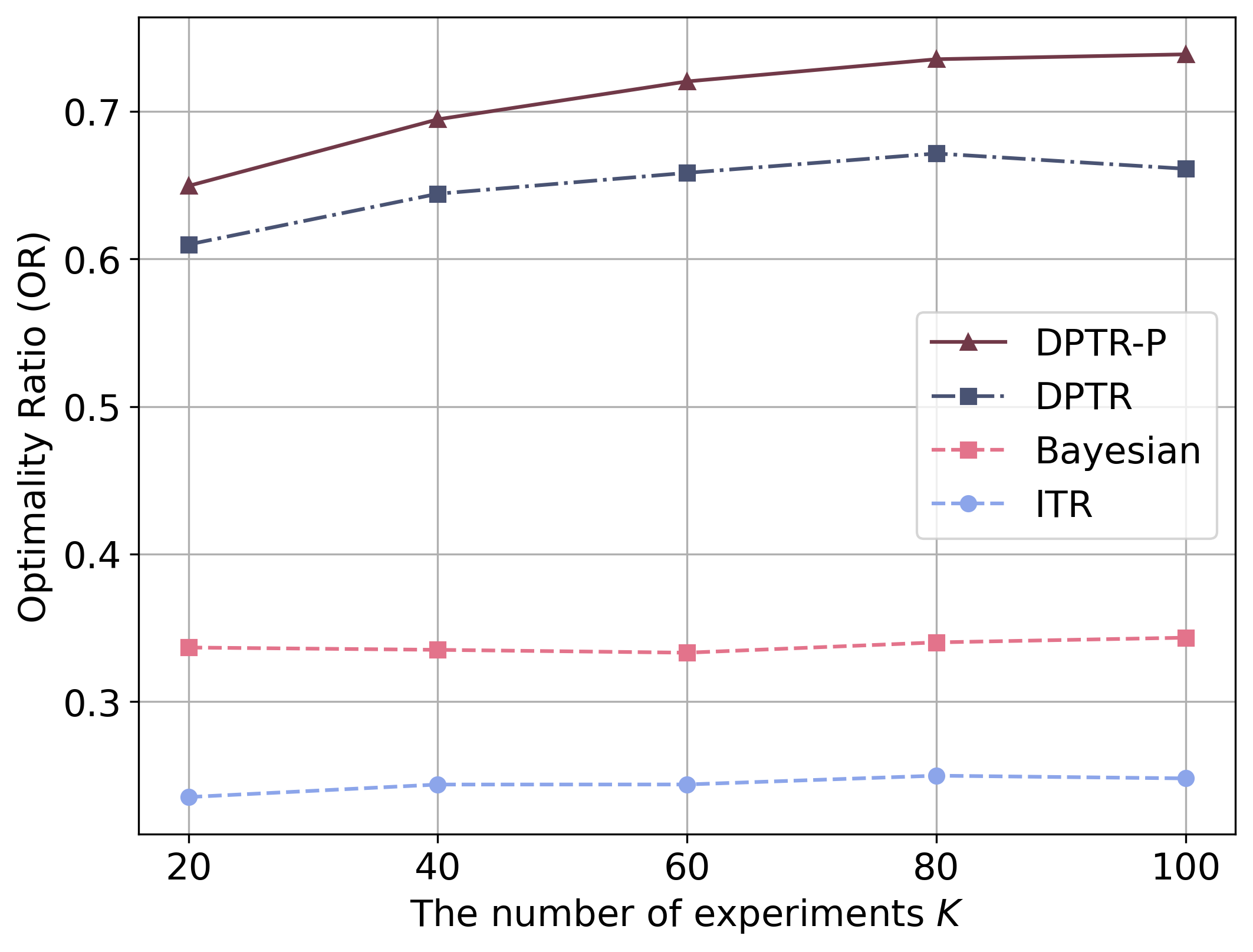}
\end{minipage}%
}%
\subfigure[Different sample size $N$]{
\begin{minipage}[t]{0.5\linewidth}
\centering
\includegraphics[width=2.5in,height=1.8in]{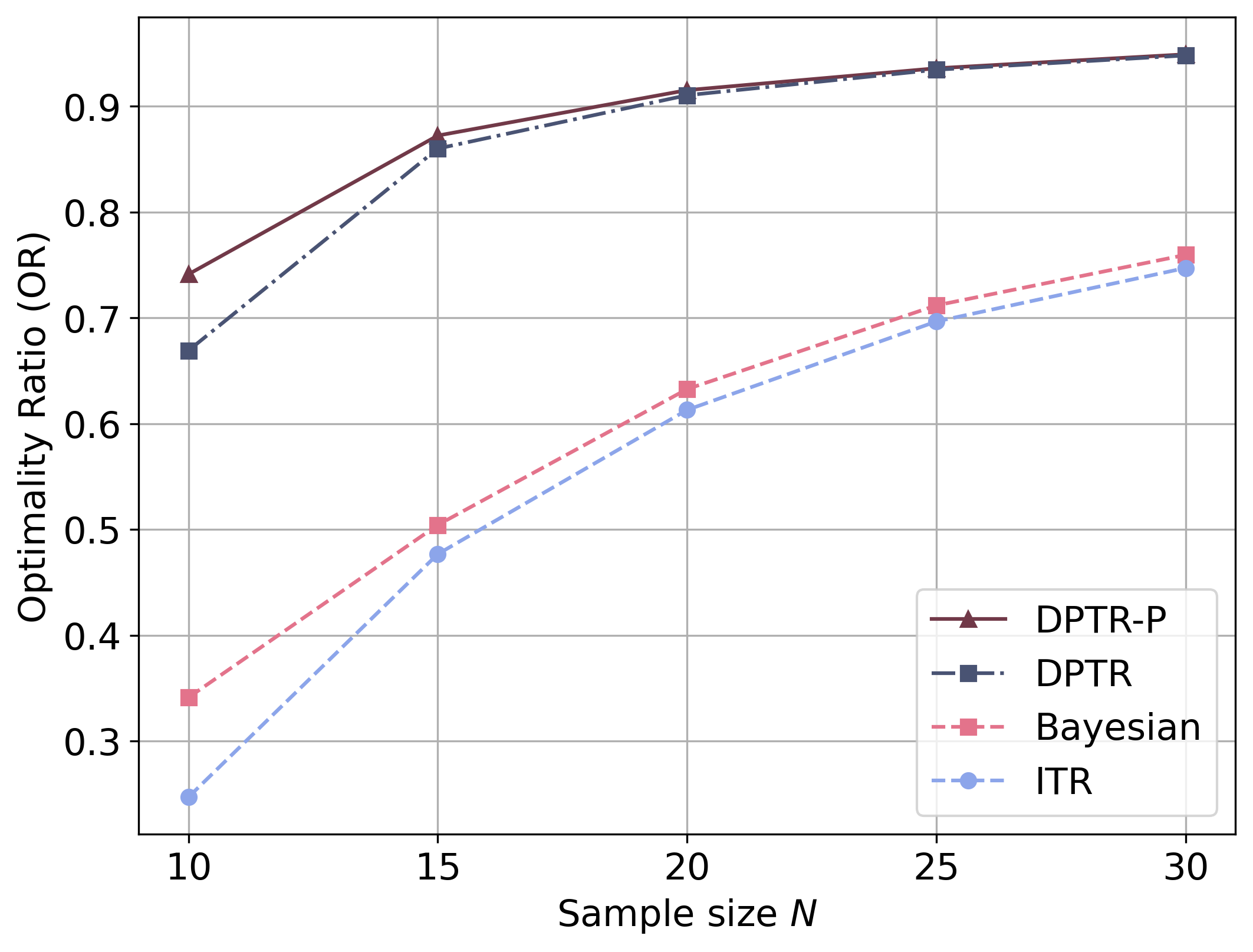}
\end{minipage}%
}%
\centering
\caption{Performance comparisons for non-overlapping experiments with covariates}
\label{fig:s1-sensitivity-feature}
\vspace{-0.15 in}
\end{figure}

In this subsection, we test two variations of our method against the \tradition\ and Bayesian benchmarks: (a) \our\ method using the common scale parameter $\hat{\beta}$ defined in Eqn.~\eqref{eq:beta_form_nonpersonlized}, and the \ourP\ method using personalized scale parameters $\hat{\beta}_k$ defined in Eqn.~\eqref{eq:beta_form_personlized}. Similar to the case without covariate information, we focus on comparing OR of different methods. The results are presented in Figure~\ref{fig:s1-sensitivity-feature}. 

On one hand, the results are consistent with those in the case without covariates shown in Figure \ref{fig:s1-sensitivity}, demonstrating the robustness of our proposed method when incorporating covariate information. On the other hand, we further show that \ourP\ could achieve an even higher performance than \our\ with a shared scale parameter, which is well aligned with our theoretical results (Theorems \ref{theorem:AP_feature} and \ref{theorem:data_driven_para_feature}), demonstrating the value of leveraging personalized information.


\subsection{Non-overlapping Experiments and Non-linear Specification}\label{subset: no-overlap-with-feature}

In this subsection, we conduct a series of numerical experiments to evaluate the performance of \our\ under non-overlapping experiments and non-linear specification, as introduced in Section~\ref{subsec:scenario-2}. We consider the following experimental setup: the platform runs $K = 100$ experiments, each with $N = 100 $ observations. The noise \( \epsilon_{k,i} \) follows a normal distribution, \( \epsilon_{k,i} \sim \mathcal{N}(0, 3^2) \). We set the number of covariates as $d_x = 4$, with covariate values sampled from $x_{k,i}^j \sim U(0, 1)$. The ground-truth response function is defined as
$g_{k}(X_{k,i}) = [\gamma_{k,0}^{\top}X_{k,i},\gamma_{k,1}^{\top}X_{k,i}]^{\top}$, where the components of $\gamma_{k,0}$ and $\gamma_{k,1}$ are independently drawn from the distribution $U(-0.3,0.5)$. To estimate the nuisance parameter, we use a two-layer fully connected neural network with 10 units per layer and ReLU activations, without dropout, as the function class $\mathcal{F}$. As illustrated in Section~\ref{subsec:scenario-2}, double machine learning and cross-fitting techniques are applied to estimate the ATE of each experiment $k$, denoted by $\hat\tau_k$.

\begin{figure}[htbp!]
 \vspace{-0.15 in}
    \centering
    \includegraphics[width=0.4\linewidth]{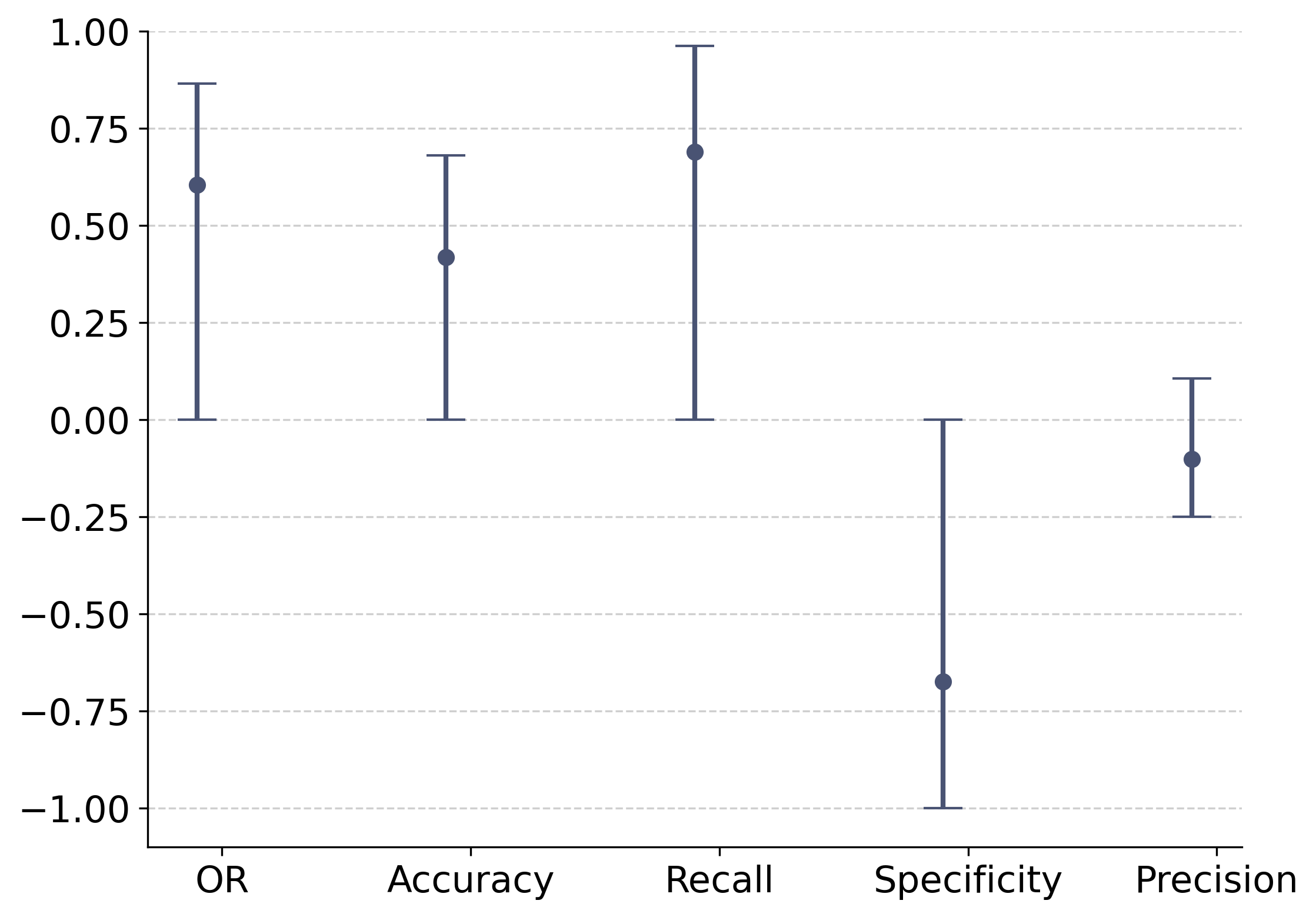}
    \caption{\our\ vs. \tradition: Non-overlapping experiments and non-linear specification}
    \label{fig:box_plot_metrics_S1_dml} 
    \vspace{-0.15 in}
\end{figure}

Similar to Section~\ref{sec:s1 - without covariate}, we also run the simulation for 1,000 iterations and show the middle 95\% of the differences in five metrics between \our\ and \tradition\ in the five metrics (OR, Accuracy, Recall, Specificity and Precision), as shown in Figure \ref{fig:box_plot_metrics_S1_dml}. It is clearly illustrated in Figure~\ref{fig:box_plot_metrics_S1_dml} that the results are consistent with those of \our\ method in the setting of non-overlapping experiments and non-linear specification (see Figure \ref{fig:s1-basic}(a)).


Next, we incorporate the personalized scale parameter $\hat{\beta}_k$ defined in Eqn.~\eqref{eq:beta_form_personlized_DML}, and evaluate the performance of our methods under two scenarios: one with a relatively large number of experiments ($K = 100$), and the other with a relatively small number ($K = 5$). Under each situation, we vary \( \sigma \) from \( 1 \) to \( 5 \) in increments of \( 1 \). For each parameter specification, we run the simulation for 1,000 iterations and report the average OR of each roll-out method in Table~\ref{tab:1}. Our simulation results show that \our\ and \ourP\ consistently yield significant reward improvements over \tradition, regardless of the number of experiments $K$ and noise variance $\sigma^2$. The improvement is noticeably greater when the number of experiments is larger. In this case, the estimation accuracy of $\hat{\beta}$, $\hat{\beta}_k$, and $\hat{\tau}_0$ is higher, rendering data pooling via our methods more effective. We also observe that \ourP\ consistently outperforms \our, except for the case where $K$ is large and $\sigma^2$ is small. Even though $b_k^2$ is constructed based on a misspecified linear model, it still provides benefits when the number of experiments is small or the variance is large.

\begin{table}[!ht]
\vspace{-0.15 in}
\centering
		{\def\arraystretch{1.1}  
\centering
\begin{tabular*}{1\textwidth}
{@{\extracolsep{\fill}}ccccccc}
\hline
\hline
 & \multicolumn{3}{c}{$K = 100$}  & \multicolumn{3}{c}{$K = 5$}  \\
\hline
Noise Variance ($\sigma^2$)  & \tradition\ & \our\ & \ourP\  &  \tradition\  & \our\  &\ourP\  \\
\hline
$1^2$  &0.4097 & 0.9214 & 0.8930 &0.3804 & 0.6768 & 0.7476  \\
$2^2$ &0.1708 & 0.8394 & 0.8417 & 0.1662 & 0.4855 & 0.6728 \\
$3^2$    &0.1003 & 0.7123 & 0.7612 &0.0990 & 0.3726 & 0.5882  \\
$4^2$ &0.0745 & 0.5831 & 0.7029 & 0.0739 & 0.3455 & 0.5357\\
$5^2$ &0.0596 & 0.5349 & 0.6858 & 0.0468 & 0.2982 & 0.4951\\
\hline
\hline
\end{tabular*}
\caption{Performance comparison under OR: Non-overlapping experiments and non-linear specification}
\label{tab:1}
}
\vspace{-0.3 in}
\end{table}

\subsection{Overlapping Experiments and Linear Specification}\label{subset: with-overlap-no-feature}

We now numerically test the performance of \our\ under overlapping experiments and linear specification, as described in Section~\ref{subsec:scenario-3}. The experimental setup is as follows: the platform runs $K = 100$ experiments, with $N= 10 + K$ observations. The $D_{k,i}$'s are i.i.d. Bernoulli random variables with $\mathbb P[D_{k,i}=1]=\mathbb P[D_{k,i}=0]=0.5$. The noise \( \epsilon_{k,i} \) follows a normal distribution, \( \epsilon_{k,i} \sim \mathcal{N}(0, 3^2) \). The ground-truth ATE, \( \tau_k \), is randomly sampled from a normal distribution \( \mathcal{N}(1, 3^2) \).

\begin{figure}[htbp!]
 \vspace{-0.15 in}
    \centering
     \includegraphics[width=0.4\linewidth]{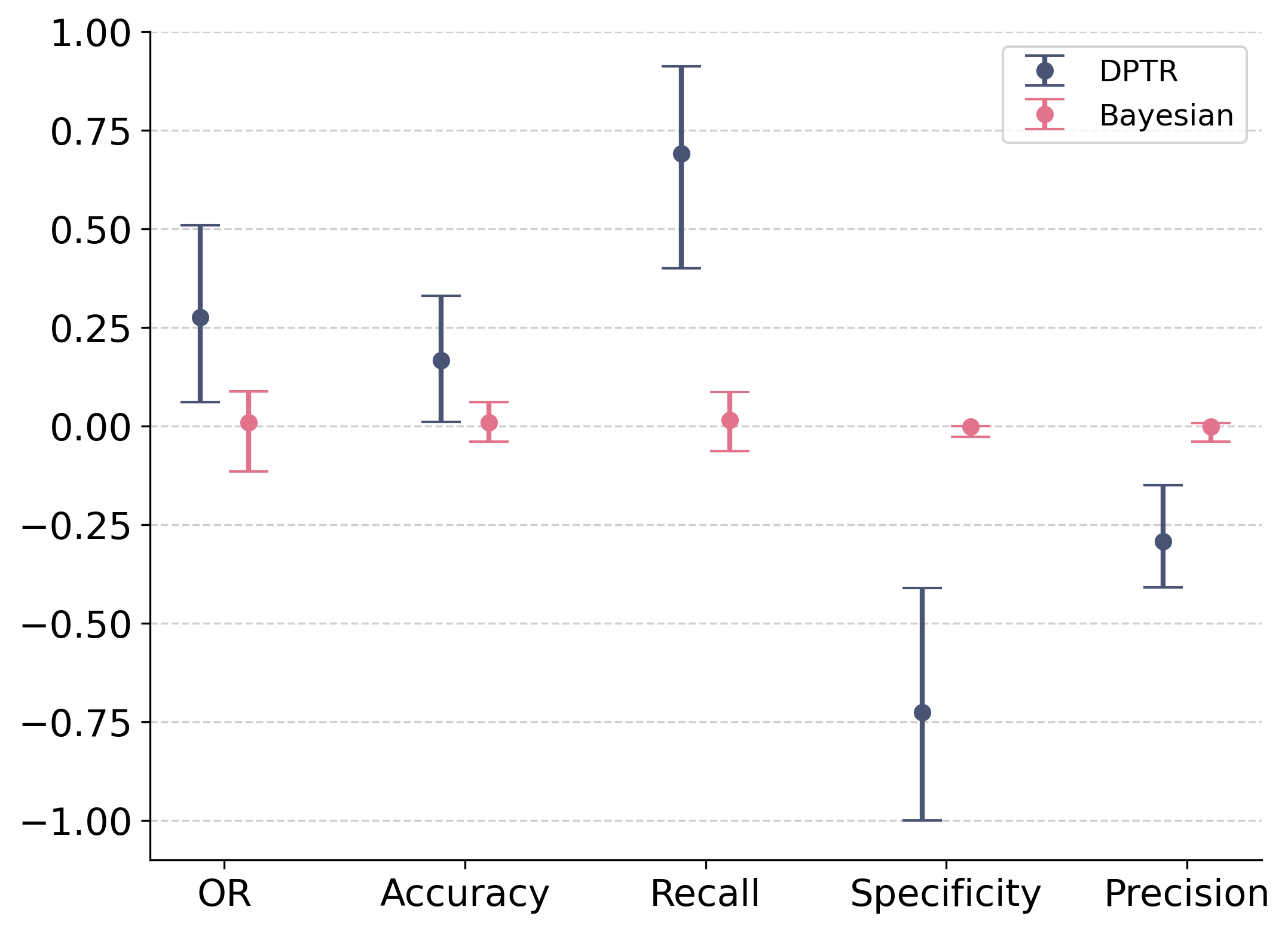}
    \caption{\our\ (Bayesian method) vs. \tradition: Overlapping experiments and linear specification}
    \label{fig:box_plot_metrics_S3_overlap}
    \vspace{-0.15 in}
    
\end{figure}

We repeat the experiment 1,000 times using the Bayesian, \our\ and \tradition\ methods, and present the middle 95\% of the differences in five metrics (OR, Accuracy, Recall, Specificity, and Precision) to compare their performance, as shown in Figure \ref{fig:box_plot_metrics_S3_overlap}. The results closely mirror those in Figure~\ref{fig:s1-basic} (a), even with overlapping experiments.

\begin{figure}[!ht]
\vspace{-0.15 in}
\centering
\subfigure[$K = 100$]{
\begin{minipage}[t]{0.5\linewidth}
\centering
\includegraphics[width=2.5in,height=1.8in]{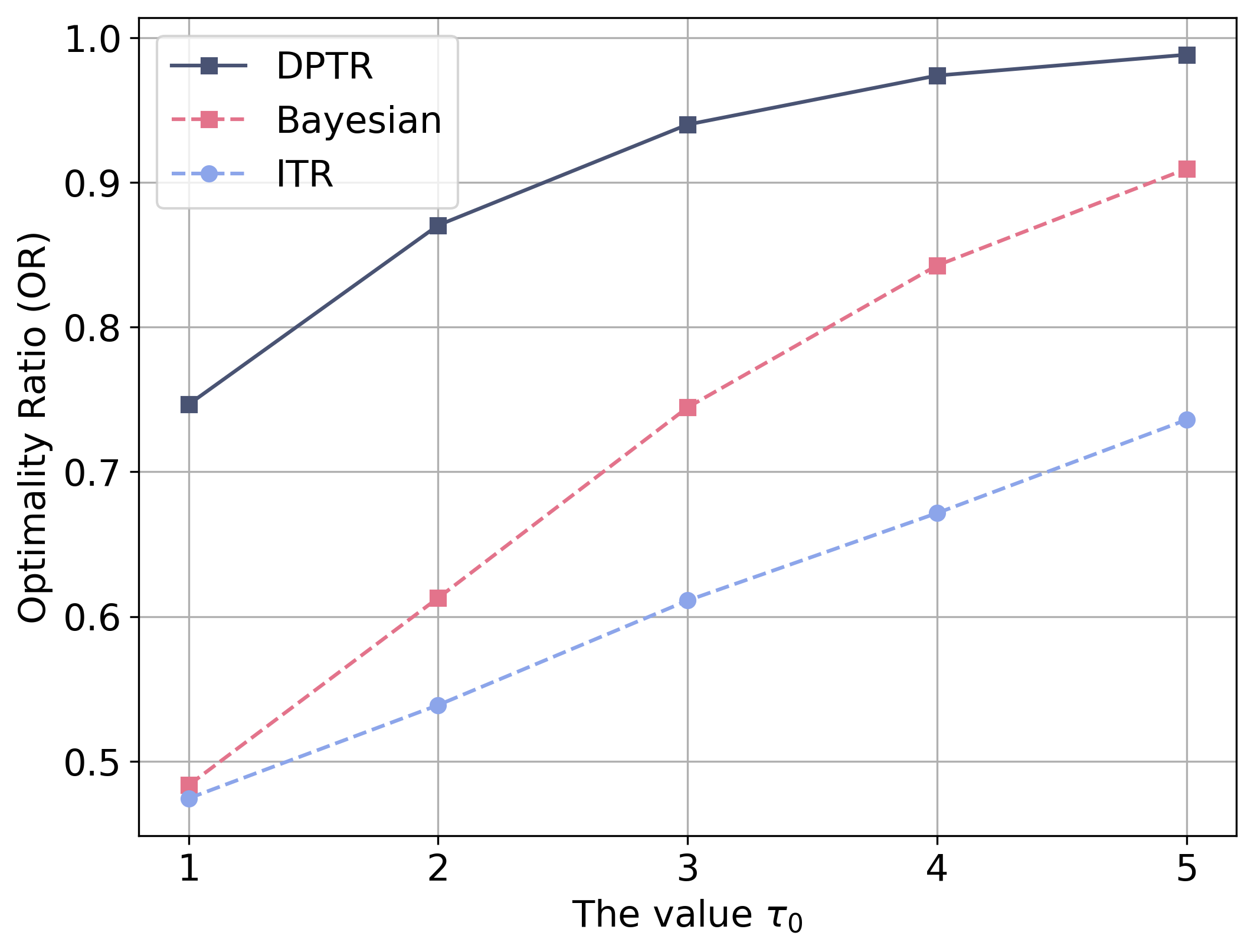}
\end{minipage}
}%
\subfigure[$K = 5$]{
\begin{minipage}[t]{0.5\linewidth}
\centering
\includegraphics[width=2.5in,height=1.8in]{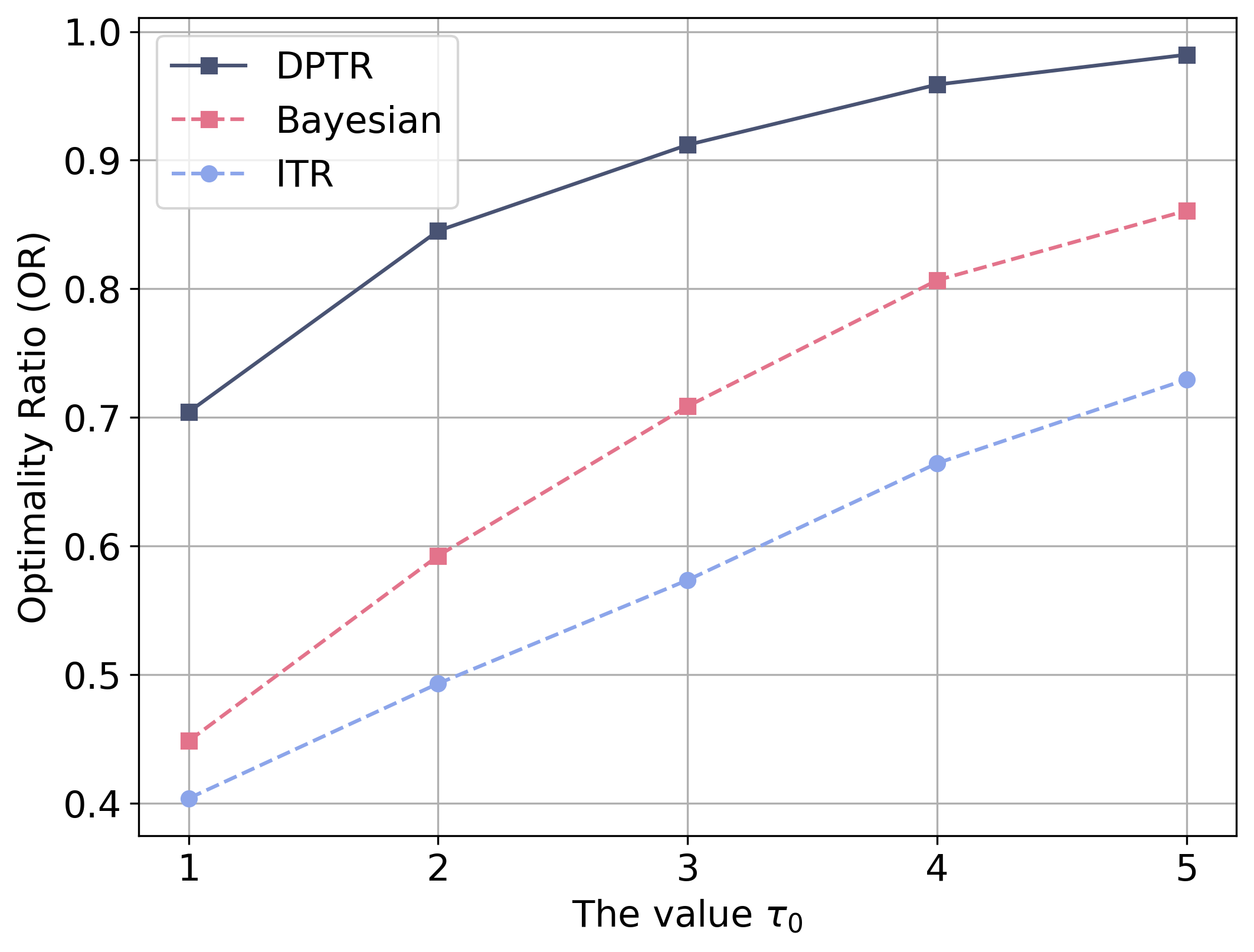}
\end{minipage}%
}%

\subfigure[$K = 100$]{
\begin{minipage}[t]{0.5\linewidth}
\centering
\includegraphics[width=2.5in,height=1.8in]{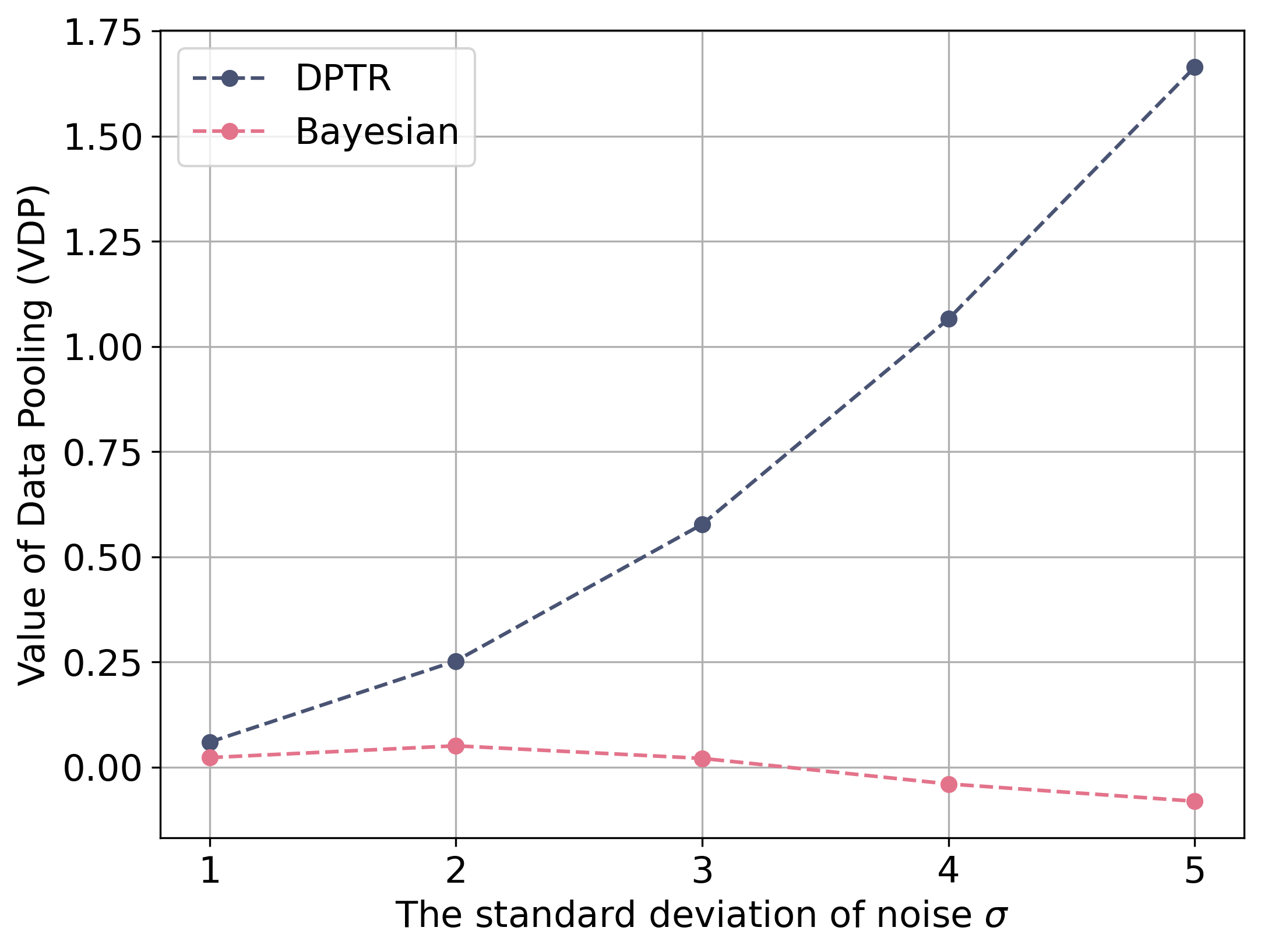}
\end{minipage}%
}%
\subfigure[$K = 5$]{
\begin{minipage}[t]{0.5\linewidth}
\centering
\includegraphics[width=2.5in,height=1.8in]{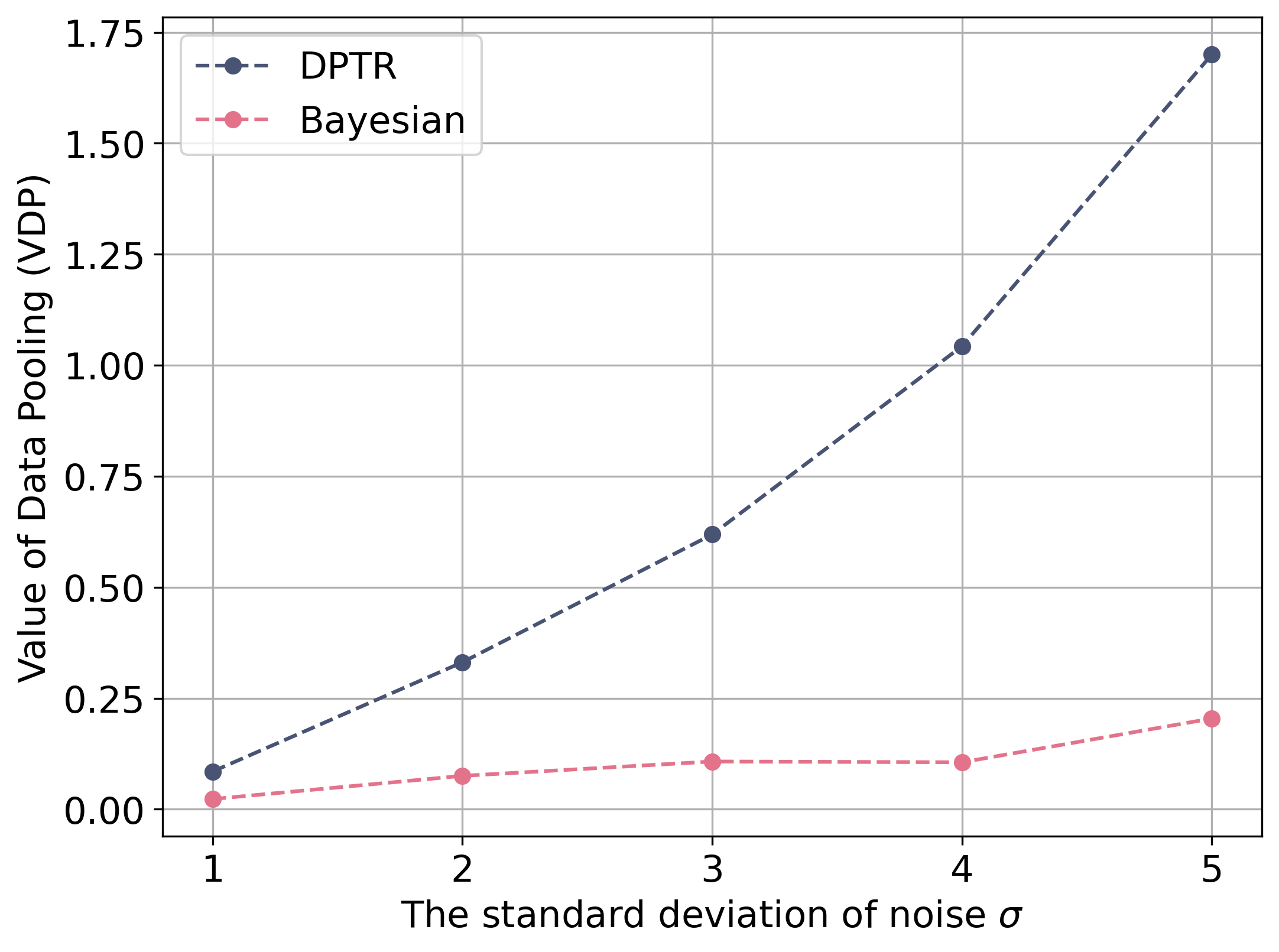}
\end{minipage}%
}%
\centering
\caption{Performance comparisons under overlapping experiments and linear specification}
\label{fig:2}
\vspace{-0.15 in}
\end{figure}

We compare \our, \tradition, and Bayesian methods and report OR and VDP in Figure~\ref{fig:2}. The findings are closely aligned with those reported in Section~\ref{subsec:no-overlap-no-feature}. The results show that our \our\ method remains competitive even when the number of experiments is small. In summary, the proposed \our\ method has a robust performance under linear specification even when users are treated by multiple experiments simultaneously.

Furthermore, we include additional comparisons between the \our\ and \tradition\ methods with model misspecifications. Suppose the underlying data-generating process is unknown, and the treatment effects are estimated using both OLS and DM approaches, where the latter represents the case of model misspecification. We then report the performance in Table \ref{tab:ols_dim}, allowing for a systematic comparison across estimation strategies and specification settings.

The results reveal a clear contrast between the two approaches. The \tradition\ method is highly sensitive to model misspecification: when the assumed model deviates from the true data-generating process, its achieved reward deteriorates sharply. This highlights its reliance on correct model specification and its vulnerability in practical settings where such assumptions are often difficult to verify. In contrast, the \our\ method exhibits strong stability and robustness. Its performance remains consistently high across different estimation methods and is only mildly affected by misspecification. Notably, even under model misspecification, \our\ often outperforms the \tradition\ method under correct model specification (with only one minor exception).

This robustness can be attributed to the design of \our, which reduces dependence on specific functional form assumptions and is therefore less prone to estimation bias arising from misspecification. Overall, these findings provide further empirical support for the reliability and broad applicability of the \our\ method, particularly in realistic environments where the true data-generating process is complex or unknown. 

\begin{table}[!ht]
\centering
		{\def\arraystretch{1.1}  
\centering
\begin{tabular*}{1\textwidth}
{@{\extracolsep{\fill}}ccccccccc}
\hline
\hline
 & \multicolumn{4}{c}{$K = 100$}  & \multicolumn{4}{c}{$K = 5$}  \\
\hline
 & \multicolumn{2}{c}{OLS}  &  \multicolumn{2}{c}{DM} & \multicolumn{2}{c}{OLS} & \multicolumn{2}{c}{DM}  \\
\hline
Anchor Value ($\tau_0$)  & \tradition\ & \our\ & \tradition\ & \our\  & \tradition\ & \our\ & \tradition\ & \our\ \\
\hline
$1$  &0.4743 & 0.7466 & 0 &0.7934 & 0.4037 & 0.7042 &0.0070 &0.5838  \\
$2$ &0.5388& 0.8704 & 0 & 0.8703 & 0.4930 & 0.8451 &0.0026 &0.6878 \\
$3$    &0.6111 & 0.9400& 0 &0.9253 & 0.5735 & 0.9119 &0.0026 &0.7091  \\
$4$ &0.6715 & 0.9739 & 0 & 0.9532 & 0.6643 & 0.9589 &0.0031&0.7117\\
$5$ &0.7360 & 0.9893 & 0 & 0.9341 & 0.7295 & 0.9821 &0.0016 &0.7146\\
\hline
\hline
\end{tabular*}
\caption{Performance comparison under OR: model misspecification vs. model specification}
\label{tab:ols_dim}
}
\end{table}

\subsection{Overlapping Experiments and Non-linear Specification}\label{subset: with-overlap-with-feature}

In this subsection, we test the performance of our method in the scenario introduced in Section~\ref{subsec:scenario-4}. We evaluate the performance of our method only when \( K \) is relatively small. The experimental setup is as follows: the platform runs $K = 5$ experiments, with $N = 100 + K$ observations. The noise \( \epsilon_{k,i} \) follows a normal distribution, \( \epsilon_{k,i} \sim \mathcal{N}(0, 3^2) \). We set the number of covariates as $d_x = 4$, the covariate distribution as $x_{k,i}^j \sim U(0,1)$, and the true response function as $g(X_{i}) = [\gamma_{0}^{\top}X_{i},\gamma_{1}^{\top}X_{i},\cdots,\gamma_{K}^{\top}X_{i}]^{\top}$, where the coefficients $\gamma_{0},\gamma_{1},\cdots,\gamma_K$ are randomly drawn from the distribution $U(-0.3,0.5)$.  The estimation is also based on double machine learning and cross-fitting. To estimate the nuisance parameter $g(\cdot)$, we adopt the two-layer fully connected neural network with $K + 10$ units per layer and ReLU activations, without dropout, to estimate $g(\cdot)$. Similar to Section \ref{subset: with-overlap-no-feature}, we also include additional comparisons which use DM approach. 

First, as reported in Table~\ref{tab:2}, under the DML method, \our\ consistently outperforms \tradition, demonstrating that our proposed data pooling technique effectively combines data from multiple experiments even under overlapping experiments and nonlinear specification. Secondly,  the performance ranking is approximately
“DML + \our\ \(>\) DM + \our\ \(>\) DM + \tradition\ \(>\) DML + \tradition”. This overall ordering indicates that our \our\ method consistently outperforms the \tradition\ method regardless of the underlying estimation approach, which aligns with the discussion in the previous section. A more nuanced pattern also emerges. Under the \tradition\ method, the DM estimator outperforms DML. In contrast, within our \our\ framework, the DML-based estimator achieves substantially better performance. This reversal suggests that \our\ effectively enhances DML by mitigating its variance, thereby unlocking its potential advantages. Moreover, as reported in Table \ref{tab:real_overlapping} using the real experimental dataset from \cite{ye2023deep} in Section \ref{subset:real_ovelapping}, the same pattern persists when the sample size \(N\) is relatively small, further supporting our conclusion.


\begin{table}[!ht]
\centering
		{\def\arraystretch{1.1}  
\centering
\begin{tabular*}{0.9\textwidth}
{@{\extracolsep{\fill}}ccccc}
\hline
\hline
 & \multicolumn{2}{c}{DML}  & \multicolumn{2}{c}{DM} \\
 \hline
Noise Variance ($\sigma^2$)  &  \tradition\  & \our\ &  \tradition\  & \our\ \\
\hline
$1^2$  &0.3590& 0.6897&0.4365& 0.7130  \\
$2^2$  & 0.1499 & 0.5257& 0.1718 & 0.4628 \\
$3^2$   &0.0992 & 0.5008 &0.0953 & 0.3610 \\
$4^2$  & 0.0650&0.4438& 0.0768&0.3396\\
$5^2$  & 0.0512& 0.4344& 0.0607& 0.3158\\
\hline
\hline
\end{tabular*}
\caption{The comparison of OR under different $\sigma^2$ under Scenario 4.}
\label{tab:2}
}
\end{table}


\section{Applications to Real-world A/B Tests}\label{sec:realdata}

In this section, we evaluate the performance of the \our\ method using the real-world A/B testing data, covering both non-overlapping and overlapping scenarios. Section \ref{subset:real_non_overlapping} reports experiment results based on a publicly available dataset from Criteo \citep{Diemert2018} to demonstrate the performance of our proposed method in the non-overlapping scenario. In Section \ref{subset:real_ovelapping}, we utilize the experimental data from \cite{ye2023deep}, which includes multiple experiments, to demonstrate the \our\ method's effectiveness in the overlapping scenario.

\subsection{Non-Overlapping A/B Testing}\label{subset:real_non_overlapping}


In this subsection, we will evaluate the performance of the \our\ method in a non-overlapping scenario. The dataset\footnote{The dataset can be accessed via the link \url{https://ailab.criteo.com/criteo-uplift-prediction-dataset/}.} used in this analysis originates from a randomized controlled trial (RCT), conducted by Criteo, an advertising platform, as part of a large-scale randomized ad-targeting campaign. In this RCT, a randomly selected portion of the population was deliberately excluded from being targeted by advertisements. This RCT was initially released to benchmark uplift modeling methods \citep{Diemert2018} with “visits” as the outcome of interest. The dataset comprises 13,979,592 rows, each representing a user characterized by 12 covariates, a treatment indicator for advertisement exposure, and a binary label indicating whether the user visited the advertised site. The treatment rate is 85\% and the average visit rate is 4.70\%.

Unlike the synthetic data setting explored in Section~\ref{sec:Synthetic_valid}, this dataset is from a single experiment and contains user covariate information. Consequently, our focus shifts to determining customized treatment rollouts, specifically, deciding whether personalized recommendations should be offered to users. 
To validate the proposed approach using this dataset, we first group all samples based on user covariates. Then, we estimate the full-sample benchmark HTEs across different groups using the full dataset. These full-sample estimates serve as a high-precision benchmark for evaluation; they are not known ground-truth causal effects. Next, we evaluate both \our\ and \tradition\ methods, using random samples from the dataset associated with each group. The detailed validation procedure is as follows:

\begin{enumerate}
    \item \underline{Group Generation}: We partition the dataset based on the medians of user covariates. Since the covariate information in this dataset is encrypted, we cannot group the data based on specific covariate values. Instead, we categorize each covariate into two groups based on its median value. As many covariates have a large number of values equal to the median, we randomly assign samples to ensure that both groups have equal sample sizes. Thus, we partition the entire dataset based on the combinations formed by the realizations of the 12 covariates. Since the features are not completely independent, the number of data points in the groups formed by splitting on the median of each feature is not necessarily the same. Thus, to ensure each group has enough data for reliable analysis, we exclude groups with fewer than 1,000 data entries. This results in 1,744 groups, that is, \( K = 1,744 \). 
    \item \underline{Full-Sample Benchmark HTE Computation}: For group $k$, let \( D_{k,i} \in \{0,1\} \) represent the treatment variable for subject \( i \), where \( D_{k,i} = 1 \) indicates the implementation of the personalized recommendation, and \( D_{k,i} = 0 \) indicates no implementation. Let \( Y_{k,i} \) denote the outcome, indicating whether the user visits the recommended advertisement. The full-sample benchmark HTE for group \( k \) is then given by: $\tau_k = \frac{1}{N_{k,1}} \sum_{D_{k,i} = 1} Y_{k,i} - \frac{1}{N_{k,0}} \sum_{D_{k,i} = 0} Y_{k,i}$ where \( N_{k,1} \) and \( N_{k,0} \) are the total number of subjects who experienced and did not experience the treatment, respectively.
    The histogram of the full-sample benchmark HTEs across all groups, estimated from the entire dataset, is shown in Figure~\ref{fig:criteo_true_ate}.
    \begin{figure}[htbp!]
    \vspace{-0.15 in}
    \centering
     \includegraphics[width=0.4\linewidth]{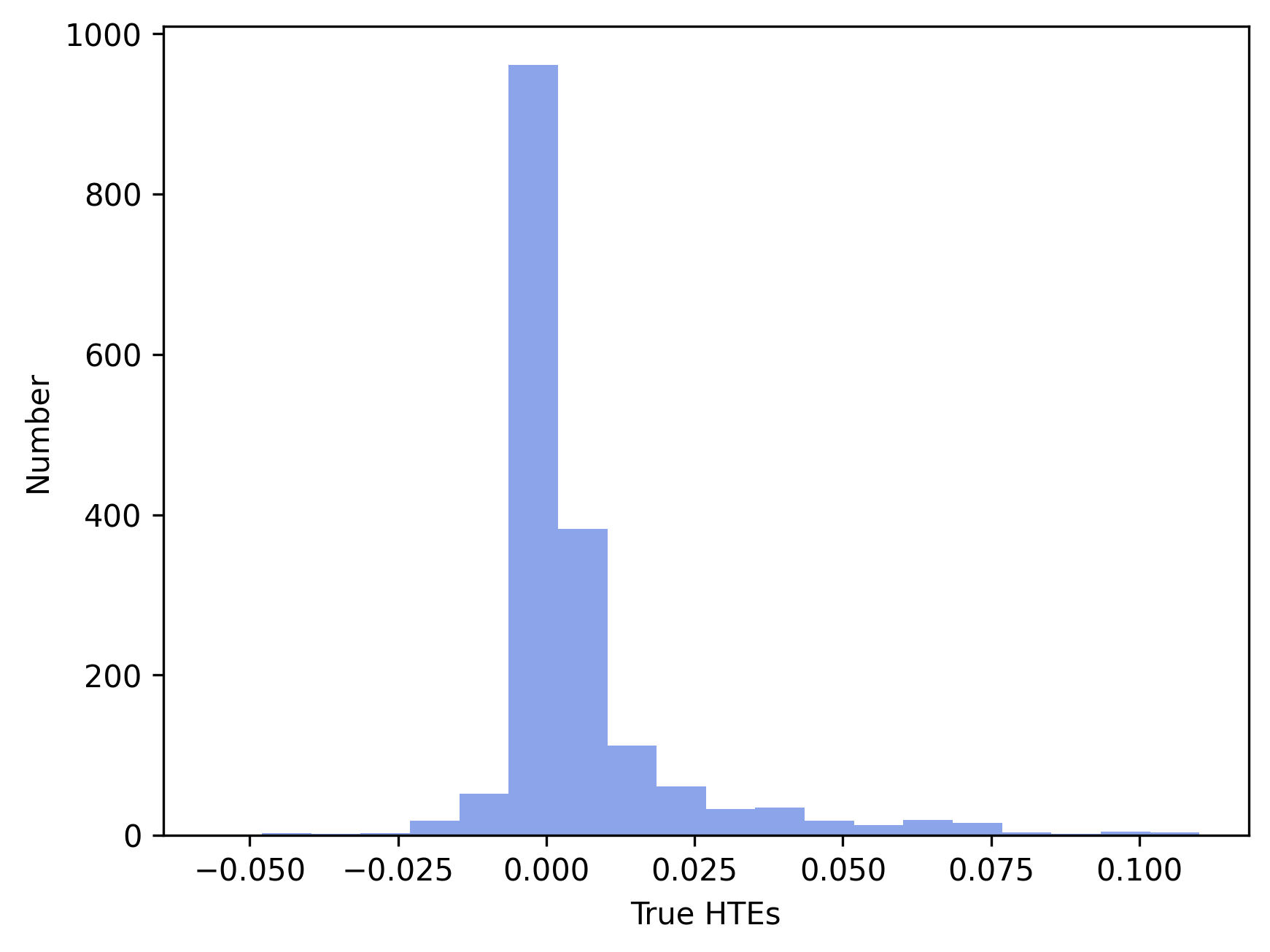}
    \caption{Histogram of full-sample benchmark HTEs for all groups for the Criteo dataset.}
    \label{fig:criteo_true_ate}
    \vspace{-0.15 in}
\end{figure}
\item \underline{Random Sampling and Evaluation}: For each group, we randomly select \( N \) users, with \( N/2 \) users drawn from the treatment group $( D_{k,i} = 1 )$ and \( N/2 \) users drawn from the control group $( D_{k,i} = 0 )$ for every $k$. Using this randomly sampled sub-dataset, we apply both the \tradition\ and \our\ methods (Algorithms \ref{alg:general_dm} and \ref{alg:general_dp}, respectively) to generate roll-out decisions. Here, we select the difference-in-means method as $\mathcal{M}(\cdot)$ and set the significance level $\alpha = 0.05$. Finally, we can compare their performance by measuring different metrics. 
\end{enumerate}

\begin{figure}[!ht]
\vspace{-0.15 in}
\centering
\subfigure[OR]{
\begin{minipage}[t]{0.3\linewidth}
\centering
\includegraphics[width=2.0in,height=1.5in]{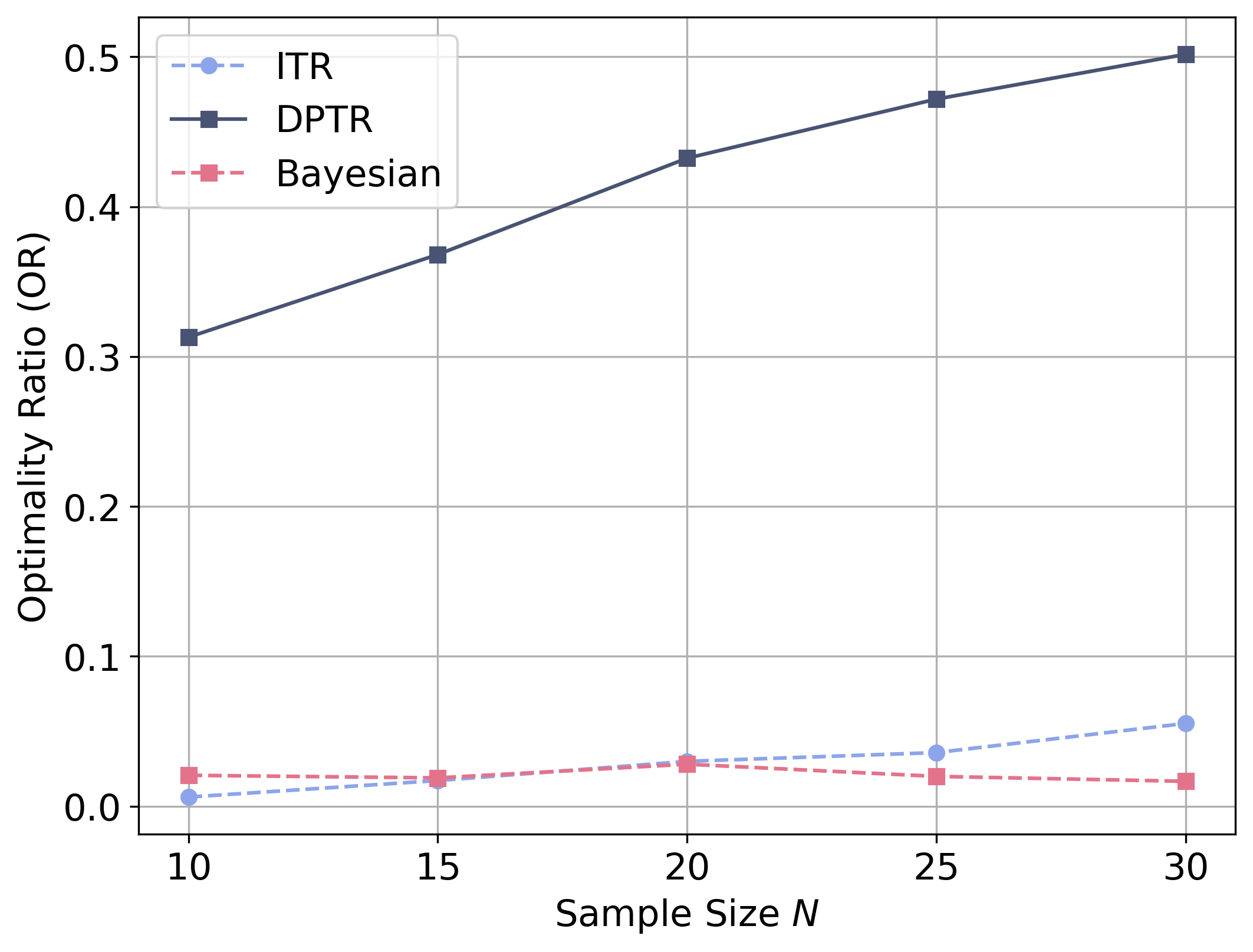}
\end{minipage}
}%
\subfigure[VDP]{
\begin{minipage}[t]{0.3\linewidth}
\centering
\includegraphics[width=2.0in,height=1.5in]{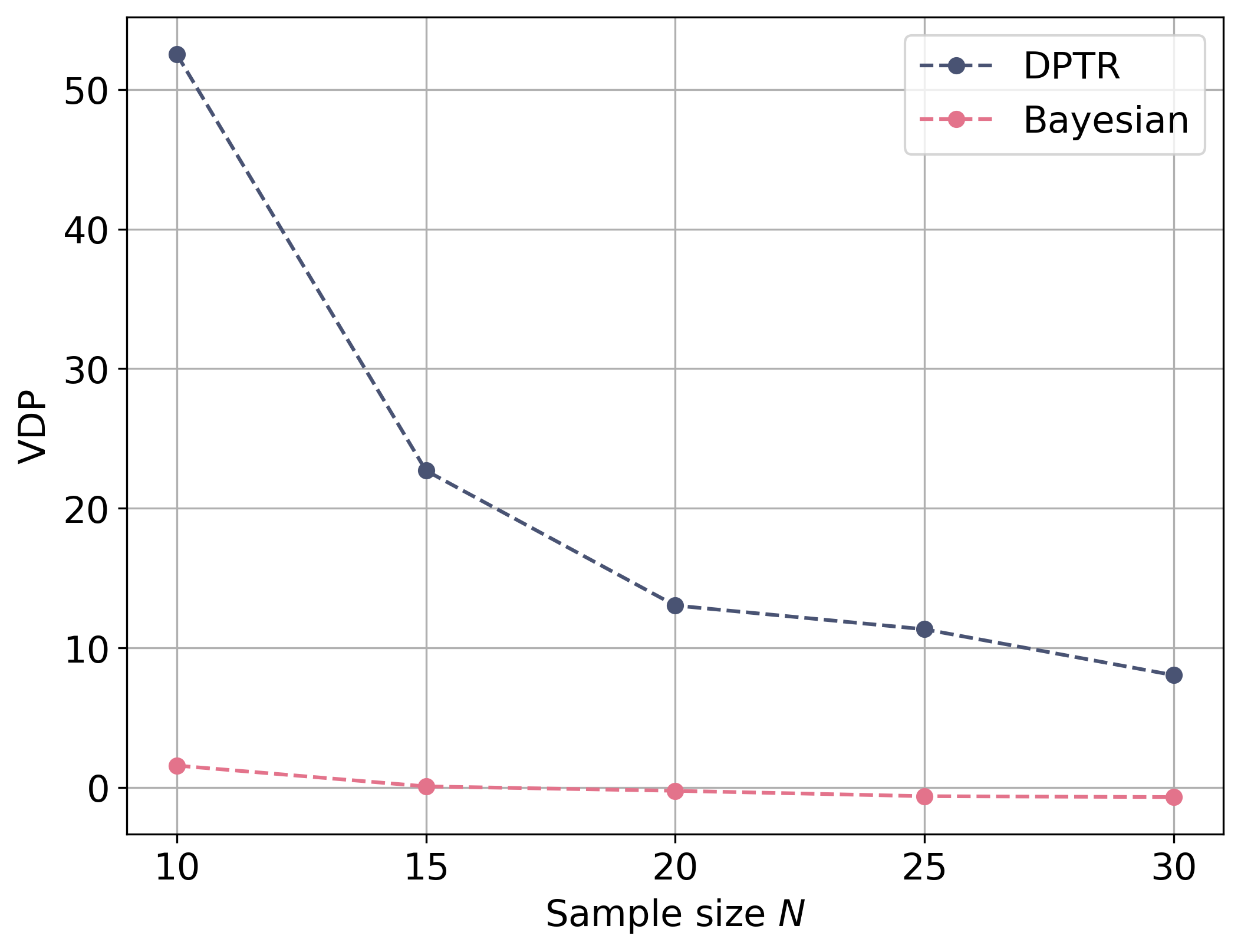}
\end{minipage}%
}%
\subfigure[Recall and Specificity]{
\begin{minipage}[t]{0.33\linewidth}
\centering
\includegraphics[width=2.0in,height=1.5in]{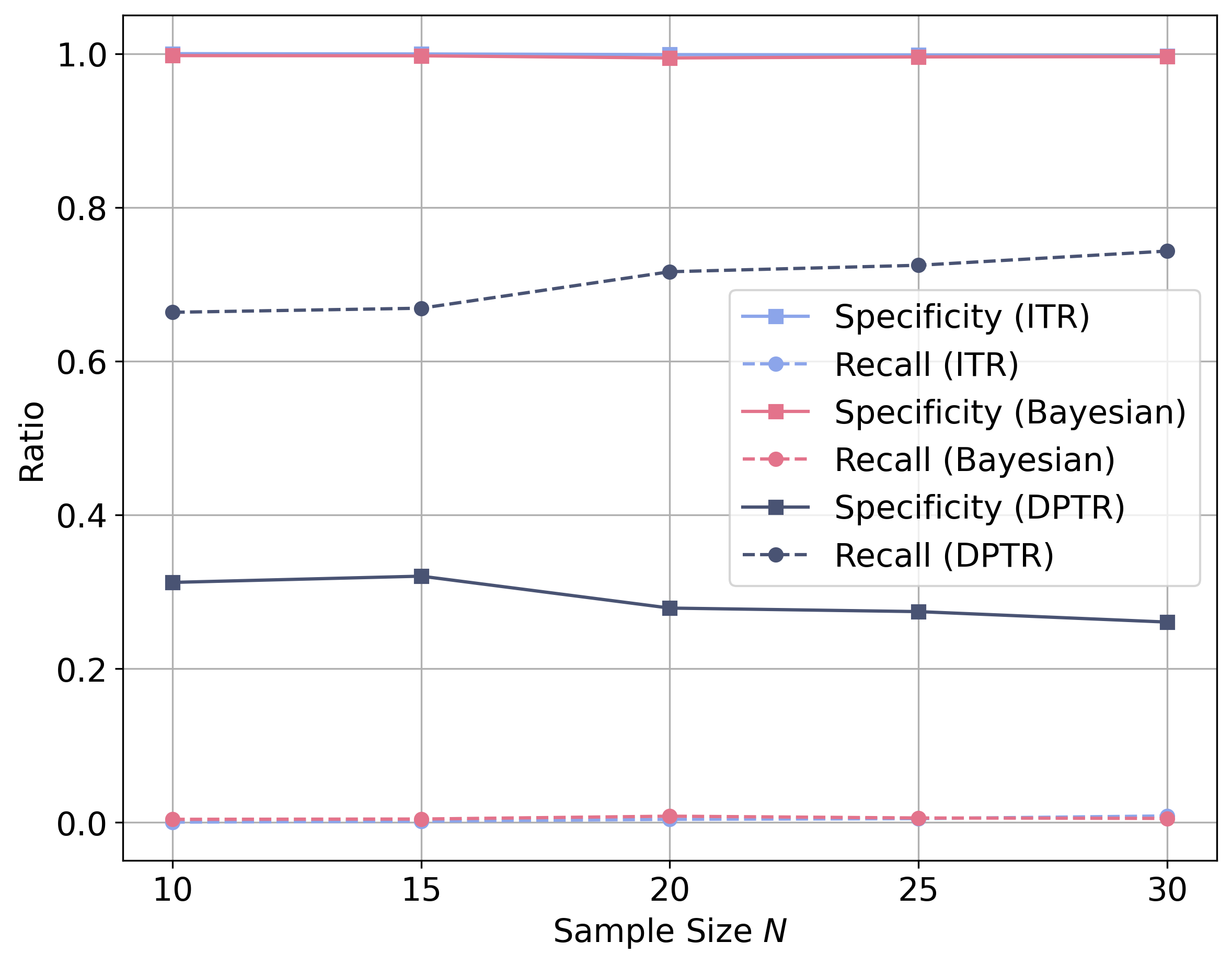}
\end{minipage}%
}%
\centering
\caption{Performance comparisons in different sample size $N$ in Criteo dataset.}
\label{fig:results_criteo}
\vspace{-0.15 in}
\end{figure}

Since the sample sizes vary across groups, we normalize \(\tau_k \) when calculating the OR and VDP values by multiplying it with the normalizing factor \(\frac{N_{k,0}+N_{k,1}}{\sum_{k}(N_{k,0}+N_{k,1})}\), which represents the proportion of group \(k\)'s data size relative to the total data size of all groups. We vary the sample size $N$ from 10 to 30 in increments of 5 and repeat the experiment 1,000 times. The averaged results are shown in Figure~\ref{fig:results_criteo}.

According to Figure~\ref{fig:results_criteo}(a), the \our\ method consistently outperforms the \tradition\ method, irrespective of the sample size $N$. Additionally, as the sample size increases, the performance difference between \our\ and \tradition\ becomes larger. However, 
the Bayesian method performs almost on par with the \tradition\ method. The significant performance difference between \our\ method and the Bayesian method hinges on how the shrinkage parameter $\beta$ is constructed: while the one for \our\ requires the average sample variance across experiments (Eqn.~\ref{eq:data-driven-beta-S1}), the one for the Bayesian method utilizes the individual-level sample variance (Eqn.~\ref{eq:data_driven_baye}). As reported in Figure \ref{fig:criteo_true_ate}, most full-sample benchmark HTEs are centered around zero, which may benefit most from shrinkage, yet their sample variances tend to be relatively small. Thus, the associated shrinkage parameters $\hat{\beta}_k^{\text{bayes}}$ are closer to zero, resulting in a similar performance of the \tradition\ method. In contrast, \our\ leverages the average sample variance, which may significantly deviate from zero. This allows the shrinkage parameter to more effectively guide decision-making and achieve superior OR values. In sum, these results further highlight that our \our\ method is more decision-aware and robust than the Bayesian method.


Figure~\ref{fig:results_criteo}(b) further reveals that the relative performance advantage of the \our\ method over the \tradition\ method increases significantly as the amount of experimental data decreases. This finding highlights that, when experimental data is limited, the \our\ method provides substantial performance improvements.


Figure~\ref{fig:results_criteo}(c) shows that when $N$ is small, the Recall value under the \tradition\ method and Bayesian method is closer to 0, indicating that few or no personalized recommendations are rolled out to specific groups, even when these groups may be associated with positive ATEs. Because few groups are treated with the personalized recommendation, the Specificity value is also higher under the \tradition\ and Bayesian methods. On the other hand, despite large variation in the estimated ATEs across groups, the \our\ method mitigates this issue by pooling the estimates, which reduces variance and shifts the estimate towards a more positive range. As a result, the Recall value under the \our\ method increases significantly. This comes at the cost of mistakenly selecting some groups for the personalized recommendation. However, by balancing Recall and Specificity, the \our\ method generates a much higher reward compared to the \tradition\ and Bayesian methods when the personalized roll-out decisions are involved.
Finally, we remark that additional evaluation of the \our\ method using Expedia’s experimental data in a similar non-overlapping setting is provided in Appendix \ref{app:real_expedia_data}.

\subsection{Overlapping A/B Tests}\label{subset:real_ovelapping}

In this subsection, we will use the 
experimental data from \cite{ye2023deep} to demonstrate the performance of our \our\ method in the overlapping scenario. The dataset was collected from a large-scale online short-video-sharing platform, which serves hundreds of millions of users globally each day. 

The dataset comprises a unique set of three A/B tests or treatments, each of which examines the effect of a major adjustment to the video recommendation algorithm on one of three main pages of the online platform: (i) the Discover Page (DP), (ii) the Live Page (LP), and (iii) the For You Page (FYP). As with most A/B tests conducted on online platforms, the primary objective is to enhance user engagement, which is well approximated by the amount of screen time a user spends on the platform each day. Each experiment is randomized using a distinct hash function of user IDs, ensuring that the treatment assignment mechanisms across experiments are mutually independent. \cite{ye2023deep} have demonstrated the presence of interaction effects across experiments in this dataset. As a result, the three experiments yield eight distinct treatment combinations. Using stratified sampling, \cite{ye2023deep} construct a new dataset with approximately 258,325 users in each treatment combination. We refer to the total data as population data. A detailed description of the dataset and experiments can be found in \cite{ye2023deep}.

First, using the full population data, we compute the full-sample benchmark ATEs for all treatment combinations. The relative ATE values for the eight treatment combinations are reported in Table 2 of \cite{ye2023deep}. For combinations with statistically insignificant treatment-combination effects, we follow a zero-imputation convention and set the corresponding ATE to zero. Because lack of statistical significance is not evidence of a zero effect, this convention should be viewed as a working benchmark rather than ground truth. Next, we randomly sample $N$ users for each treatment combination from the corresponding population data to construct a new dataset, which we refer to as the historical experimental data. We then apply both the \our\ and \tradition\ methods to this dataset to make roll-out decisions, and evaluate their performance using OR values calculated based on the full-sample benchmark ATEs. This sampling procedure is repeated 1,000 times, and we report the average OR values across these iterations.

To more robustly demonstrate the advantage of the \our\ method over the \tradition\ method, we vary $N$ across the values 1000, 5000, 10000, and 20000. For the method $\mathcal{M}$, we consider the following four methods: (i) DM, (ii) OLS without covariate information, (iii) OLS with covariate information, and (iv) DML. Next, we provide a detailed description of the implementation procedures for each method as follows:
\begin{enumerate}
    \item [(i)] \textbf{DM}: For this method, we focus on a non-overlapping scenario without covariate information. Accordingly, we use only the data corresponding to the three treatment combinations: $(1,0,0)$, $(0,1,0)$, and $(0,0,1)$. The model assumed for this method follows the form of Eqn. \eqref{eq:simple ols}.

    \item [(ii)] \textbf{OLS without covariate information}: For this method, we consider an overlapping scenario that utilizes data from all treatment combinations. However, covariate information is not included, and the assumed model follows the form of Eqn.~\eqref{eq:DGP}.
    \item [(iii)] \textbf{OLS with covariate information}: For this method, we consider an overlapping scenario and incorporate covariate information. The assumed model follows the form of Eqn.~\eqref{eq:DGP_OLS_feature}.

    \item [(iv)] \textbf{DML }: For this method, we consider an overlapping scenario and use the DML approach to estimate the ATEs. We assume that the policy ATEs are linearly additive, and the model follows the form of Eqn.~\eqref{eq:DML_DGP_overlapping}.
\end{enumerate}

The averaged OR values of 1,000 instances for different methods $\mathcal{M}$ and sample size $N$  are reported in the Table \ref{tab:real_overlapping}. First, we observe that our \our\ method consistently outperforms the \tradition\ method with different methods and sample sizes. This confirms the effectiveness of our method for prescribing roll-out decisions using real-world datasets. Second, for any given method $\mathcal{M}$, increasing the sample size $N$ improves the performance for both the \our\ and \tradition\ methods. This is consistent with our intuition: a larger sample size typically results in a smaller variance. Furthermore, for a fixed sample size, when adopting \tradition\ method, we may observe that ``OLS without covariate" outperforms ``OLS with covariate", which, in turn, results in higher ORs than DML. This may imply the potential risk of model misspecification. More importantly, regardless of whether model misspecification exists, \our\ method outperforms the \tradition\ method, demonstrating its robustness, especially when model misspecification may arise in analyzing real-world datasets.

\begin{table}[!ht]
\vspace{-0.1in}
\centering
		{\def\arraystretch{1.1}  
\centering
\begin{tabular*}{1\textwidth}
{@{\extracolsep{\fill}}cccccc}
\hline
\hline
Method ($\mathcal{M}$) & Roll-out method  &  $1,000$  & $5,000$ & $10,000$ & $20,000$ \\
\hline
\multirow{2}{*}{DM} 
&\tradition  &0.0207  &0.0408 & 0.0584 &0.0732 \\
&\our  &0.1610 &0.2482 & 0.2780 & 0.3080 \\
\hline
\multirow{2}{*}{OLS without covariate} 
& \tradition  & 0.0568  & 0.1070 & 0.2281 & 0.3564 \\
 & \our  & 0.2878  & 0.4213 & 0.5257 & 0.6926\\
 \hline
\multirow{2}{*}{OLS with covariate}    
&\tradition &0.0161  & 0.0853 & 0.1499 &0.2667 \\
&\our &0.4342  & 0.5213 & 0.5437 & 0.6060 \\
\hline
\multirow{2}{*}{DML }
&\tradition & 0.0080  & 0.0398 & 0.1058 & 0.2145\\
&\our & 0.3654  & 0.4208 & 0.4845 & 0.5474\\
\hline
\hline
\end{tabular*}
\caption{The comparison of OR under different methods $\mathcal{M}$ and sample size $N$.}
\label{tab:real_overlapping}
}
\vspace{-0.2in}
\end{table}


\section{Conclusion}\label{sec:Conclusion}
In conclusion, we introduce the Data-Pooling Treatment Roll-Out (\our) framework to improve decision-making in online experiments by aggregating data across multiple experiments. The framework balances variance reduction against shrinkage-induced bias and aligns the scale of pooling with the downstream roll-out decision. We provide formal theoretical guarantees for non-overlapping experiments under linear specifications and complement them with synthetic experiments and real-world applications covering heterogeneous treatment effects, overlapping experiments, and nonlinear specifications. Across these evaluations, \our\ is especially useful when per-experiment sample sizes are small and many related experiments are available. This study highlights the value of data pooling for policy roll-outs, while also clarifying the settings, such as tight capacity constraints or strong sign-reversing interactions, in which the method may not offer an advantage.

In this paper, we focus on the case of binary treatments and propose a method that pools data across multiple experiments to improve the per-experiment reward by shrinking the estimators derived from individual datasets. Although our primary analysis is centered on binary treatments, the underlying idea is broadly applicable. For example, it can be extended to settings with continuous treatments, such as pricing interventions in A/B testing, where the goal is to estimate dose-response relationships \citep{Zhang2025}. Moreover, the approach can be applied to observational (non-experimental) data, where confounding may arise, offering a principled way to stabilize estimates in the presence of such complexities \citep{jiang2025instrumenting, chitla2025nonparametric}. Finally, extensions to optimization problems with uncertain objectives are promising, but settings with binding capacity constraints require additional analysis because the unconstrained roll-out rule studied here need not be optimal in such environments \citep[][]{Natarajan2011}. 



%
%
%




\bibliographystyle{informs2014} 
\bibliography{reference} 





  



\ECSwitch
\begin{APPENDIX}{}
\section{Proofs}
\proof{\bf \noindent Proof of Theorem \ref{theorem:AP}.}
    First of all, because true ATE $\tau_k$ follows the normal distribution $\mathcal{N}(\tau_0,\sigma_0^2)$, $\mathcal{R}(\beta,\tau_0)$ can be written as:
    \begin{align*}
        \mathcal{R}(\beta,\tau_0) &= \lim_{K \to \infty}\frac{1}{K}\sum_{k \in [K]}\tau_k \mathbb{P}\Big( \bar\tau_k>\frac{N}{N+\beta}\frac{2\sigma z_{1 - \alpha/2}}{\sqrt{N}}\Big) \\
        &=\int_{-\infty}^{+\infty}\frac{1}{\sqrt{2\pi}\sigma_0}e^{-\frac{(\tau_k - \tau_0)^2}{2\sigma_0^2}}\Phi\Big(\frac{N\tau_k + \tau_0\beta}{2\sqrt{N} \sigma} -  z_{1 - \alpha/2}\Big)\tau_k d(\tau_k).
    \end{align*}
We take the derivative of $\mathcal{R}(\beta,\tau_0)$ with respect to $\beta$ as follows:
\begin{align*}
    \frac{d(\mathcal{R}(\beta,\tau_0))}{d(\beta)} = \frac{\exp\Big(
-\frac{\left(N \tau_0 + \beta \tau_0 - 2 \sqrt{N} z_{1 - \alpha/2} \sigma \right)^2}{2 N \left(N \sigma_0^2 + 4 \sigma^2\right)} 
\Big)\tau_0 \left(
-\beta \tau_0 \sigma_0^2 + 2 \sigma \big(\sqrt{N} z_{1 - \alpha/2} \sigma_0^2 + 2 \tau_0 \sigma\big)
\right)} {
\sqrt{N} \sqrt{2 \pi} \sigma_0 \sqrt{\frac{4}{\sigma_0^2} + \frac{N}{\sigma^2}} \sigma \left(N \sigma_0^2 + 4 \sigma^2\right)
}.
\end{align*}
The derivative factors as
\begin{equation*}
    \frac{d\mathcal{R}(\beta,\tau_0)}{d\beta}
    \;=\;
    \frac{\phi(H)\,\tau_0}{2\sigma\sqrt{N}\,\nu}\cdot
    \frac{\tau_0\bigl(4\sigma^2-\sigma_0^2\beta\bigr)+2\sigma\sigma_0^2\sqrt{N}\,z_{1-\alpha/2}}{2\sigma\nu^2},
\end{equation*}
where $\nu^2:=1+N\sigma_0^2/(4\sigma^2)$ and $\phi(H)\in\mathbb{R}^+$ depends smoothly on $\beta$. Setting the derivative to zero yields the unconstrained critical point
\begin{equation*}
    \beta_{\mathrm{int}} \;=\; \frac{4\sigma^2}{\sigma_0^2} + \frac{2\sqrt{N}\,z_{1-\alpha/2}\,\sigma}{\tau_0}.
\end{equation*}
We analyze three cases.

\textbf{Case 1 ($\tau_0>0$).} The numerator in the second part of the derivative equals $4\sigma^2\tau_0+2\sigma\sigma_0^2\sqrt{N}\,z_{1-\alpha/2}>0$ at $\beta=0$, with a slope $-\tau_0\sigma_0^2<0$, and vanishes at $\beta_{\mathrm{int}}>0$; combined with the strictly positive multiplicative factor, the derivative is strictly positive on $[0,\beta_{\mathrm{int}})$ and strictly negative on $(\beta_{\mathrm{int}},\infty)$. Hence $\beta^*=\beta_{\mathrm{int}}$ is the unique interior maximizer and $\mathcal{R}(\beta^*,\tau_0)>\mathcal{R}(0,\tau_0)$.

\textbf{Case 2 ($\tau_0=0$).} The selection event $\bar\tau_k>(N/(N+\beta))(2\sigma z_{1-\alpha/2}/\sqrt{N})$ reduces to $(N/(N+\beta))\hat\tau_k>(N/(N+\beta))(2\sigma z_{1-\alpha/2}/\sqrt{N})$, in which the strictly positive factor $N/(N+\beta)$ cancels. The roll-out rule therefore coincides with that of $\beta=0$ for every $\beta\ge 0$. Thus, we have, $\mathcal{R}(\beta,0)=\mathcal{R}(0,0)$. We adopt $\beta^*=0$ as the canonical optimum.

\textbf{Case 3 ($\tau_0<0$).} The feasibility constraint $\beta_{\mathrm{int}}\ge 0$ requires $|\tau_0|\ge \sqrt{N}\,z_{1-\alpha/2}\,\sigma_0^2/(2\sigma)$, equivalently $\tau_0\le-\sqrt{N}\,z_{1-\alpha/2}\,\sigma_0^2/(2\sigma)$. When this holds, the numerator in the second part of the derivative is non-positive at $\beta=0$, with a positive slope, and vanishes at $\beta_{\mathrm{int}}\ge 0$; combined with the strictly negative multiplicative factor, the derivative is strictly positive on $[0,\beta_{\mathrm{int}})$ and strictly negative on $(\beta_{\mathrm{int}},\infty)$, so $\beta^*=\beta_{\mathrm{int}}$ and $\mathcal{R}(\beta^*,\tau_0)>\mathcal{R}(0,\tau_0)$. Instead, when  $-\sqrt{N}\,z_{1-\alpha/2}\,\sigma_0^2/(2\sigma)<\tau_0<0$, the bracket is strictly positive on $[0,\infty)$ and the multiplicative factor is strictly negative. Thus, the derivative is strictly negative, consequently, the constrained maximizer is $\beta^*=0$ and $\mathcal{R}(\beta^*,\tau_0)=\mathcal{R}(0,\tau_0)$.

Combining the three cases yields
\begin{equation}\label{eq:beta_star_proof}
    \beta^* = \begin{cases}
 \dfrac{4\sigma^2}{\sigma_0^2} + \dfrac{2\sqrt{N}\,z_{1 - \alpha/2}\,\sigma}{\tau_0}, & \tau_0 > 0\ \text{or}\ \tau_0 < -\dfrac{\sqrt{N}\,z_{1-\alpha/2}\,\sigma_0^2}{2\sigma}, \\[6pt]
0, & -\dfrac{\sqrt{N}\,z_{1-\alpha/2}\,\sigma_0^2}{2\sigma} \le \tau_0 < 0,
\end{cases}
\end{equation}
with the strict-inequality clause $\mathcal{R}(\beta^*,\tau_0)>\mathcal{R}(0,\tau_0)$ active in Cases~1 and~3 (interior regimes) only.
This concludes the proof.
\QED
\\

\proof{\bf \noindent Proof of Theorem \ref{theorem:mse_optimal}.}
We first rewrite $\text{MSE}(\beta,\tau_0)$ as follows:
\begin{align*}
    \text{MSE}(\beta,\tau_0) &= \lim_{K \to +\infty} \frac{1}{K}\sum_{k\in [K]}\mathbb{E}\Big[\Big(\frac{N}{N + \beta}\hat{\tau}_k + \frac{\beta}{N + \beta}\tau_0 - \tau_k\Big)^2 \Big] \\
    &=  \lim_{K \to +\infty} \frac{1}{K}\sum_{k\in [K]}\Big[\Big(\frac{N}{N+\beta}\Big)^2\mathbb{E}[(\hat{\tau}_k - \tau_k)^2] + \Big(\frac{\beta}{N + \beta}\Big)^2(\tau_0 - \tau_k)^2 \Big] \\
    & =  \lim_{K \to +\infty} \frac{1}{K}\sum_{k\in [K]}\Big[\Big(\frac{N}{N+\beta}\Big)^2 \frac{4\sigma^2}{N} + \Big(\frac{\beta}{N + \beta}\Big)^2(\tau_0 - \tau_k)^2 \Big] \\
    & = \Big(\frac{N}{N+\beta}\Big)^2 \frac{4\sigma^2}{N} + \Big(\frac{\beta}{N + \beta}\Big)^2 \lim_{K \to +\infty} \frac{1}{K}\sum_{k\in [K]}(\tau_0 - \tau_k)^2 \\
    &= \Big(\frac{N}{N+\beta}\Big)^2 \frac{4\sigma^2}{N} + \Big(\frac{\beta}{N + \beta}\Big)^2 \sigma_0^2.
\end{align*}
Applying the first-order condition with respect to $\beta$ yields
$\beta_{\mathrm{MSE}}^* = \frac{4\sigma^2}{\sigma_0^2}$,
which completes the proof.
\QED
\\

\proof{\bf \noindent Proof of Theorem \ref{theorem:data_driven_para}.} We have, by definition, $\hat{\tau}_0 = \frac{1}{K}\sum_k \hat\tau_k$ and first prove that $\hat{\tau}_0 \to_p \tau_0$. Specifically, we have,
    \begin{align*}
        \Big|\frac{1}{K}\sum_k \hat\tau_k - \tau_0 \Big| \le \Big|\frac{1}{K}\sum_k \hat\tau_k - \frac{1}{K}\sum_k \tau_k\Big| + \Big|\frac{1}{K}\sum_k \tau_k - \tau_0\Big|.
    \end{align*}
Since $\hat\tau_k-\tau_k \sim \mathcal{N}(0,4\sigma^2/N)$, the estimation error is sub-Gaussian. Thus, by the concentration theorem of sub-Gaussian random variables (Proposition 2.5 in \cite{Wainwright2019}), for any $t > 0$, we have,
\begin{align*}
    \mathbb{P}\Big(\Big|\frac{1}{K}\sum_k \hat\tau_k - \frac{1}{K}\sum_k \tau_k\Big|\ge t\Big) \le 2\exp\big(-KNt^2/(8\sigma^2)\big).
\end{align*}
Consequently, we have, $\Big|\frac{1}{K}\sum_k \hat\tau_k - \frac{1}{K}\sum_k \tau_k\Big| \to_p 0$ when $K \to \infty$.

In addition, we have, $\tau_k \sim \mathcal{N}(\tau_0,\sigma_0^2)$, by the law of large numbers, we have $\big|\frac{1}{K}\sum_k \tau_k - \tau_0\big| \to_p 0$. Combining the above results, we can conclude that $\big|\frac{1}{K}\sum_k \hat\tau_k - \tau_0 \big| \to_p 0 $, and equivalently, $\hat\tau_0 \to_p \tau_0$. Then, we can conclude that $\hat\tau_0 \to_p \tau_0$.

In order to prove $\hat\beta^* \to_p \beta^*$, we first prove the following convergence results:
\begin{align*}
    &\frac{1}{K}\sum_{k}s_k^2 \to_p \sigma^2,\\
    &\frac{1}{K}\sum_{k}(\hat{\tau}_k - \hat{\tau}_0)^2 \to_p \sigma_0^2 + \frac{4\sigma^2}{N}.
\end{align*}

We will first prove the convergence of variance for the treatment group as follows:
\begin{equation} \label{eq:variance_convergence}
    \frac{1}{K}\sum_{k}\frac{1}{\frac{N}{2}-1}\sum_{D_{k,i} = 1}\Big(Y_{k,i} - \frac{2}{N}\sum_{D_{k,i} = 1}Y_{k,i}\Big)^2 \to_p \sigma^2.
\end{equation}
Because the error term $\epsilon_k$ follows the normal distribution, $Y_{k,i} - \frac{2}{N}\sum_{D_{k,i} = j}Y_{k,i}$ also follows the normal distribution which makes that $(Y_{k,i} - \frac{2}{N}\sum_{D_{k,i} = j}Y_{k,i})^2$ is a sub-exponential random variable according to Lemma 5.14 of \cite{Vershynin2010}. Combining the fact that $\mathbb{E}\big[\frac{1}{\frac{N}{2}-1}\sum_{D_{k,i} = 1}(Y_{k,i} - \frac{2}{N}\sum_{D_{k,i} = j}Y_{k,i})^2\big] = \sigma^2$ and the independence across all experiments, by applying the concentration theorem of sub-exponential random variables (Proposition 2.9 in \cite{Wainwright2019}), we can prove that Eqn.~\eqref{eq:variance_convergence} holds. 
The analysis in the control group is the same as the above one for the treatment group. Thus, we have,
\begin{equation} \label{eq:numerator}
    \frac{1}{K}\sum_{k}s_k^2 = \frac{1}{N-2}\frac{1}{K}\sum_{k}\sum_{j \in \{0,1\}}\sum_{D_{k,i} = j}\Big(Y_{k,i} - \frac{2}{N}\sum_{D_{k,i} = j}Y_{k,i}\Big)^2 \to_p \sigma^2.
\end{equation}
For the term $\frac{1}{K}\sum_{k}(\hat{\tau}_k - \hat{\tau}_0)^2$, we can decompose it as follows:
\begin{align*}
    \frac{1}{K}\sum_{k}(\hat{\tau}_k - \hat{\tau}_0)^2 &= \frac{1}{K}\sum_{k}(\hat{\tau}_k - \tau_k)^2 + \frac{1}{K}\sum_{k}(\tau_k - \hat\tau_0)^2 + \frac{1}{K}\sum_{k}2(\hat{\tau}_k - \tau_k)(\tau_k - \hat\tau_0) \\
    &= \frac{1}{K}\sum_{k}(\hat{\tau}_k - \tau_k)^2 + \frac{1}{K}\sum_{k}(\tau_k - \tau_0)^2 + \frac{1}{K}\sum_{k}2(\tau_k - \tau_0)(\tau_0 - \hat\tau_0) \\
    &~~~~~~~+(\tau_0 - \hat\tau_0)^2+\frac{1}{K}\sum_{k}2(\hat{\tau}_k - \tau_k)(\tau_k - \hat\tau_0) \\
    &= \frac{1}{K}\sum_{k}(\hat{\tau}_k - \tau_k)^2 + \frac{1}{K}\sum_{k}(\tau_k - \tau_0)^2 + \frac{1}{K}\sum_{k}2(\tau_k - \tau_0)(\tau_0 - \hat\tau_0) \\
    &~~~~~~~+(\tau_0 - \hat\tau_0)^2+\frac{1}{K}\sum_{k}2\tau_k(\hat{\tau}_k - \tau_k) - 2\hat\tau_0 \frac{1}{K}\sum_{k}(\hat{\tau}_k - \tau_k).
\end{align*}

Since we have $\mathbb{E}[(\hat{\tau}_k - \tau_k)^2] = \frac{4\sigma^2}{N}$ and $\mathbb{E}[(\tau_k - \tau_0)^2] = \sigma_0^2$, by the concentration theorem of sub-exponential random variables (Proposition 2.9 in \cite{Wainwright2019}), we obtain,
\begin{align*}
&\frac{1}{K}\sum_{k}(\hat{\tau}_k - \tau_k)^2 \to_p \frac{4\sigma^2}{N},\\
&\frac{1}{K}\sum_{k}(\tau_k - \tau_0)^2 \to_p \sigma_0^2.
\end{align*}
The remaining three  parts of the decomposition, $(\tau_0 - \hat\tau_0)^2$, $\frac{1}{K}\sum_{k}2\tau_k(\hat{\tau}_k - \tau_k)$, and  $2\hat\tau_0 \frac{1}{K}\sum_{k}(\hat{\tau}_k - \tau_k)$ converge to zero in probability due to the fact $\hat\tau_0 \to_p \tau_0$ and $\mathbb{E}[\hat\tau_k - \tau_k] = 0$. Thus, we can show that,
\begin{equation}
    \frac{1}{K}\sum_{k}(\hat{\tau}_k - \hat{\tau}_0)^2 \to_p \sigma_0^2 + \frac{4\sigma^2}{N}.
\end{equation}
Furthermore, combining the result in Eqn.~\eqref{eq:numerator}, we can conclude that:
\begin{equation}
    \frac{1}{K}\sum_{k}(\hat{\tau}_k - \hat{\tau}_0)^2 -  \frac{1}{KN}\sum_{k}4s_k^2 \to_p \sigma_0^2.
\end{equation}
Finally, by Slutsky’s Theorem, we have,
\begin{align*}
    &\frac{\frac{1}{K}\sum_{k}4s_k^2}{\frac{1}{K}\sum_{k}(\hat{\tau}_k - \hat{\tau}_0)^2 - \frac{1}{KN}\sum_{k}4s_k^2} \to_p \frac{4\sigma^2}{\sigma_0^2},\\
    &\frac{z_{1 - \alpha/2}\sqrt{N\frac{1}{K}\sum_{k}4s_k^2}}{\hat\tau_0} \to_p \frac{2\sqrt{N}z_{1 - \alpha/2}\sigma}{\tau_0}.
\end{align*}
Thus, we have, $\hat\beta^* \to_p \beta^*$. This completes the proof.\QED \\

\proof{\bf \noindent Proof of Theorem \ref{thm:cost}.}
Let
\[
\hat c=\frac{2\hat\sigma z_{1-\alpha/2}}{\sqrt N}-\frac{\hat\beta^*\hat\tau_0}{N},
\qquad
c^*=\frac{2\sigma z_{1-\alpha/2}}{\sqrt N}-\frac{\beta^*\tau_0}{N}.
\]
By Theorem~\ref{theorem:data_driven_para} and the preceding variance convergence, $\hat c\to_p c^*$.
For any threshold $c\in\mathbb{R}$ define
\[
g_c(\tau,\hat\tau)=\tau\,\mathds{1}\{\hat\tau>c\},\qquad
P_K g_c=\frac1K\sum_{k=1}^K g_c(\tau_k,\hat\tau_k),\qquad
Pg_c=\mathbb{E}[g_c(\tau_k,\hat\tau_k)].
\]
The class $\mathcal{G}=\{g_c:c\in\mathbb{R}\}$ is a VC-subgraph threshold class multiplied by the envelope $|\tau|$. Since $\tau_k\sim\mathcal{N}(\tau_0,\sigma_0^2)$, the envelope has finite second moment. Therefore $\mathcal{G}$ is Glivenko--Cantelli and
\[
\sup_{c\in\mathbb{R}}|P_K g_c-Pg_c|\to_p0.
\]
This uniform convergence is the step that handles the fact that the implemented threshold $\hat c$ is estimated from the same collection of experiments as the summands. Consequently,
\[
\begin{aligned}
|\bar r(\hat\beta^*,\hat\tau_0)-\mathcal{R}(\beta^*,\tau_0)|
&= |P_K g_{\hat c}-Pg_{c^*}|\\
&\le \sup_c |P_Kg_c-Pg_c|+|Pg_{\hat c}-Pg_{c^*}|.
\end{aligned}
\]
The first term converges to zero in probability by the uniform law of large numbers. For the second term, continuity of the joint normal distribution of $(\tau_k,\hat\tau_k)$ implies
$Pg_{\hat c}\to_p Pg_{c^*}$; for instance,
$|Pg_{\hat c}-Pg_{c^*}|\le \mathbb{E}[|\tau_k|\mathds{1}\{|\hat\tau_k-c^*|\le |\hat c-c^*|\}]$ and the right-hand side converges to zero by dominated convergence and $\hat c\to_p c^*$. Hence
\[
\bar r(\hat\beta^*,\hat\tau_0)\to_p Pg_{c^*}=\mathcal{R}(\beta^*,\tau_0).
\]
Under the maintained assumption $\tau_0>0$, Theorem~\ref{theorem:AP} gives the strict inequality $\mathcal{R}(\beta^*,\tau_0)>\mathcal{R}(0,\tau_0)$, completing the proof.
\QED
\\

\proof{\bf \noindent Proof of Theorem \ref{theorem:select_pro}.}
Under Assumptions~\ref{assum:linear_additive} and~\ref{assum:1}, the probability of selecting experiment $k$ can be rewritten as
\begin{equation*}
    \mathbb{P}(\bar{\tau}_k^{\mathrm{lb}}>0) = \mathbb{P}\Big(\hat{\tau}_k>\frac{2\sqrt{N}z_{1-\alpha/2}\hat{\sigma}-\hat{\beta}\hat{\tau}_0}{N}\Big).
\end{equation*}
The proof reduces to verifying the following two identities:
\begin{align}
    \lim_{K \to \infty}\mathbb{P}\Big(\hat{\tau}_k>\frac{2\sqrt{N}z_{1-\alpha/2}\hat{\sigma}-\hat{\beta}\hat{\tau}_0}{N}\Big) &= \mathbb{P}\Big(\hat{\tau}_k>\frac{2\sqrt{N}z_{1-\alpha/2}\sigma-\beta^*\tau_0}{N}\Big), \label{eq:pro_conver}\\
    \mathbb{P}\Big(\hat{\tau}_k>\frac{2\sqrt{N}z_{1-\alpha/2}\sigma-\beta^*\tau_0}{N}\Big) &= \Phi\Big(\frac{N\tau_k + \tau_0\beta^*}{2\sqrt{N} \sigma} -  z_{1 - \alpha/2}\Big). \label{eq:pro_rewrite}
\end{align}

Eqn.~\eqref{eq:pro_rewrite} follows directly from Assumption~\ref{assum:1}, under which $\hat{\tau}_k \sim \mathcal{N}(\tau_k,\tfrac{4\sigma^2}{N})$.

To prove Eqn.~\eqref{eq:pro_conver}, define
\begin{align*}
    \Psi_1(K) &= \hat{\tau}_k - \frac{2\sqrt{N}z_{1-\alpha/2}\hat{\sigma}-\hat{\beta}\hat{\tau}_0}{N}, &
    \Psi_2 &= \hat{\tau}_k - \frac{2\sqrt{N}z_{1-\alpha/2}\sigma-\beta^*\tau_0}{N},
\end{align*}
where $\Psi_1(K)$ depends on $K$ through the estimates $\hat{\sigma},\hat{\beta},\hat{\tau}_0$ obtained from all experiments. From the proof of Theorem~\ref{theorem:data_driven_para}, $\hat{\sigma} \xrightarrow{p} \sigma$, $\hat{\beta} \xrightarrow{p} \beta^{*}$, and $\hat{\tau}_{0} \xrightarrow{p} \tau_{0}$, so Slutsky's theorem gives
\begin{equation*}
    \frac{2\sqrt{N}z_{1-\alpha/2}\hat{\sigma}-\hat{\beta}\hat{\tau}_0}{N} \to_p \frac{2\sqrt{N}z_{1-\alpha/2}\sigma-\beta^*\tau_0}{N},
\end{equation*}
and hence $\Psi_1(K) \to_p \Psi_2$. Convergence in probability implies convergence in distribution. Moreover, under Assumption~\ref{assum:1}, the cumulative distribution functions $F_{\Psi_1(K)}$ and $F_{\Psi_2}$ of $\Psi_1(K)$ and $\Psi_2$ are continuous, so $\lim_{K\to\infty} F_{\Psi_1(K)}(0) = F_{\Psi_2}(0)$, which gives
\begin{equation*}
    \lim_{K \to \infty} \mathbb{P}(\Psi_1(K) > 0) = 1 - F_{\Psi_2}(0) = \mathbb{P}(\Psi_2 > 0),
\end{equation*}
establishing Eqn.~\eqref{eq:pro_conver}.
\QED \\

\proof{\bf \noindent Proof of Theorem \ref{theorem:AP_feature}.}
Condition on the design matrices $\{\bm t_k\}_{k=1}^K$ and hence on $\{b_k\}_{k=1}^K$. Under Assumption~\ref{assum:1},
\[
\hat\tau_k\mid \tau_k,b_k\sim\mathcal{N}(\tau_k,\sigma^2b_k^2/N),
\qquad
\tau_k\sim\mathcal{N}(\tau_0,\sigma_0^2),
\]
with $\tau_k$ independent of the design. Because the scale function may vary with $b_k$, the expected reward can be maximized pointwise in $b_k$. For a fixed $b_k>0$, the contribution to the expected reward is
\[
\int_{-\infty}^{+\infty}
\frac{1}{\sqrt{2\pi}\sigma_0}
e^{-\frac{(\tau-\tau_0)^2}{2\sigma_0^2}}
\Phi\left(\frac{N\tau+\tau_0\beta(b_k)}{\sqrt N\,\sigma b_k}-z_{1-\alpha/2}\right)
\tau\,d\tau .
\]
Differentiating this expression with respect to $\beta(b_k)$ gives a strictly positive multiplicative factor times
\[
-\beta(b_k)\tau_0\sigma_0^2+\sigma b_k\bigl(\sqrt N z_{1-\alpha/2}\sigma_0^2+\tau_0\sigma b_k\bigr).
\]
Under the maintained regime $\tau_0>0$, this linear term is positive at $\beta(b_k)=0$, has negative slope, and crosses zero once. Therefore the unique maximizer over $\beta(b_k)\ge0$ is
\[
\beta^*(b_k)=\frac{\sigma^2b_k^2}{\sigma_0^2}+\frac{\sqrt N z_{1-\alpha/2}\sigma b_k}{\tau_0}.
\]
Since the objective is an average of these conditional contributions and the feasible scale can be chosen as a function of $b_k$, pointwise maximization yields the stated optimal scale function. This concludes the proof.
\QED \\

\proof{\bf \noindent Proof of Theorem \ref{theorem:data_driven_para_feature}.}
Condition on the design matrices and hence on $\{b_k\}_{k=1}^K$. Assumption~\ref{assum:design_regular} implies that the triangular array of estimation errors
\[
u_k:=\hat\tau_k-\tau_k
\]
has conditional mean zero, conditional variance $\sigma^2 b_k^2/N$, and uniformly bounded fourth moments. Therefore
\[
\frac1K\sum_{k=1}^K u_k\to_p0,\qquad
\frac1K\sum_{k=1}^K u_k^2-\frac{\sigma^2}{N}\frac1K\sum_{k=1}^K b_k^2\to_p0.
\]
Together with the law of large numbers for $\tau_k\sim\mathcal{N}(\tau_0,\sigma_0^2)$, this gives
\[
\hat\tau_0=\frac1K\sum_{k=1}^K\hat\tau_k\to_p\tau_0.
\]
Similarly, since the OLS residual variance estimator satisfies
\[
s_k^2\sim\frac{\sigma^2}{N-d_x-2}\chi^2_{N-d_x-2}
\]
conditional on the design, and the $s_k^2$'s are independent across experiments with finite variance,
\[
\frac1K\sum_{k=1}^K s_k^2\to_p\sigma^2.
\]
Using the decomposition
\[
\hat\tau_k-\hat\tau_0=(\tau_k-\tau_0)+u_k+(\tau_0-\hat\tau_0),
\]
the preceding convergences, Assumption~\ref{assum:design_regular}, and Cauchy's inequality imply
\[
\frac1K\sum_{k=1}^K(\hat\tau_k-\hat\tau_0)^2
\to_p
\sigma_0^2+\frac{\sigma^2}{N}\frac1K\sum_{k=1}^K b_k^2.
\]
Consequently,
\[
\frac1K\sum_{k=1}^K(\hat\tau_k-\hat\tau_0)^2
-
\frac1N\left(\frac1K\sum_{k=1}^Ks_k^2\right)
\left(\frac1K\sum_{k=1}^K b_k^2\right)
\to_p \sigma_0^2.
\]
Since $\tau_0>0$ and $\sigma_0^2>0$, Slutsky's theorem yields, for any design value $b_k$ satisfying Assumption~\ref{assum:design_regular},
\[
\hat\beta^*(b_k)\to_p
\frac{\sigma^2b_k^2}{\sigma_0^2}
+
\frac{\sqrt N z_{1-\alpha/2}\sigma b_k}{\tau_0}
=
\beta^*(b_k).
\]

It remains to justify the reward convergence with the data-dependent thresholds. Define the parameter vector collecting the nuisance quantities in the threshold by
\[
\theta=(\sigma,\tau_0,\beta^*(\cdot)),\qquad
\hat\theta=(\hat\sigma,\hat\tau_0,\hat\beta^*(\cdot)),
\]
and write
\[
c(b;\theta)=\frac{\sigma b z_{1-\alpha/2}}{\sqrt N}-\frac{\beta^*(b)\tau_0}{N},\qquad
c(b;\hat\theta)=\frac{\hat\sigma b z_{1-\alpha/2}}{\sqrt N}-\frac{\hat\beta^*(b)\hat\tau_0}{N}.
\]
By the convergence established above and the boundedness of $b_k$, $\max_k|c(b_k;\hat\theta)-c(b_k;\theta)|\to_p0$. Consider the class
\[
\mathcal{G}_b=\left\{(\tau,\hat\tau,b)\mapsto \tau\,\mathds{1}\{\hat\tau>c(b;\eta)\}:\eta\in\Theta,\ b\in[\underline b,\bar b]\right\},
\]
where $\Theta$ is a compact neighborhood of the probability limit of $\hat\theta$. This is a VC-subgraph threshold class with envelope $|\tau|$, which has finite second moment. Hence, conditional on the designs, the class is Glivenko--Cantelli for the triangular array:
\[
\sup_{g\in\mathcal{G}_b}\left|\frac1K\sum_{k=1}^Kg(\tau_k,\hat\tau_k,b_k)-\mathbb{E}\{g(\tau_k,\hat\tau_k,b_k)\mid b_k\}\right|\to_p0.
\]
The same continuity argument used in the proof of Theorem~\ref{thm:cost}, now uniformly over bounded $b_k$, then gives
\[
\bar r(\hat\beta^*(\cdot),\hat\tau_0)-\tilde r(\beta^*(\cdot),\tau_0)\to_p0.
\]
Finally, the conditional law of large numbers for the independent summands
\[
\tau_k\mathds{1}\left\{\hat\tau_k>\frac{\sigma b_k z_{1-\alpha/2}}{\sqrt N}-\frac{\beta^*(b_k)\tau_0}{N}\right\}
\]
implies $\tilde r(\beta^*(\cdot),\tau_0)\to_p\mathcal{R}(\beta^*(\cdot),\tau_0)$. Combining the two displays proves the theorem.
\QED

\section{Numerical Results with Expedia Experiment Data} \label{app:real_expedia_data}
In this dataset, we complement our analysis with an additional dataset to assess the performance of the \our\ method in a non-overlapping setting. 
The dataset comes from a field experiment conducted by Expedia, the world’s largest online travel agency. This publicly available dataset\footnote{\url{www.kaggle.com/c/expedia-personalized-sort/data}} includes data from consumers searching for hotels, who were randomly assigned to one of two groups: (i) those who viewed a personalized ranking, where hotels were ordered according to their suitability for consumers based on Expedia's internal ranking algorithm, and (ii) those who viewed a random ranking, where hotels were listed in no particular order.

Following the data cleaning process outlined in \citet{ursu2018power}, we obtained a total of approximately 166 thousands of  consumer queries for hotels, along with their corresponding choices (clicks and purchases), spanning an eight-month period ending in June 2013. This dataset includes approximately 4.5 million observations of hotels displayed on Expedia.

In this dataset, we use the origin-destination pair (the country of the customer and the country of the hotels being searched) to partition the whole dataset. To ensure each group has enough data for reliable analysis, we exclude groups with fewer than 250 queries. This results in 119 groups, that is, \( K = 119 \). The histogram of the full-sample benchmark HTE across all groups, estimated from the entire dataset, is shown in Figure~\ref{fig: true_ate}. As in the Criteo analysis, these full-sample estimates serve as a high-precision benchmark for evaluation rather than as known ground-truth causal effects.

    \begin{figure}[htbp!]
    \centering
     \includegraphics[width=0.4\linewidth]{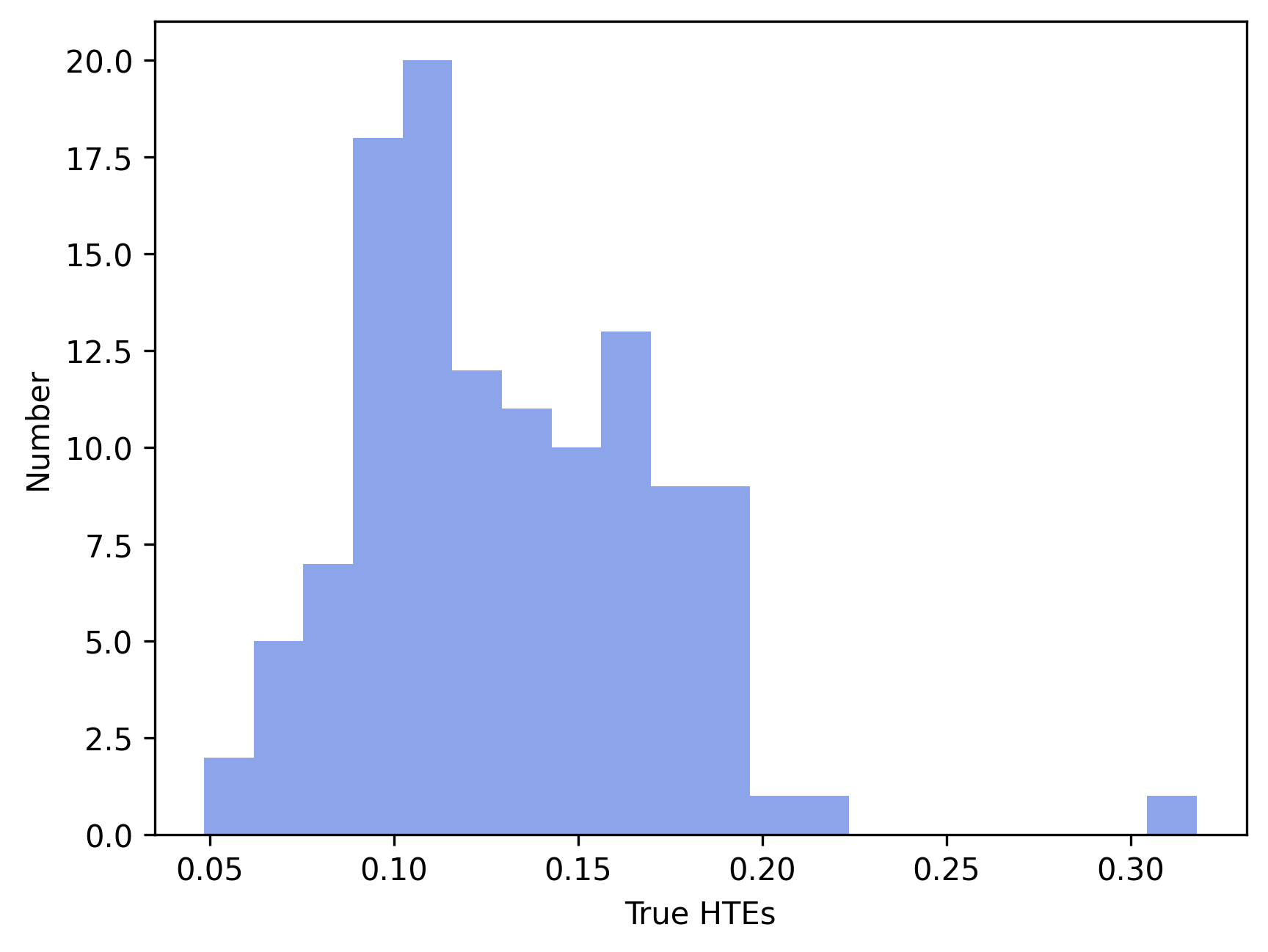}
    \caption{Histogram of full-sample benchmark HTEs for all groups of the Expedia dataset.}
    \label{fig: true_ate}
    
\end{figure}

Considering that in real-world scenarios, platforms incur costs $\tau_{\text{min}}$ for both the design and maintenance of algorithms. Thus, for any policy $k$, the actual reward obtained by the platform is given by, $(\tau_k - \tau_{\text{min}})$. We assess the impact of implementation costs in this example by considering two frictional cost values, \(\tau_{\text{min}} = 0.05\) and \(0.1\)\footnote{This does not imply that the actual implementation cost of recommendation algorithms reaches this level, although sophisticated recommendations may increase webpage response times, potentially leading to long-term negative outcomes for the platform, as webpage speed is an important factor for online consumers \citep{GallinoKaracaogluMoreno2023}.}. Since the data sizes vary across groups, we normalize \((\tau_k - \tau_{\text{min}})\) when calculating the OR and VDP values by multiplying it by the parameter \(\frac{N_{k,0}+N_{k,1}}{\sum_{k}(N_{k,0}+N_{k,1})}\), which represents the proportion of group \(k\)'s data size relative to the total data size of all groups. 

Besides the OR and VDP metrics, we also report Recall rate which measures the ratio of correct decisions made by the roll-out method (\tradition\ or \our) across all experiments where the roll-out should occur and Specificity which measures the ratio of correct decisions made by the roll-out method (\tradition\ or \our) across all experiments where the roll-out should not occur.

\begin{figure}[!ht]
\centering
\subfigure[OR ($\tau_{\text{min}} = 0.05$)]{
\begin{minipage}[t]{0.3\linewidth}
\centering
\includegraphics[width=2.0in,height=1.5in]{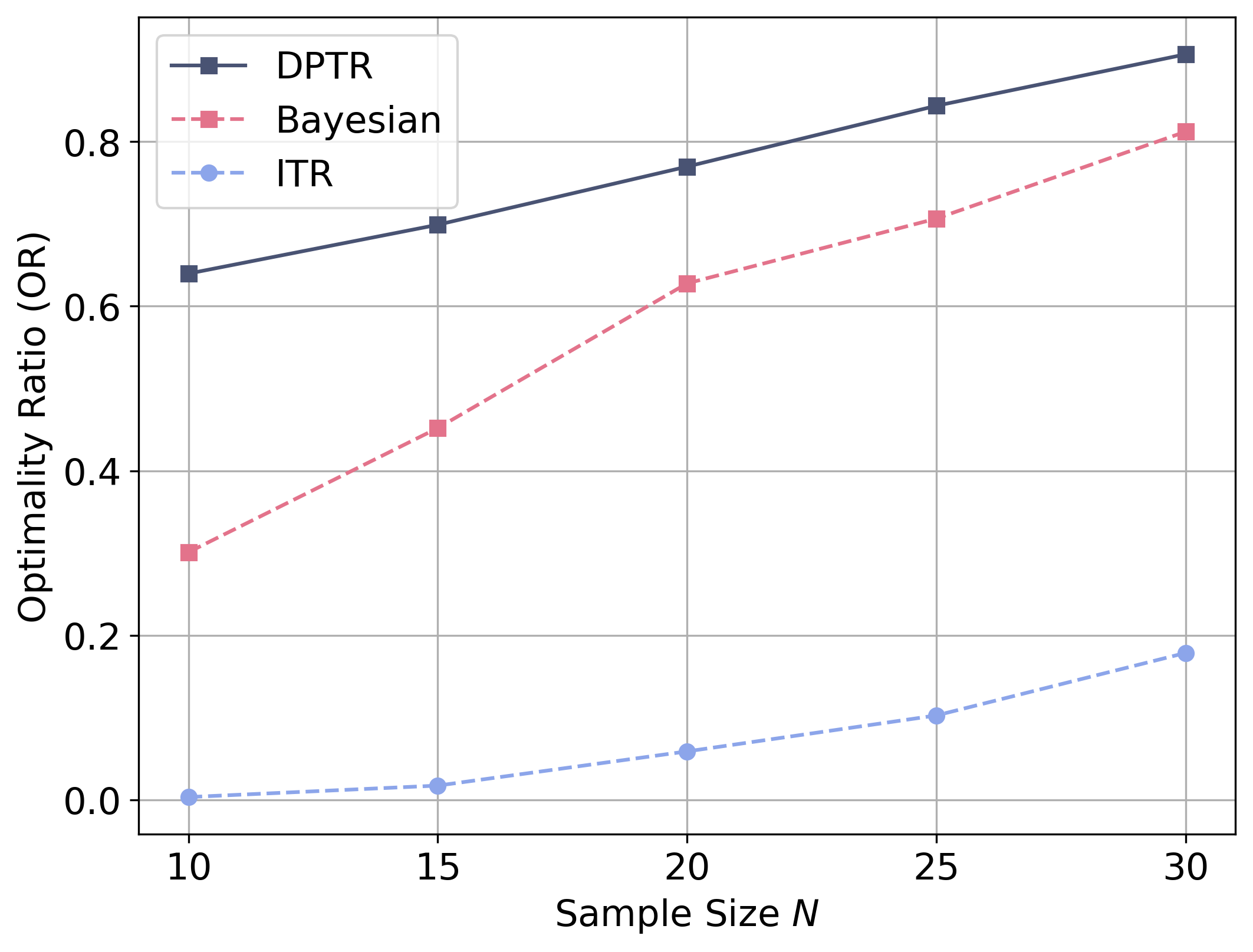}
\end{minipage}
}%
\subfigure[VDP ($\tau_{\text{min}} = 0.05$)]{
\begin{minipage}[t]{0.3\linewidth}
\centering
\includegraphics[width=2.0in,height=1.5in]{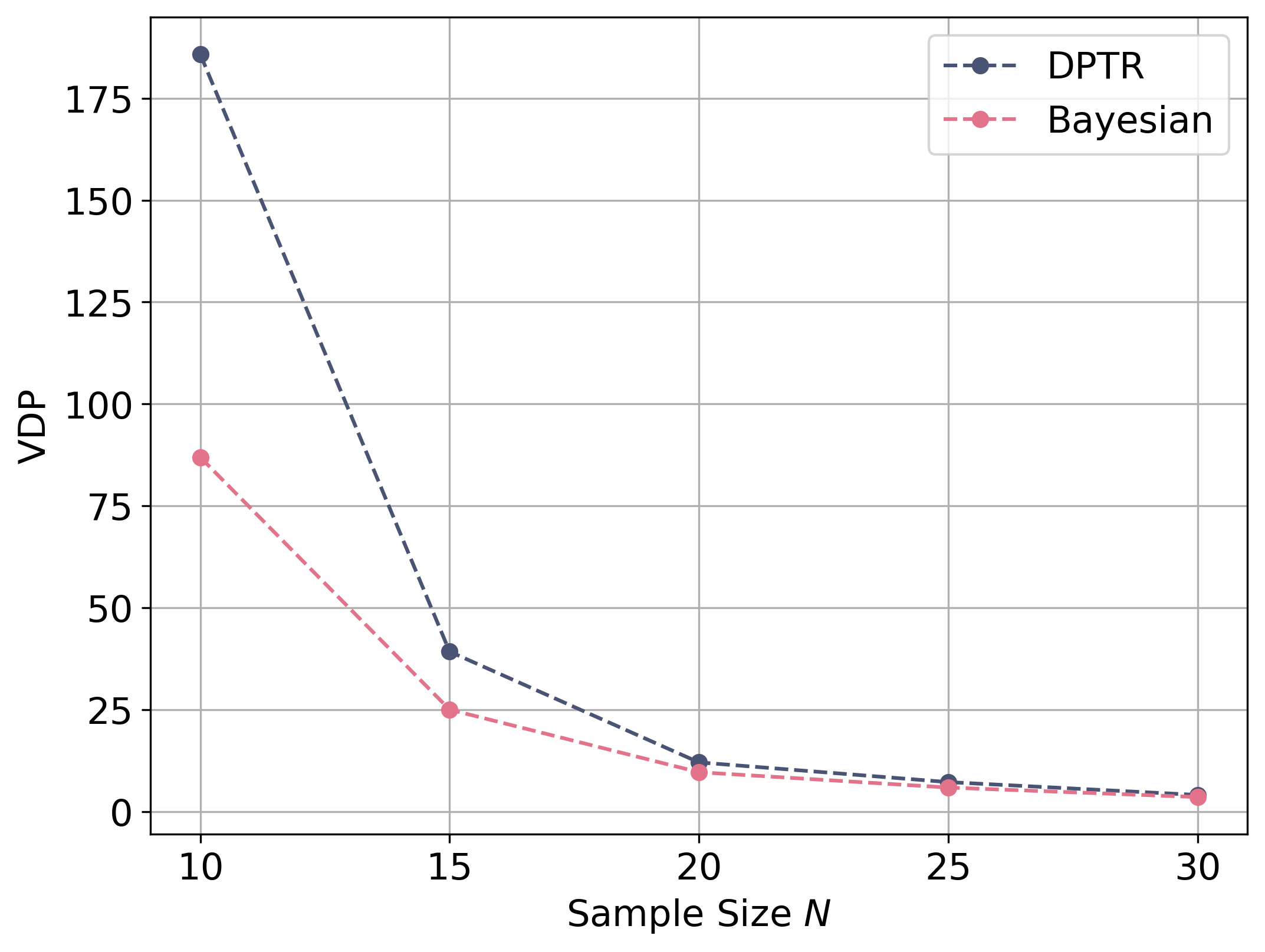}
\end{minipage}%
}%
\subfigure[Recall and  Specificity  ($\tau_{\text{min}} = 0.05$)]{
\begin{minipage}[t]{0.34\linewidth}
\centering
\includegraphics[width=2.0in,height=1.5in]{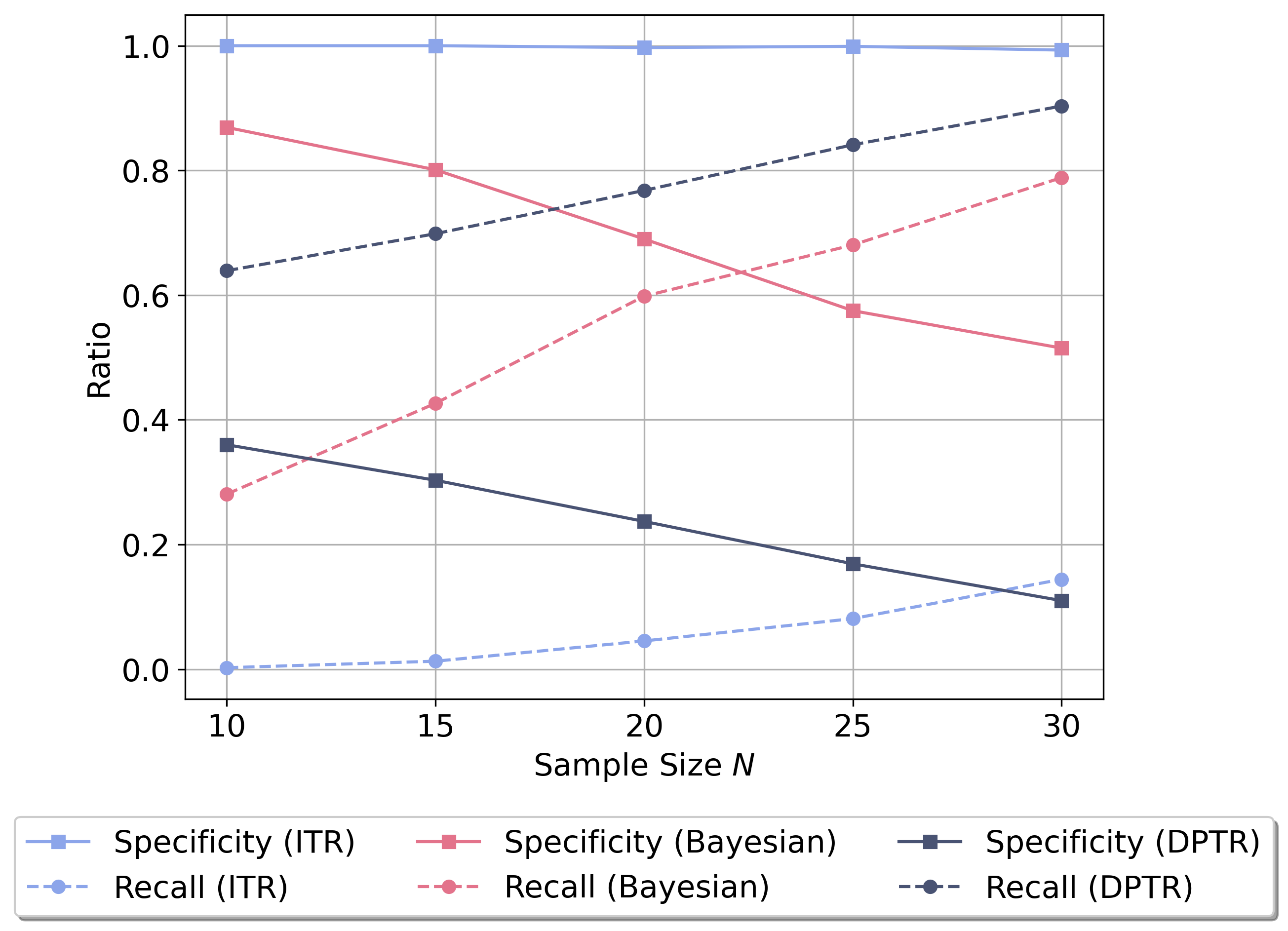}
\end{minipage}%
}%

\subfigure[OR ($\tau_{\text{min}} = 0.1$)]{
\begin{minipage}[t]{0.3\linewidth}
\centering
\includegraphics[width=2.0in,height=1.5in]{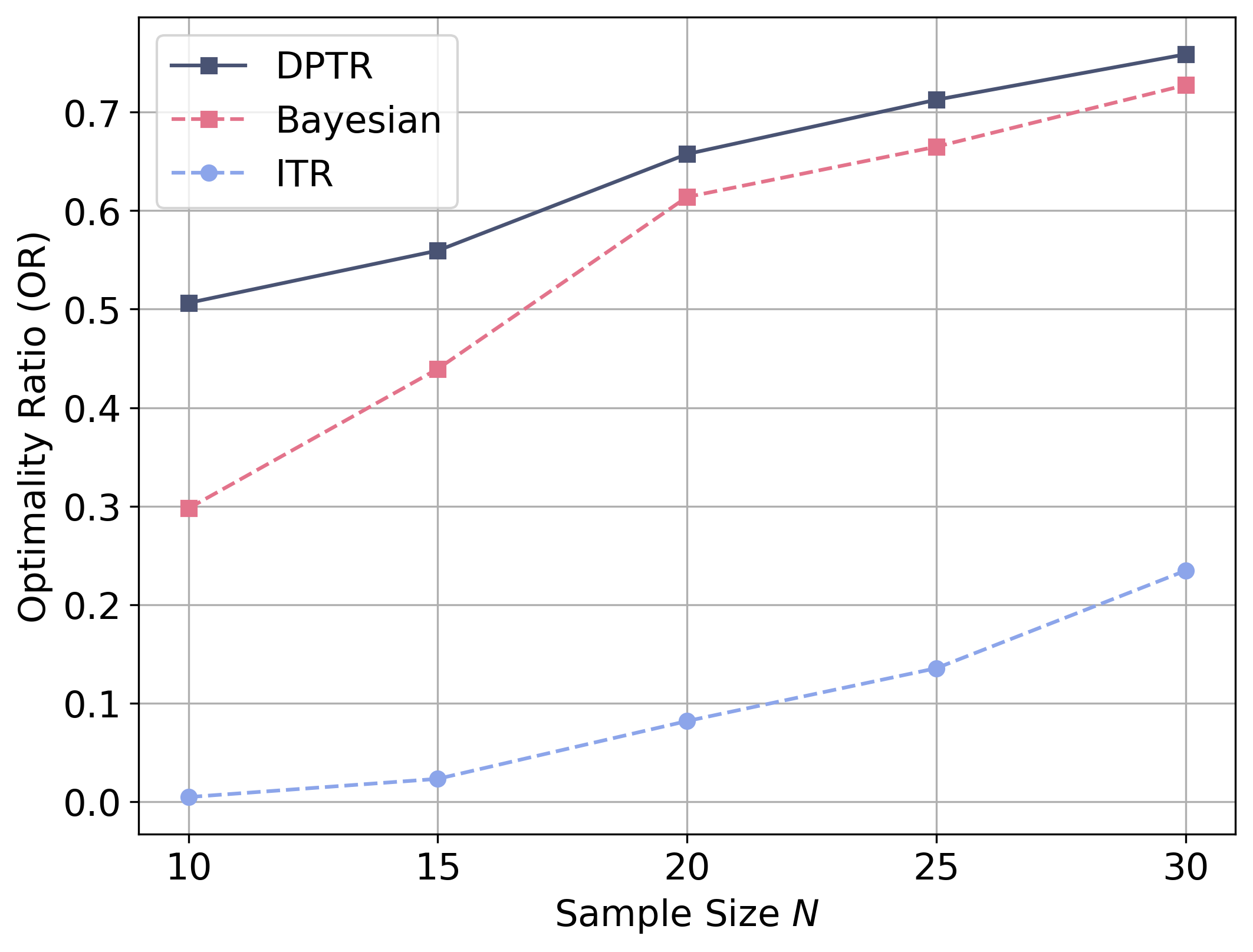}
\end{minipage}
}%
\subfigure[VDP ($\tau_{\text{min}} = 0.1$)]{
\begin{minipage}[t]{0.3\linewidth}
\centering
\includegraphics[width=2.0in,height=1.5in]{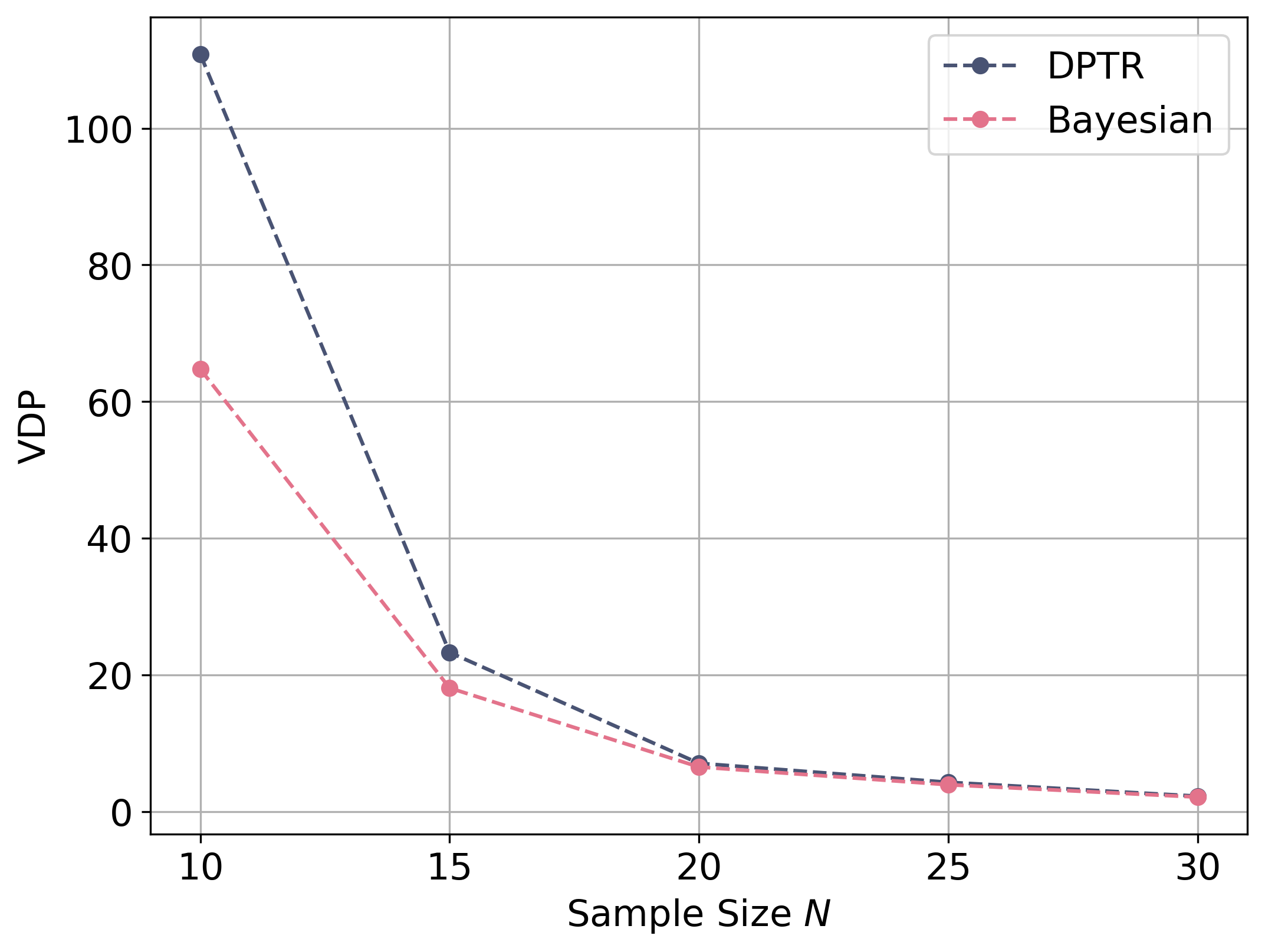}
\end{minipage}%
}%
\subfigure[Recall and Specificity ($\tau_{\text{min}} = 0.1$)]{
\begin{minipage}[t]{0.33\linewidth}
\centering
\includegraphics[width=2.0in,height=1.5in]{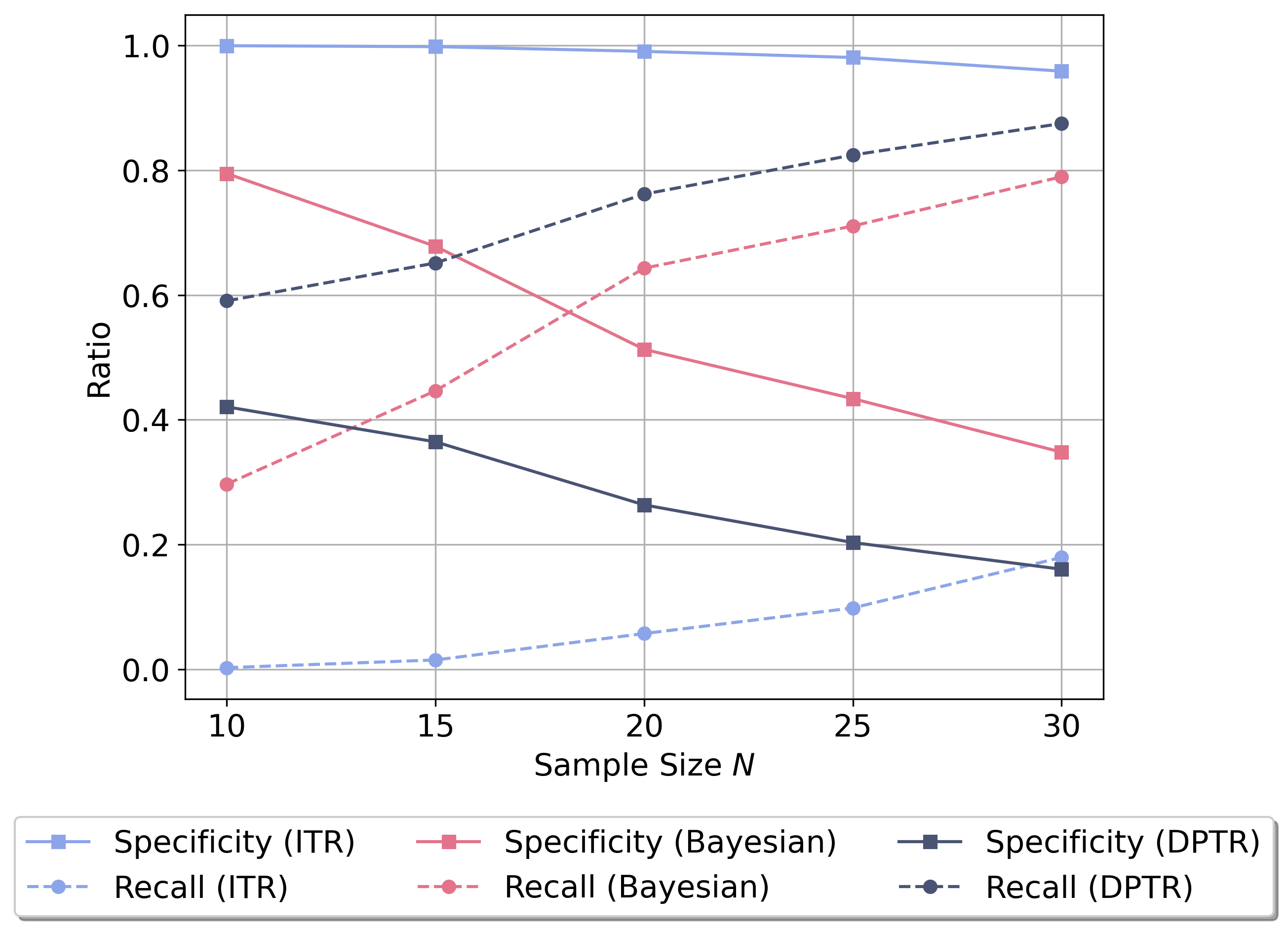}
\end{minipage}%
}%
\centering
\caption{Performance comparisons under different sample size $N$ and  $\tau_{\text{min}}$ in Expedia dataset.}
\label{fig:results_expedia}
\end{figure}

Furthermore, for any $\tau_{\text{min}}$, we vary the sample size $N$ from 10 to 30 in increments of 5 and repeat the experiment 1,000 times. The averaged results are shown in Figure~\ref{fig:results_expedia}. As depicted in the figure, we first observe that, regardless of the value of $\tau_{\text{min}}$, the trends across all indicators remain largely consistent, with no significant fluctuations. This suggests that the choice of $\tau_{\text{min}}$ has minimal impact on the comparison between the \tradition\ and \our\ methods.

According to Figure~\ref{fig:results_expedia}(a) and (d), the \our\ method consistently outperforms the \tradition\ and Bayesian methods, irrespective of the sample size $N$. Additionally, as the sample size increases, the reduction in estimator variance leads to a corresponding increase in OR values for both methods. Figure~\ref{fig:results_expedia}(b) and (e) further reveal that the relative performance advantage of the \our\ and Bayesian methods over the \tradition\ method increases significantly as the amount of experimental data decreases. This finding highlights that, when experimental data is limited, the data-pooling methods, such as \our\ and Bayesian methods, can provide substantial performance improvements.

Finally, Figure~\ref{fig:results_expedia}(c) and (f) show that when $N$ is small, the Recall value under the \tradition\ method is closer to 0, indicating that few or no personalized recommendations are rolled out to specific groups. Since few or no groups launch the personalized recommendation, the Specificity value is higher under the \tradition\ method. On the other hand, despite large variation in the estimate of the benefit of personalized recommendations in each group, the pooled estimate in the \our\ and Bayesian methods reduce this variation, pushing the estimate towards a more positive region. As a result, the Recall value under the \our\ and Bayesian methods increases significantly. This comes at the cost of some groups being incorrectly selected for the personalized recommendation. However, overall, by balancing Recall and Specificity, the \our\ and Bayesian methods generate a much higher reward compared to the \tradition\ method in this tailored roll-out decision situation.

We observe that the \tradition\ method has a very low Recall value and a high Specificity value. This phenomenon suggests that traditional methods (\tradition) tend to exclude experiments that should not be rolled out, even if this tendency results in the exclusion of many experiments that should actually be rolled out. In contrast, our proposed \our\ method maximizes the reward from the final roll-out decision by balancing the trade-off between Recall and Specificity.

Furthermore, we test the performance of our method when different groups have different sample sizes. We denote \(\rho\) as the sample proportion with respect to the total data size, which means we will randomly select \(\rho \times (N_{k,1} + N_{k,0})\) customer queries as sample size for group \(k\). We vary \(\rho\) from 0.05 to 0.25 in increments of 0.05 while keeping \(\tau_{\text{min}} = 0.1\) fixed. Similarly, we repeat the experiment 1,000 times and present the averaged results in Figure \ref{fig:results_expedia_proportion}. We observe that variations in sample sizes across different groups do not affect the performance of our \our\ method.

\begin{figure}[!ht]
\centering
\subfigure[OR]{
\begin{minipage}[t]{0.3\linewidth}
\centering
\includegraphics[width=2.0in,height=1.5in]{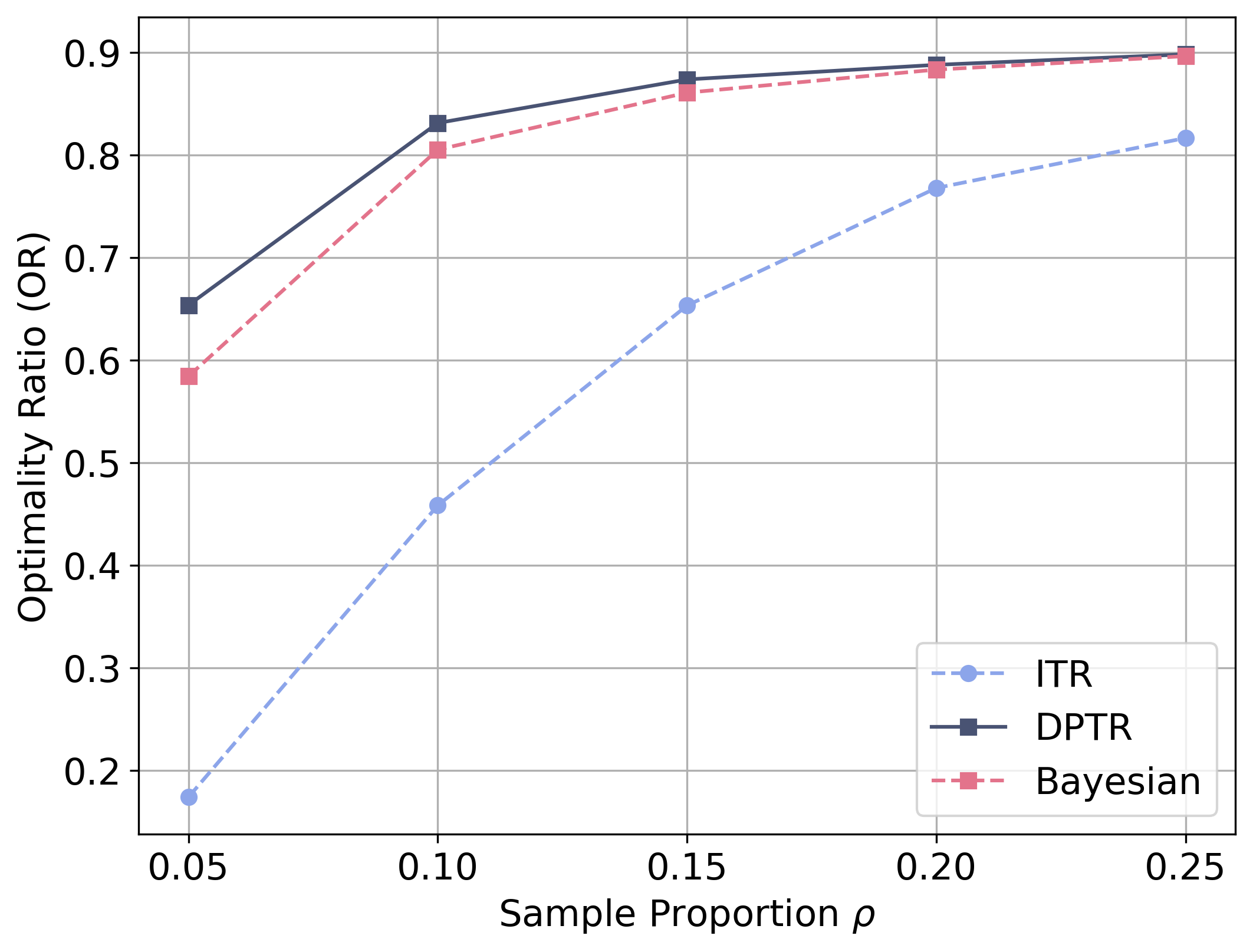}
\end{minipage}
}%
\subfigure[VDP]{
\begin{minipage}[t]{0.3\linewidth}
\centering
\includegraphics[width=2.0in,height=1.5in]{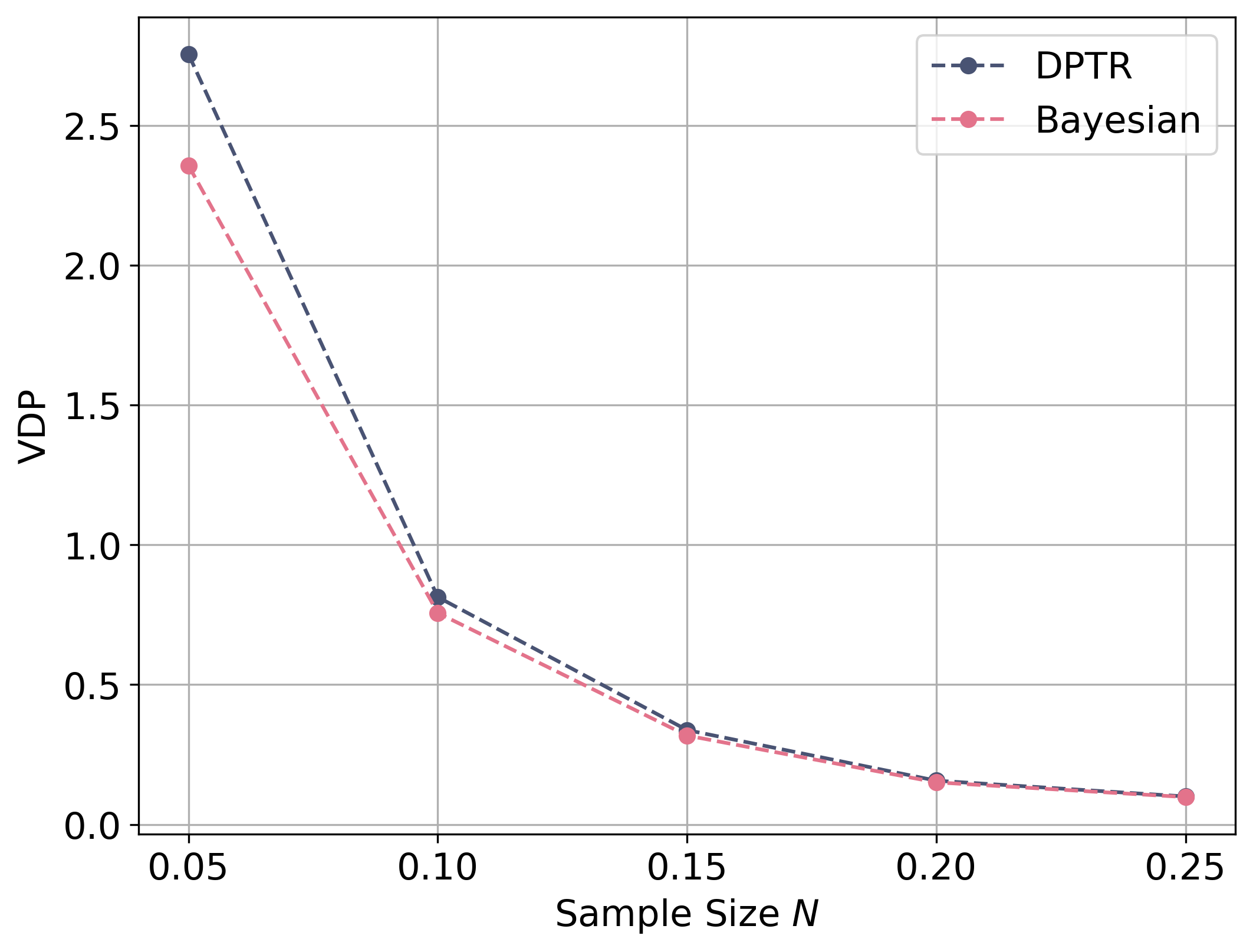}
\end{minipage}%
}%
\subfigure[Recall and Specificity]{
\begin{minipage}[t]{0.33\linewidth}
\centering
\includegraphics[width=2.0in,height=1.5in]{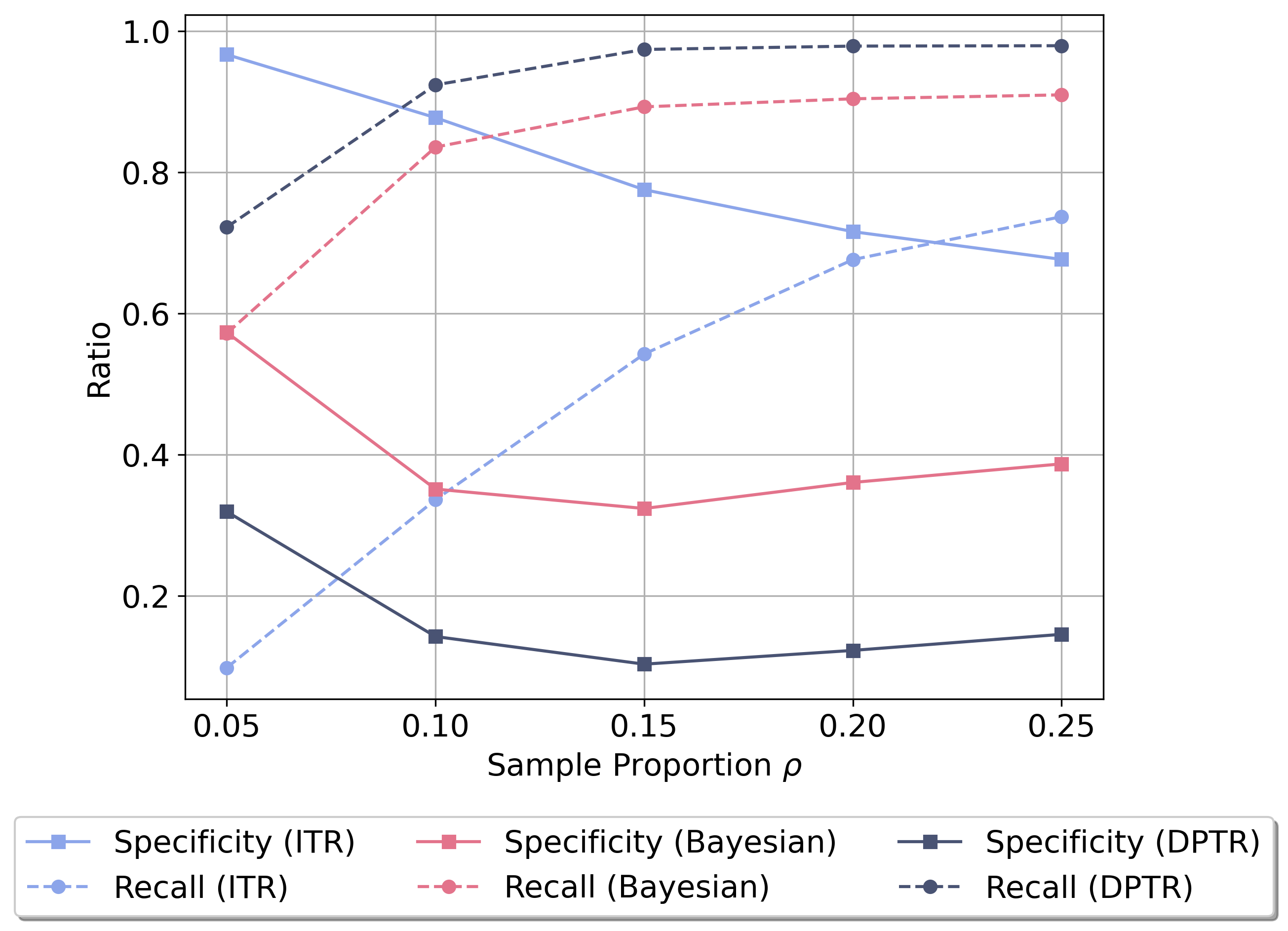}
\end{minipage}%
}%
\centering
\caption{Performance comparisons under different sample proportion $\rho$ in Expedia dataset.}
\label{fig:results_expedia_proportion}
\end{figure}

\section{Detailed Estimation Procedures for the DML method}
\label{app:dml_scenario2}

In this section, we provide a detailed description of the estimation procedure for the DML method in our setting. Based on the notation of Section \ref{subsec:scenario-2}, to begin, we define the loss function for estimation and inference function as:
\begin{equation*}
    \ell(Y,\boldsymbol{t},f(X)) = (Y - f(X)^{\top}\boldsymbol{t})^2. \quad H(X,f(X):\boldsymbol{t}^*) = f(X)^{\top} \boldsymbol{t}^*.
\end{equation*}
Here, for each experiment \( k \), we apply the cross-fitting techniques \citep[][]{chernozhukov2018double,farrell2020deep} to obtain the estimator \( \hat{\tau}_k \). Specifically, we define,
\begin{equation} \label{eq:psi}
    \psi(Y,X,\bm t,f(\cdot),\Lambda) = H(X,f(X):\boldsymbol{t}^*) - H_f(X,f(X):\boldsymbol{t}^*)\Lambda(X)^{-1}\ell_f(Y,\boldsymbol{t},f(X)),
\end{equation}
where $H_f$ and $\ell_f$ are the gradients of $H$ and $\ell$ with respect to $f$, and $\Lambda(X) = \mathbb{E}[\ell_{ff}(Y,\boldsymbol{t},f(X))|X]$ represents the conditional expectation of the Hessian of $\ell$. The expectation of $H(X, f(X); \boldsymbol{t}^*)$ represents the true ATE we aim to estimate. However, in practice, due to the complexity and regularization of $f(X)$, the sample mean of $H(X, f(X); \boldsymbol{t}^*)$ does not yield an unbiased estimator. Therefore, the second term in Eqn. \eqref{eq:psi} was introduced to correct the bias.

The estimation and experiment roll-out process can be summarized as follows. First, the data set \( \mathcal{S}_k \) is split into $S$ subsets with equal size, denoted by \( \mathcal{S}_{k,s} \) where $s \in \{1,\cdots,S\}$. Let \( \mathcal{S}_{k,s}^c \) be the complement of \( \mathcal{S}_{k,s} \). Then, for each $s \in \{1,\cdots,S\}$, we use \( \mathcal{S}_{k,s}^c \) to estimate $\Lambda_k(X_k) = \mathbb{E}[\ell_{ff}(Y_k,\boldsymbol{t}_k,g_k(X_k))|X_k]$ and  $g_k(\cdot)$. We denote $\hat{g}_{k,s}(\cdot)$ and $\hat\Lambda_{k,s}(\cdot)$ as the estimators for ${g}_{k,s}(\cdot)$ and $\Lambda_{k,s}(\cdot)$, respectively. Then, based on Eqn.~\eqref{eq:psi}, the final estimator $\hat\tau_k$ can be written as:
\begin{equation*}
    \hat\tau_k = \frac{1}{S}\sum_{s = 1}^{S}\hat\tau_{k,s},\quad \hat\tau_{k,s} = \frac{1}{|\mathcal{S}_{k,s}|}\sum_{i \in \mathcal{S}_{k,s}}\psi(Y_{k,i},X_{k,i},\bm t_{k,i},\hat{g}_{k,s}(X_{k,i}),\hat\Lambda_{k,s}(X_{k,i})).
\end{equation*}
The estimator for variance of $\hat\tau_k$ can be written as:
\begin{equation*}
    \hat\Psi_k = \frac{1}{S}\sum_{s = 1}^{S}\hat\Psi_{k,s},\quad \hat\Psi_{k,s} = \frac{1}{|\mathcal{S}_{k,s}|}\sum_{i \in \mathcal{S}_{k,s}}\Big(\psi(Y_{k,i},X_{k,i},\bm t_{k,i},\hat{g}_{k,s}(X_{k,i}),\hat\Lambda_{k,s}(X_{k,i})) - \hat\tau_k\Big)^2.
\end{equation*}

\section{Bayesian Benchmark Method} \label{app:bayesian_method}
In this section, we provide a detailed description of how the Bayesian framework pools information across multiple experiments within our setting. In particular, we formalize the hierarchical structure that enables information sharing across experiments, specify the prior and posterior formulations, and clarify how such pooling improves estimation efficiency under limited per-experiment data. We also highlight the key differences between this approach and our method, with an emphasis on how the extent of pooling is determined and how it impacts the resulting decision-making performance.

In Scenario 1, the ATE of experiment $k$, $\tau_k$, follows the prior distribution $\mathcal{N}(\tau_0, \sigma_0^2)$. The outcome $Y_k$ is sampled from
\begin{equation*}
Y_{k,i} \sim \mathcal{N}(a_k+\tau_kD_{k,i},\sigma_k^2),\ i = 1,\cdots,N.
\end{equation*}
For each experiment $k$, the platform randomly assigns  \( N/2 \)  to the treatment condition and \( N/2 \) to the control condition. Direct application of the Bayes rule implies the posterior distribution of $\tau_k$ given the DM estimator $\hat\tau_k$: 
\begin{equation*} \label{eq:bayesian_posterior}
    \tau_k | \hat{\tau}_k \sim \mathcal{N}\Bigg(\hat{\tau}_k\frac{N}{N + \beta^{\text{bayes}}_k} + \tau_0\frac{\beta^{\text{bayes}}_k}{N+\beta^{\text{bayes}}_k},\frac{N}{N + \beta^{\text{bayes}}_k}\frac{4\sigma_k^2}{N}\Bigg),\ \text{where}\  \beta^{\text{bayes}}_k = \frac{4\sigma_k^2}{\sigma_0^2}.
\end{equation*}
Similarly, in the data-driven setting, we apply the data pooling technique to estimate the prior mean $\tau_0$ and variance $\sigma_0^2$. As shown in the proof of Theorem \ref{theorem:data_driven_para}, we use $\hat\tau_0 = \frac{1}{K}\sum_{k=1}^K \hat \tau_k$ and the variance estimate $\hat\sigma^2=\frac{1}{K}\sum_{k}(\hat{\tau}_k - \hat{\tau}_0)^2 - \frac{1}{KN}\sum_{k}4s_k^2$, where $s_k^2$ is the unbiased estimator for $\sigma_k^2$ defined in Theorem \ref{theorem:data_driven_para}. Then the implemented data-driven Bayesian scale parameter is written as:
\begin{equation} \label{eq:data_driven_baye}
    \hat\beta_{k,+}^{\text{bayes}} = \max(0,\frac{4s_k^2}{\hat\sigma^2}).
\end{equation}

Comparing the scale parameters in Eqn. \eqref{eq:data-driven-beta-S1} and Eqn. \eqref{eq:data_driven_baye} reveals insights on how our \our\ method differs from the Bayesian method. First, in the Bayesian method, the shrinkage parameter varies across different experiments, while in our method, it remains uniform, enhancing the effect of pooling data from different experiments. Second, the shrinkage parameter in our method includes an additional term that accommodates the significance level $\alpha$, making it decision-aware. To empirically compare the \our\ method with the Bayesian method, we follow Table 6 in \cite{Raftery1995}  and assumes the platform rolls out treatment $k$ if the posterior probability of $\tau_k > 0$ is at least $1 - \alpha/2$, in line with a two-sided test with significance level $\alpha$ in the frequentist framework.

\section{Robustness under Model Misspecification} \label{subset: robust_model_misspecification}
Our analysis so far has focused on Assumption \ref{assum:linear_additive} that the policy ATEs are linearly additive. In practice, however, this assumption does not hold in general \citep{ye2023deep}. To understand how well our proposed data pooling method works when the treatment effects of different policies are not linearly additive, we consider the \textit{Generalized Sigmoid
Form II} DGP in \citet{ye2023deep}:
\begin{equation} \label{eq:non_linear_model}
    Y_{i} = \frac{\upsilon}{1 + \exp(-g(X_{i})^{\top}\bm t_i)}  + \epsilon_{i},\ i = 1,\dots,N,
\end{equation}
where $g(\cdot): \mathbb{R}^{d_x}\to \mathbb{R}^{K+1}$ is the true response function, $\bm t_i$ is the treatment vector which includes a constant term, and  $\epsilon_{i}$ denotes the i.i.d. random noise. Thus, the optimal reward the platform can obtain is given by:
\begin{equation*}
    r^* = \max_{\bm t} \mathbb{E}[Y|\bm t] - \mathbb{E}[Y|\bm t_0],
\end{equation*}
where $\bm t_0$ is the base treatment vector in which the treatment indicators for all experiments equal to zero.
While the data generating process follows Eqn. \eqref{eq:non_linear_model}, we deliberately ignore the non-linear model specifications and apply the same method in Section~\ref{subset: with-overlap-with-feature} to decide whether the experiment $k$ should be implemented. After implementing Algorithm \ref{alg:general_dm} and \ref{alg:general_dp}, we obtain the roll-out decisions $\hat{\mathcal{A}}_{\tradition}$ and $\hat{\mathcal{A}}_{\our}$. The reward obtained by the \tradition\ and \our\ can be written as:
\begin{equation*}
\hat{r}_{\tradition} = \mathbb{E}[Y|\hat{\mathcal{A}}_{\tradition}]- \mathbb{E}[Y|\bm t_0],\quad \bar{r}(\beta,\tau) = \mathbb{E}[Y|\hat{\mathcal{A}}_{\our}]- \mathbb{E}[Y|\bm t_0].
\end{equation*}

The experimental setup is as follows: the error term is sampled from a normal distribution \( \mathcal{N}(0, 3^2) \), and the number of experiments is set to \( K = 4 \). We draw \(\upsilon\) from the uniform distribution \( U(10, 20) \), and define the function \( g(X_i) \) as  
$g(X_i) = \left\{\gamma_0^{\top}X_i, \gamma_1^{\top}X_i, \ldots, \gamma_K^{\top}X_i \right\}$
where each vector \(\gamma_0, \gamma_1, \ldots, \gamma_K\) consist of components which are independently drawn from \( U(-0.3, 0.5) \). The covariate \( X_i \) is of dimension \( d_x = 4 \), where each component is sampled independently from \( U(0, 1) \). To provide a more comprehensive comparison of the performance of the \tradition\ and \our\ methods, we vary the number of experiments from 4 to 7 with an increment of 1, and also vary $\sigma$ from 3 to 5 in increments of 1. We repeat each setting 1000 times and report the average OR values in Table~\ref{tab:3}.

\vspace{-0.1in}
\begin{table}[!ht]
\centering
		{\def\arraystretch{1.1}  
\centering
\begin{tabular*}{1\textwidth}
{@{\extracolsep{\fill}}ccccccc}
\hline
\hline
 & \multicolumn{2}{c}{$\mathcal{N}(0,3^2)$}  & \multicolumn{2}{c}{$\mathcal{N}(0,4^2)$}  & \multicolumn{2}{c}{$\mathcal{N}(0,5^2)$}  \\
\hline
 $K$  & \tradition\   & \our\ &  \tradition\  & \our\  &  \tradition\   & \our\ \\
\hline
$4$ &0.0269 &0.3754 &0.0259& 0.3862 & 0.0217 & 0.3862  \\
$5$ &0.0142 &0.4267 & 0.0151 & 0.3913 & 0.0104 & 0.4043   \\
$6$ &0.0065  & 0.4641 &0.0057  & 0.4346 &0.0058  & 0.4301 \\
$7$ &0.0023  & 0.5133 & 0.0013  & 0.4864 & 0.0032  & 0.4945\\ 
\hline
\hline
\end{tabular*}
\caption{The performance comparison with OR under different $K$ and variance of error term.}
\label{tab:3}}
\vspace{-0.2in}
\end{table}

First, we observe that the \our\ method consistently outperforms the \tradition\ method, regardless of the number of experiments or the magnitude of the error term. Second, due to the effect of nonlinearity, the performance of both the \our\ and \tradition\ methods does not show a strictly monotonic decline as the variance of error term increases. Finally, an interesting phenomenon is that when nonlinearity is ignored, increasing the number of experiments tends to amplify the degree of nonlinearity. This, in turn, leads to deteriorating performance of the \tradition\ method, while the \our\ method continues to improve. This finding further demonstrates that the \our\ method is capable of rolling out high-reward treatments even under model misspecification, highlighting its robustness.

\section{The Limitations of \our\ Method}

\subsection{Nonlinear Additive Treatment Effects}  
  \label{app:limit_nonlinear}
The theoretical and simulation results in the main text primarily rely on Assumption \ref{assum:linear_additive}. While Section \ref{subset: robust_model_misspecification} demonstrates that our method continues to outperform the \tradition\ approach under the Generalized Sigmoid Form II data-generating process (DGP) of \cite{ye2023deep}, it is important to more precisely characterize the nonlinear settings in which \our\ performs well and those in which its performance deteriorates. In this section, we provide additional analysis and supporting numerical evidence.

We denote by $Y(\boldsymbol{D})$  the expected outcome  under treatment vector $\boldsymbol{D} \in \mathbb{R}^K$. Let $e_i \in \mathbb{R}^K$ denote the unit vector corresponding to treatment $i$, and let $e_0$ denote the all-zero vector. We consider the following condition. 

\textbf{Condition 1 (Sign Consistency of Marginal Effects)}: For any $i \in [K]$, suppose the baseline marginal effect \( Y(e_i) - Y(e_0) \) is either strictly positive or strictly negative. Then, for any treatment vector \( \boldsymbol{D}_{(-i)} \in \mathbb{R}^K\) with \(i\)-th component equals zero, the marginal effect of activating the \(i\)-th treatment preserves the same sign; that is,
\[
Y(\boldsymbol{D}_{(-i)} + e_i) - Y(\boldsymbol{D}_{(-i)} + e_0)
\]
has the same sign as \( Y(e_i) - Y(e_0) \).

Condition 1 requires that treatment effects exhibit a consistent direction across different treatment configurations. It is straightforward to verify that both the linear additive model and the Generalized Sigmoid Form II DGP in \cite{ye2023deep} satisfy this condition.

Our key conjecture is that \our\ remains effective under nonlinear additive treatment effects as long as Condition 1 holds, since the method primarily relies on the relative ordering of treatment effects rather than strict linearity. However, when this condition is violated, i.e., when treatment interactions induce sign reversals, the performance of \our\ may deteriorate. We currently lack a formal proof of this claim and instead provide empirical evidence to support this intuition.

To illustrate, consider the following nonlinear data generating process (DGP):
\begin{equation}
    Y_i = a + \sum_{k = 1}^K\tau_kD_{i,k} - \sum_{k=1}^K\sum_{j = k+1}^K\tau_{k,j}D_{i,k}D_{i,j} + \epsilon_{i}.
\end{equation}
In this setting, pairwise interaction terms introduce nonlinearities. For example, when $K=2$, $\tau_1 > 0$, and $\tau_2 > 0$, Condition 1 holds if  \( \tau_{1,2} < \min(\tau_1, \tau_2) \). However, if the interaction term becomes large which \( \tau_{1,2} > \tau_2 \) or \( \tau_{1,2} > \tau_1 \), the marginal effect of a treatment may change sign depending on the presence of other treatments, thereby violating Condition 1. Thus, in the following experiment, we will fix the distribution of $\tau_1,\cdots,\tau_K$ and change the distribution of the interaction term $\tau_{k,j}$. 

In this numerical experiment, we fix $K=5, N =  10$, draw $\tau_{1}, \ldots, \tau_{5} \sim \mathcal{N}(1,3^2)$, and let interaction terms $\tau_{k,j} \sim \mathcal{N}(\tau^{\text{cross}}, 1)$, where $\tau^{\text{cross}}$ controls the strength of nonlinearity. The noise term satisfies $\epsilon_i \sim \mathcal{N}(0,5^2)$. We compare three methods: \our, \tradition, and a Bayesian benchmark. For each value of $\tau_{\text{cross}} \in [0,3]$, we repeat the experiment 1,000 times and report the average Optimality Ratio (OR) in Figure \ref{fig:case_study_nonlinear}.

\begin{figure}[htbp!]
    \centering
     \includegraphics[width=0.5\linewidth]{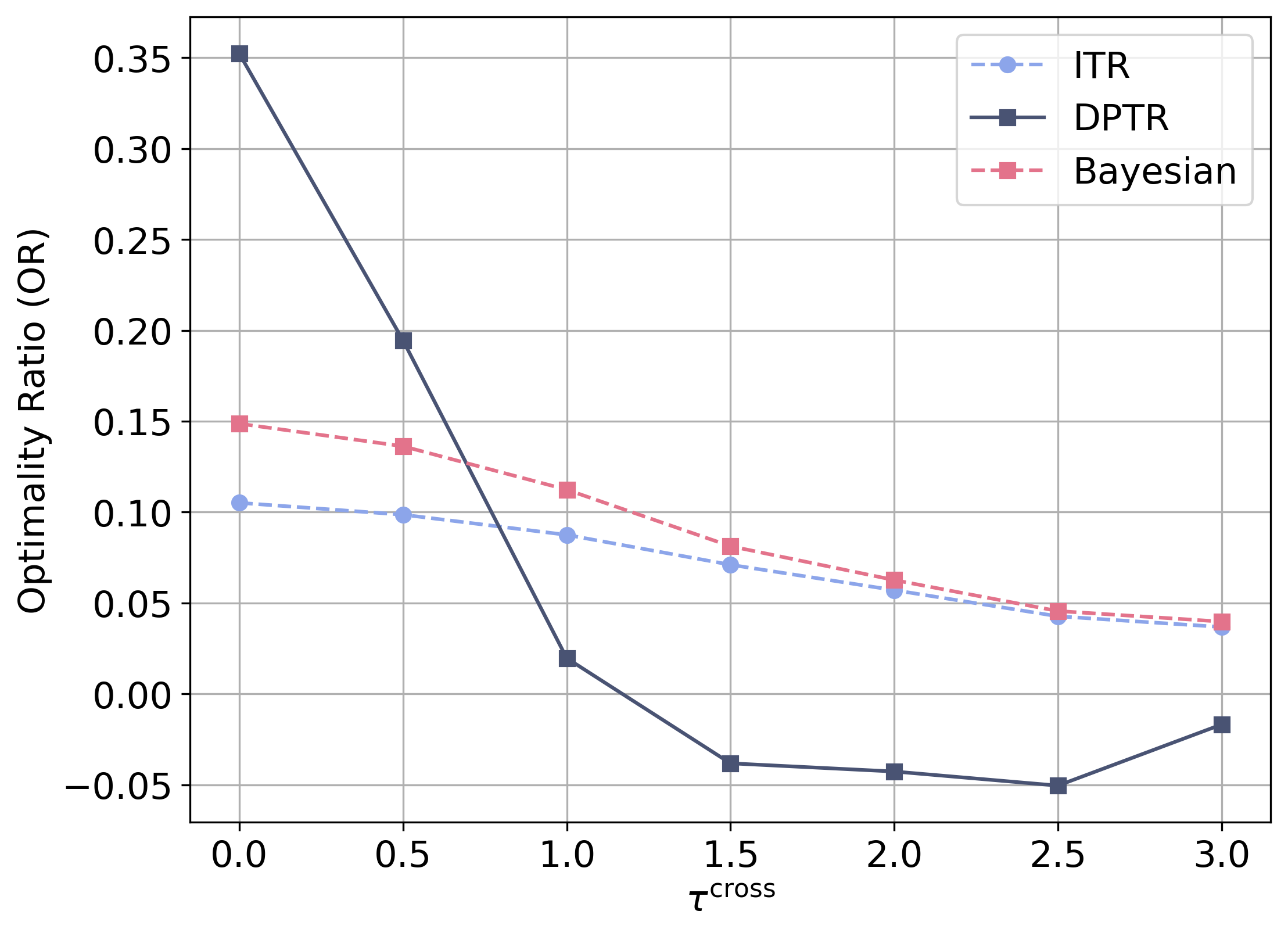}
    \caption{A simple numerical case study under nonlinear additive treatment effects.}
    \label{fig:case_study_nonlinear}
\end{figure}

As shown in Figure \ref{fig:case_study_nonlinear}, when $\tau_{\text{cross}} < 1$, \our\ consistently outperforms both \tradition\ and the Bayesian approach. In this regime, interaction effects are moderate and Condition 1 approximately holds. In contrast, when $\tau_{\text{cross}} \geq 1$, the performance of \our\ deteriorates, reflecting the increasing prevalence of sign reversals in marginal treatment effects.

In summary, although \our\ can be robust to certain forms of nonlinearity, its effectiveness critically depends on the consistency of treatment effect directions. This analysis highlights both the scope and the limitations of our approach: \our\ performs well when nonlinearities preserve the qualitative structure of treatment effects (i.e., satisfy Condition 1), but may fail when strong interactions fundamentally alter this structure.

\subsection{Roll-out Decisions under Capacity Constraints} \label{app:limit_capacity}
In this section, we discuss another limitation of our \our\ method. The theoretical and simulation results in the main text do not impose any capacity constraint on the number of experiments that can be rolled out. However, in practical online platforms, capacity constraints on the number of implemented experiments are also common. We therefore examine the performance of our method under capacity-constrained settings.

To better characterize performance under such constraints, we have added a numerical experiment. Specifically, we consider a baseline setting with $ K = 1000, N = 10, \tau_k \sim \mathcal{N}(1, 3^2)$, and $\epsilon_{k,i} \sim \mathcal{N}(0, 3^2)$. In addition, we introduce a constraint ratio $\omega \in (0,1)$, such that at most $\omega K$ experiments can be deployed. For each method, we first identify the set of experiments selected in the unconstrained case, then rank them by estimated effects and select the top subset subject to the constraint. We vary $\omega$ from 0.5 to 0.9 and repeat each configuration 1,000 times. The results (Figure \ref{fig:per_with_con}) show that \our\ underperforms benchmark methods when the constraint is stringent, but performs competitively when the constraint is moderate. This result indicates that our method is not universally optimal across all scenarios; in particular, it may fail to deliver superior rewards under tight capacity constraints. However, the severity of the capacity constraint is typically known prior to decision-making, allowing practitioners to readily assess whether the setting is highly restrictive. Therefore, even in the presence of capacity constraints, our method can still serve as a viable alternative, offering the potential to improve overall rewards when the constraint is not overly stringent.

\begin{figure}[htbp!]
    \centering
     \includegraphics[width=0.5\linewidth]{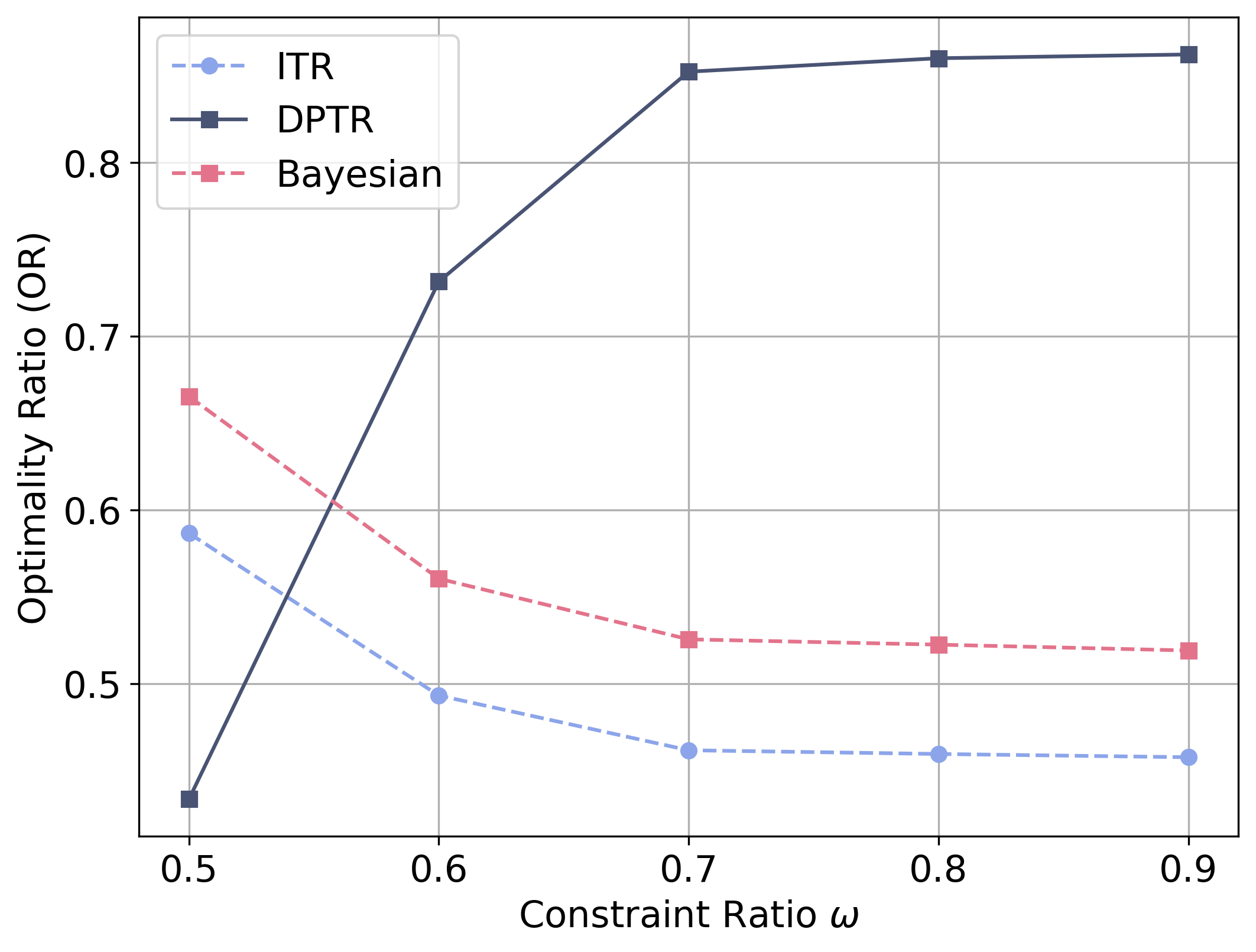}
    \caption{Performance comparison of methods with constraints.}
    \label{fig:per_with_con}
\end{figure}

\section{Other Synthetic Experiments}

\subsection{Variance Analysis of $\hat{\tau}_0$ under Overlapping Experiments}

In non-overlapping experiments, it is straightforward to see that the variance of $\hat{\tau}_0$ decreases as the number of experiments $K$ increases, since observations across experiments are independent. In contrast, under overlapping experiments, the estimators $\hat{\tau}_k$ and $\hat{\tau}_0$ are constructed from the same underlying data, making this variance reduction less immediate. In this section, we provide a detailed analysis showing that, even in the overlapping setting, the variance of $\hat{\tau}_0$ still decreases with $K$, thereby supporting the effectiveness of our method.

In the following analysis, we focus on the setting described in Section \ref{subsec:scenario-3} and adopt all notations introduced therein. For any experiment $k$, the variance of $\hat{\tau}_k$ is given by $\mathbb{V}(\hat{\tau}_k|\mathcal{T}) = I_k^{\top} \sigma^2 (\mathcal{T}^{\top} \mathcal{T})^{-1}I_k$ which corresponds to the $(k+1)$-th diagonal element of the matrix $\sigma^2 (\mathcal{T}^{\top} \mathcal{T})^{-1}$. Then, the anchor estimator is $\hat{\tau}_0 = \frac{1}{K}\sum_{k\in [K]}\hat{\tau}_k = \frac{1}{K} \boldsymbol{I}^{\top}\hat{\boldsymbol{\tau}}$ where $\boldsymbol{I}$ is the vector where all elements are one except the first component. Thus, the variance can be:
    \begin{equation}
        \mathbb{V}(\hat{\tau}_0|\mathcal{T}) = \frac{\boldsymbol{I}^{\top} \sigma^2 (\mathcal{T}^{\top} \mathcal{T})^{-1}\boldsymbol{I}}{K^2} = \underbrace{\frac{\sigma^2\sum_{k\in[K]}(\mathcal{T}^{\top} \mathcal{T})^{-1}_{k+1,k+1}}{K^2}}_{\text{Part 1}} + \underbrace{\frac{\sigma^2\sum_{i\in [K]}\sum_{j\in [K],j\neq i}(\mathcal{T}^{\top} \mathcal{T})^{-1}_{i+1,j+1}}{K^2}}_{\text{Part 2}}.
    \end{equation}
For Part 1, we can readily see that its value decreases as the number of experiments $K$ increases, since the numerator is of order $K$ while the denominator is of order $K^2$. Before analyzing Part 2, we first present the following theorem, which we use as a heuristic guide: it applies to Gaussian designs, whereas our actual setting has Bernoulli treatment indicators. The formal theoretical guarantees in Section~\ref{sec:theoretical_valid} do not rely on the calculation below.
\begin{theorem} \label{theorem:gaussian}
Let $\hat X \in \mathbb{R}^{n \times p}$ be a random matrix whose rows are i.i.d.
\[
\hat x_i \sim \mathcal{N}(0,\Sigma),
\]
where $\Sigma \in \mathbb{R}^{p \times p}$ is a diagonal covariance matrix. If $n > p+1$, then
\[
\mathbb{E}\big[(\hat X^\top \hat X)^{-1}\big]
=
\frac{1}{n - p - 1}\,\Sigma^{-1}.
\]
In particular, $\mathbb{E}[(\hat X^\top \hat X)^{-1}]$ is a diagonal matrix.
\end{theorem}

\textit{Proof}:
Since the rows of $\hat X$ are i.i.d. Gaussian with distribution $\mathcal{N}(0,\Sigma)$, it follows that
\[
\hat X^\top \hat X \sim W_p(n,\Sigma),
\]
where $W_p(n,\Sigma)$ denotes the $p$-dimensional Wishart distribution with $n$ degrees of freedom and scale matrix $\Sigma$. A standard result for the Wishart distribution states that if $W \sim W_p(n,\Sigma)$ and $n > p+1$, then
\[
\mathbb{E}[W^{-1}] = \frac{1}{n - p - 1}\,\Sigma^{-1}.
\]
Applying this result with $W = \hat X^\top \hat X$, we obtain
\[
\mathbb{E}\big[(\hat X^\top \hat X)^{-1}\big]
=
\frac{1}{n - p - 1}\,\Sigma^{-1}.
\]

Since $\Sigma$ is diagonal, its inverse $\Sigma^{-1}$ is also diagonal. Therefore, \( \mathbb{E}\big[(\hat X^\top \hat X)^{-1}\big] \) is a diagonal matrix. \QED \\

 In our setting, the treatment allocations across experiments are independent. Thus, the only difference between our setting and Theorem \ref{theorem:gaussian} is that the random variables follow a Bernoulli distribution rather than a Gaussian one, so Theorem~\ref{theorem:gaussian} does not directly apply. Guided heuristically by the intuition from Theorem \ref{theorem:gaussian}, the off-diagonal entries $(\mathcal{T}^{\top}\mathcal{T})^{-1}_{i+1,j+1}$ are expected to be small under approximately orthogonal Bernoulli designs, and we treat their contribution to Part 2 as negligible. The simulation evidence below confirms that Part 2 is approximately zero when $K$ is large.

Then the variance of $\hat{\tau}_0$ can be approximated as:
    \begin{equation} \label{eq:var_anchor_with_K}        \mathbb{V}(\hat{\tau}_0|\mathcal{T}) \approx \frac{\sigma^2\sum_{k\in[K]}(\mathcal{T}^{\top} \mathcal{T})^{-1}_{k+1,k+1} }{K^2} \le \frac{\max_{k\in [K]}(\mathcal{T}^{\top} \mathcal{T})^{-1}_{k+1,k+1}\sigma^2}{K}.
    \end{equation}
   In conclusion, even though the anchor estimate and the individual estimates are drawn from essentially the same pool of data, the anchor estimator attains a much smaller variance when the number of experiments \(K\) is large, and therefore remains highly useful and stable.

    \begin{figure}[htbp!]
    \centering
     \includegraphics[width=0.5\linewidth]{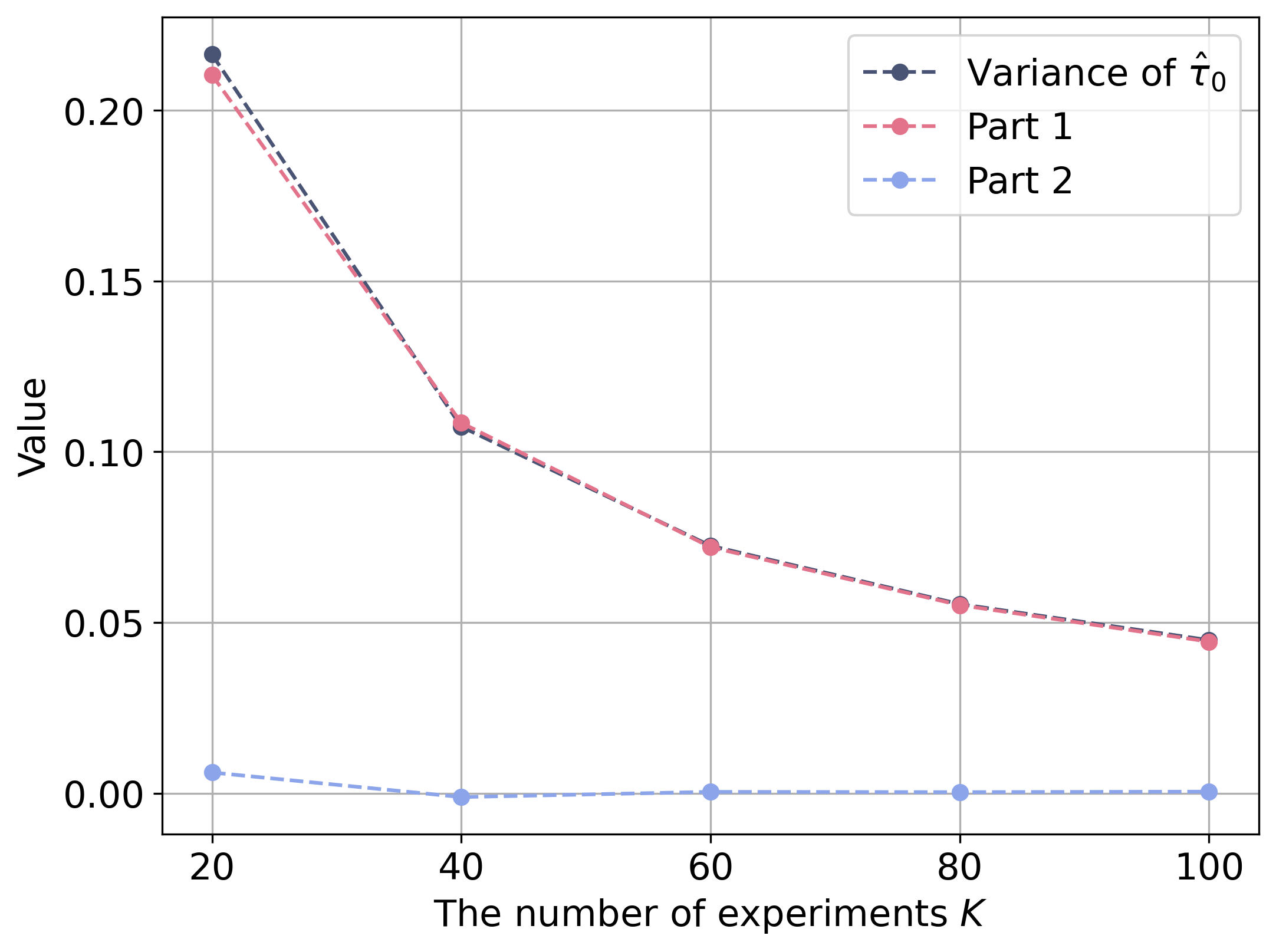}
    \caption{The variance analysis of $\hat{\tau}_0$ across 1,000 instances }
    \label{fig:variance_of_anchor}
\end{figure}

Next, we conduct a simple simulation to demonstrate the phenomenon described in Eqn.~\eqref{eq:var_anchor_with_K}, which shows that the variance of $\hat{\tau}_0$ decreases as the number of experiments $K $ increases. We adopt the same numerical setting as in Section 5.3, except that we vary the number of experiments \(K\) from 20 to 100 in increments of 20. For each value of \(K\), we repeat the simulation 1,000 times and report the resulting variance in Figure \ref{fig:variance_of_anchor}. As shown in Figure \ref{fig:variance_of_anchor}, consistent with our analysis, Part 2 is approximately zero, and the variance of $\hat{\tau}_0$ decreases as $K$ increases. This result further demonstrates that, although the estimation of $\hat{\tau}_k$ and $\hat{\tau}_0$ relies on the same dataset, the independence across experiments leads to a reduction in the variance of $\hat{\tau}_0$.

\subsection{Comparison with False Discovery Rate Control Method}
 In this section, we introduce an additional benchmark based on false discovery rate (FDR) control, namely the Benjamini–Hochberg procedure proposed by \cite{BenjaminiHochberg1995}. The motivation for including this benchmark stems from the structural similarity between our setting and the classical multiple testing framework: both involve simultaneously evaluating a collection of hypotheses and making selection decisions under uncertainty. To facilitate a direct comparison, we adopt the same numerical setting as in Section \ref{subsec:no-overlap-no-feature} and vary $\tau_0$ from 1 to 5.  For the BH procedure, we control the false discovery rate (FDR) at level $0.05$, consistent with the settings used for the other methods.  The results are reported in Figure \ref{fig:compare_with_BH}.

    \begin{figure}[htbp!]
    \centering
     \includegraphics[width=0.5\linewidth]{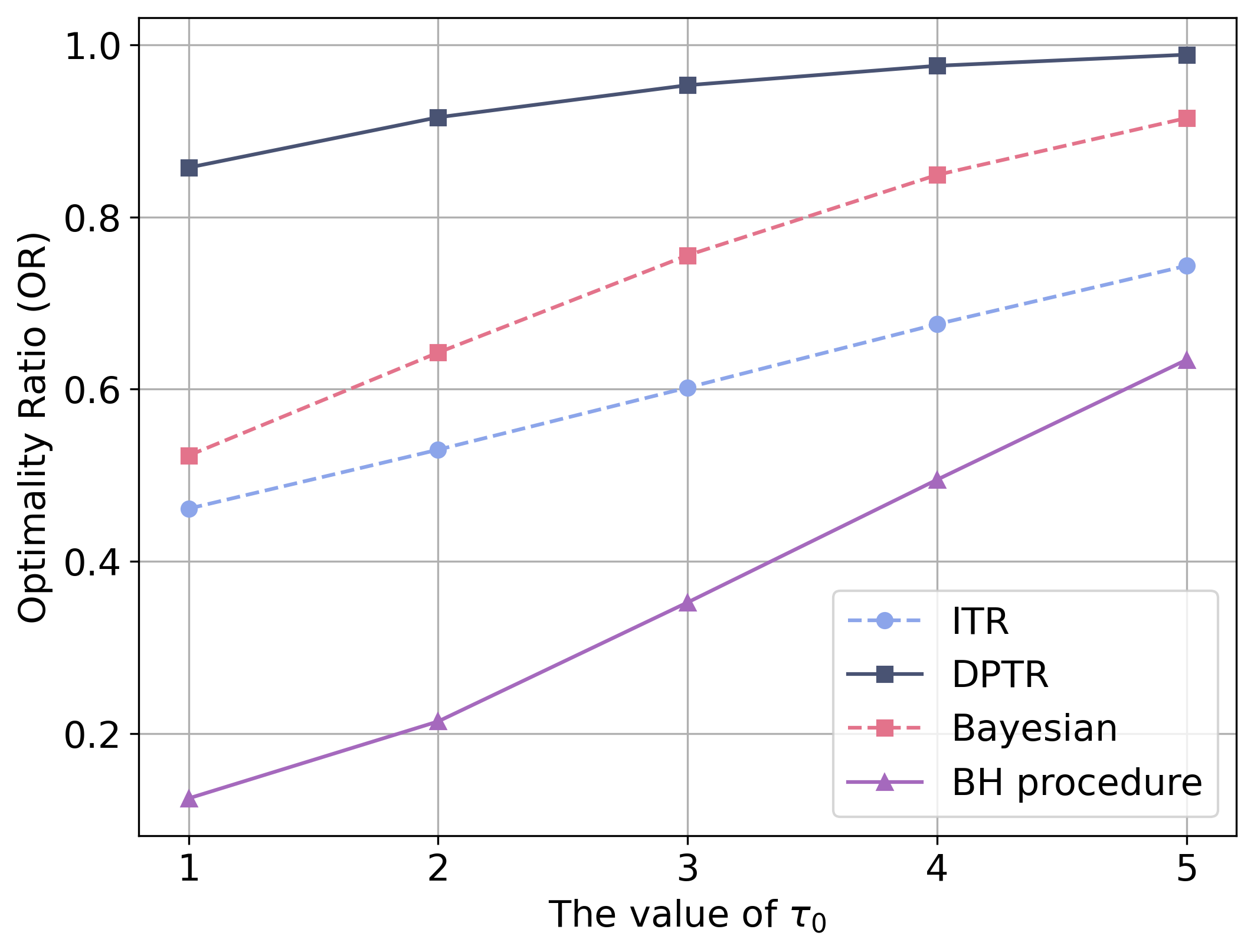}
    \caption{Performance comparisons for non-overlapping experiments without covariates.}
    \label{fig:compare_with_BH}
\end{figure}

It is important to emphasize, however, a fundamental distinction in problem formulation. The BH procedure is designed to control the expected false discovery proportion across a family of simultaneous hypothesis tests. Unlike \our, it does not pool effect estimates across experiments and does not optimize the downstream reward objective. In contrast, our setting focuses on aggregating information across multiple experiments, with the goal of maximizing downstream decision reward rather than controlling a statistical error rate. This difference in objectives leads to markedly different decision rules: FDR-based methods are inherently conservative, as they prioritize error control, whereas our method explicitly balances estimation accuracy and reward optimization.

Consistent with this distinction, Figure \ref{fig:compare_with_BH} shows that the BH procedure performs the worst among all considered methods, while our \our\ method consistently outperforms all benchmarks across all values of $\tau_0$. This outcome is intuitive. Because the BH procedure is not designed to optimize reward, it tends to reject fewer hypotheses, resulting in overly cautious decisions and consequently lower overall reward in our setting.


\end{APPENDIX}

\end{document}